\documentclass[iop]{emulateapj}

\usepackage{graphicx}
\usepackage{color}
\usepackage{epsfig,graphicx,latexsym,amsmath,amssymb}

\newcommand{\Msun}{\ensuremath{M_{\odot}}}

\newcommand{\Myr}{\ensuremath{\mathrm{Myr}}}
\newcommand{\nbody}{$N-$body}

\newcommand{\mstar}{\ensuremath{m_{\star}}}
\def\apj{{\it ApJ}}
\def\apjl{{\it ApJL}}
\def\mnras{{\it MNRAS}}

\def\nat{{\it Nature}}

\begin{document}

\title{Super massive black holes in star forming gaseous circumnuclear discs}

\author{L. del Valle$^{1}$\thanks{E-mail: ldelvalleb@gmail.com (LdV)}, A. Escala$^{1}$, C. Maureira-Fredes$^{2}$, J. Molina$^{1}$ J. Cuadra$^{3}$ \& P. Amaro-Seoane$^{2}$}

\affiliation{$^{1}$Departamento de Astronom\'ia, Universidad de Chile.\\
     $^{2}$Max Planck Institut fur Gravitationsphysik (Albert-Einstein-Institut), D-14476 Potsdam, Germany\\
     $^{3}$Instituto de Astrof\'isica, Pontificia Universidad Catolica de Chile
    }

\begin{abstract}
Using \nbody/SPH simulations we study the evolution of the separation of a pair
of SMBHs embedded in a star forming circumnuclear disk (CND\footnote{ Here we use the terminology of gaseous circumnuclear discs (CND) to make reference to massive gaseous disks with sizes of the order of one kilo parsec and not to the collection of gas and dust clouds in the galactic centre of the milky way.}).
This type of disk is expected to be formed in the central kilo parsec of the remnant of gas-rich galaxy mergers.
Our simulations indicate that orbital decay of the SMBHs occurs more quickly when the mean density of the CND is higher, due to increased dynamical friction.
However, in simulations where the CND is fragmented in high density gaseous clumps (clumpy CND), the orbits of the SMBHs are erratically perturbed by the
gravitational interaction with these clumps, delaying, in some cases, the orbital decay of the SMBHs.
The densities of these gaseous clumps in our simulations and in recent studies of clumpy CNDs are two orders of magnitude higher than the observed density 
of molecular clouds in isolated galaxies or ULIRGs, thus, we expect that SMBH orbits are perturbed less in real CNDs than in the simulated CNDs of this study and other recent studies. We also find that the migration timescale has a weak dependence on the star formation rate of the CND. Furthermore, the migration timescale of a SMBH pair in a star-forming clumpy CND is at most a factor three longer than the migration timescale of a pair of SMBHs in a CND modeled with more simple gas physics. Therefore, we estimate that the migration timescale of the SMBHs in a clumpy CND is on the order of $10^7$ yrs.\\

\vspace{11pt}
\end{abstract}



\maketitle

\section{Introduction}
\label{sec.motivation}

In the context of hierarchical structure formation
(White \& Frenk 1991; Springel et al. 2005b)
galaxies are sculpted by a sequence of mergers and accretion events.
In the early evolution of these mergers, the cores of each galaxy will sink to the
central region of the new system, until they coalesce forming a new virialized core.
During the migration of the cores, the massive black holes (MBHs) that are expected to live at
the central region of every galaxy (Richstone et al. 1998; Magorrian et al. 1998; Gultekin et al. 2009; Kormendy \& Ho 2013)
will follow the center of each core.\\

If the galaxies that are involved in this merger are rich in 
gas, numerical simulations show that 60 to 90 \% of this gas 
can fall to the central kilo parsec of the remnant 
(Barnes \& Hernquist 1996; Mihos \& Hernquist 1996; Barnes 2002; Mayer et al. 2007, 2010).
This is consistent with observations of gas-rich interacting galaxies,
where it is often found that the amount of gas contained in their central regions
is comparable with the total gas content of a large gas-rich galaxy
(Sanders \& Mirabel 1996; Downes \& Solomon 1998; Medling et al. 2014; Ueda et al. 2014)
Both numerical simulations and observations suggest that this gas often settles 
in a disk-like distribution, or circumnuclear disk (CND). 

Several authors have performed numerical simulations studying the evolution of MBHs after a galaxy merger.  
In these simulations, a pair of MBHs is embedded in a gaseous CND.   
Most studies indicate that the pair of MBHs dissipate angular momentum into the ambient gas, 
which drives the formation of an MBH binary and the subsequent orbital decay of this binary 
down to separations on the order of $\sim 1-0.1$ pc.  
The timescale for this process is on the order of $10^7$ {\Myr}
(Escala et al. 2005; Dotti et al. 2006; Fiacconi et al. 2013)
In some cases, the viscous torque of the gas is not strong enough to dissipate the angular 
momentum extracted from the MBH binary and the gas is forced to move away from the 
binary (del Valle \& Escala 2012, 2014).
In these cases, the MBH binary is left inside a cavity of low gas density, and 
dissipation of its angular momentum into its environment is less effective. 
As a result, the migration timescale of such systems can be longer than the Hubble 
time (Lodato et al. 2009; Cuadra et al. 2009).
In each of these studies, it is assumed that the gas follows a simple polytropic 
equation of state and that the CND evolves without star formation.
 
More recent simulations of the evolution of MBH pairs in galaxy mergers include 
the effects of star formation, supernovae, and cooling (Hopkins \& Quataert 2010; 
Callegariet al. 2011; Van Wassenhove et al. 2012; Roskar et al. 2014).
These studies explore how the morphology of the galaxies, their gas fraction, 
and the geometry of the merger can affect the accretion and orbital evolution of the MBH pair.  
However, in these studies, the star formation rates and supernovae feedback 
are fixed in order to reproduce observed relations, such as the Kennicutt-Schmidt 
relation (Kennicutt 1998).
Little attention has been given to how different efficiencies or intensities of 
these processes can affect the orbital evolution of an MBH pair after a galaxy merger. 

In this work we investigate the evolution of two MBHs embedded in a CND and, 
for the first time, we explore how different star formation rates can result in different migration timescales of the MBH pair. 
In section \S2 we describe the code used in our simulations, and how we model star formation, cooling, and supernovae heating. 
In section \S3 we show how the MBHs' separation evolves for different star formation rates. 
In section \S4 we compute the gravitational torque exerted by the gas on the SMBHs, and we show that 
high density gas, in the form of clumps, is the primary source of this torque.   
In section \S5 we compare the mass and density of these gas clumps to observations.  
We also discuss some features missing from our simulations that are crucial to properly 
modeling these clumps and obtaining densities in better agreement with observation.  
Finally, in section \S6 we summarize the main results of our work and their implications 
on the orbital evolution of MBHs after a galaxy merger. 

\section{Code and simulation setup}
\label{sec.simulation}

\subsection{Code}
To compute the evolution of the SMBHs embedded in the CND, we use the code 
{\tt{Gadget-3}}  (Springel et al. 2001; Springel 2005).
This code evolves the system by computing gravitational forces with a hierarchical tree algorithm, 
and it represents fluids by means of smoothed particle hydrodynamics (SPH; e.g.Monaghan1992).

The SPH technique has difficulty resolving some hydrodynamical instabilities, 
such as Kelvin-Helmholtz or Rayleigh-Taylor instability, because it generates spurious forces on 
particles that are in regions with steep gradients of density (Agertz et al.2007; MckNally, Lyra \& Passy 2012). 
A proper treatment of these instabilities is important to model processes such as 
star formation and the turbulence in the ISM. These instabilities are better resolved 
by codes that use an Eulerian grid based technique, such as AMR, however, in these codes
the orbits of massive particles experiences strong and spurious perturbations,
making massive particles to follow unphysical orbits. For this reason, Eulerian grid 
codes are not suitable to study the orbital evolution of SMBHs embedded in a gaseous 
CND (Lupi, Haardt \& Dotti 2014 and reference therein) and we choose to use SPH rather 
than a Eulerian grid based code.


To study how the formation of stars in a gaseous circumnuclear disk (CND) can 
affect the orbital evolution of a pair of SMBHs embedded in the CND, we created recipes for star 
formation, gas cooling, and heating due to supernovae and implemented these recipes in the {\tt{Gadget-3}} code.  
Our recipes have certain key differences from those included in {\tt{Gadget-3}}.   
In particular, in our recipes the gas can reach lower temperatures, star formation does 
not depend exclusively on gas density, and we do not assume the stellar feedback to be instantaneous.  
The recipes we implement resemble those of Stinson et al. (2006) 
used in the SPH code Gasoline (Wadsley et al. 2004), and those of Saitoh et al. (2008, 2010) 
used in simulations of isolated galaxies and galaxy mergers. 

In our implementation, the cooling function is computed for an optically thin gas 
with solar metallicity. 
For temperatures between $10^{4.8}$ and $10^8$ K we compute the cooling essentially 
as described by  Katz et al. (1996), and for temperature below $10^{4.8}$ K we extend the 
cooling functions down to $10$ K using the parametrization of (Gerritsen \& Icke 1997). 

To model star formation, we examine every SPH gas particle and select those that, 
for some time step $\Delta t$, satisfy the following three criteria:

\begin{align}
    \centering
    n_{\rm H}\ &>\ n_{\rm min} \\
    \vec{\nabla}\,\cdot\, \vec{v}\ &<\ 0 \\
    T\ &<\ T_{\rm max},
\end{align}

\noindent where is $n_{\rm H}$ the number density, $T$ the temperature, and $\vec{v}$ the velocity.  
$n_{\rm min}$ and $T_{\rm max}$ are parameters fixed before each simulation (they are described in section \S2.2.).  
The gas particles selected in this way are treat as candidates for star formation.   

For each gas particle that satisfies these criteria, we compute the probability $p$ of giving 
birth to a star of mass $m_{\star}$ (this mass is fixed throughout the simulation) as 

\begin{equation}
    p = \frac{m_{\rm gas}}{m_\star}\left(1-e^{-\frac{C_{\star}\Delta t}{t_{\rm form}}}\right),
    \label{probability}
\end{equation}

\noindent where $m_{\rm gas}$ is the mass of the gas particle, 
$C_{\star}$ is the constant that controls the star formation efficiency, 
and $t_{\rm form}$ is the star formation timescale, computed as the maximum of the cooling time
$t_{\rm cool}=\rho\,\epsilon\,/n_{H}\,\Lambda(T)$ 
and the dynamical time $t_{\rm dyn}=({\rm G}\,\rho)^{-1/2}$. 
Finally, for each star-formation-eligible gas particle, 
we draw a random number $r$ between zero and one. 
If $r < p$ a new star particle of mass $m_{\star}$ is spawned and the new mass of the parent gas particle is computed as: 
$m_{\rm{gas,old}}-m_{\star}$. 

We assume that every newly formed star particle is a Single Stellar Population (SSP) 
with a three piece power law initial mass function (IMF) as defined in Miller \& Scalo (1979).
From this IMF, and the parametrization of Raiteri et al. (1996) for stellar lifetimes, 
we compute the number of type II supernovae (SNII) events that occur in each SSP as a function of time. 
Then we assume that for each of these SNII events the star particle deposits $10^{51}$ 
ergs of energy into the closest 32 gas particles. 
We do not include any heating by supernovae type Ia, because the typical time of integration 
of our simulations is $30$ {\Myr} and the SNIa timescale is on the order of $1$ Gyr. 

\subsection{Simulation Setup}

In order to model the pair of SMBHs embedded in a CND, we use the same initial conditions as Escala et al. (2005).
In these initial conditions, the ratio between the mass of gas and the mass of stars in the 
CND is consistent with the mean value obtained from observations of the nuclear region of ULIRGs 
(Downes \& Solomon 1998; Medling et al. 2014; Ueda et al. 2014).
However, the radial extent of the CND is comparable to the radial extent of the smallest and more concentrated observed nuclear disks. 

Initially, the CND follows a Mestel superficial density profile and has a mass 
$M_{\rm disk}=5\times10^{9}\,{\Msun}$.
The radius of the CND is $R_{\rm disc}\,=\,400$ pc and 
its thickness is $H_{\rm disk}\,=\, 40$ pc. 
The CND is modeled with $235331$ SPH particles, each with a gravitational softening of $4$ pc and a mass of $2.13\times10^{4}\,{\Msun}$. 

The stellar component is initially distributed in a spherical bulge.  
It follows a Plummer density profile having core radius $200$ pc and mass within $r =\, 400$ pc $M_{\rm bulge}(r<400\,{\rm pc})=5\,M_{\rm disc}$. This stellar 
bulge is modeled with $10^5$ collisionless particles with a gravitational softening of $4$ pc and a mass of $2.45\times10^{5}\,{\Msun}$.

The black hole pair is modeled by two collisionless particles of mass $M_{\rm BH}=5\times 10^7 {\Msun}$. 
These particles are initially symmetric about the center of the disk, in circular orbits of radius $200$ pc.  
The orbital plane is the plane of the disk.

For our recipe of star formation we need to set four free parameters:
(1) the star formation efficiency $C_{\star}$,
(2) the minimum density of gas required for star formation to occur $n_{\rm min}$,
(3) the maximum temperature a gas particle that can give birth to a star may have $T_{\rm max}$, and
(4) the mass of the newly formed star {\mstar}.

\begin{figure}
    \centering
    \includegraphics[width=0.5\textwidth]{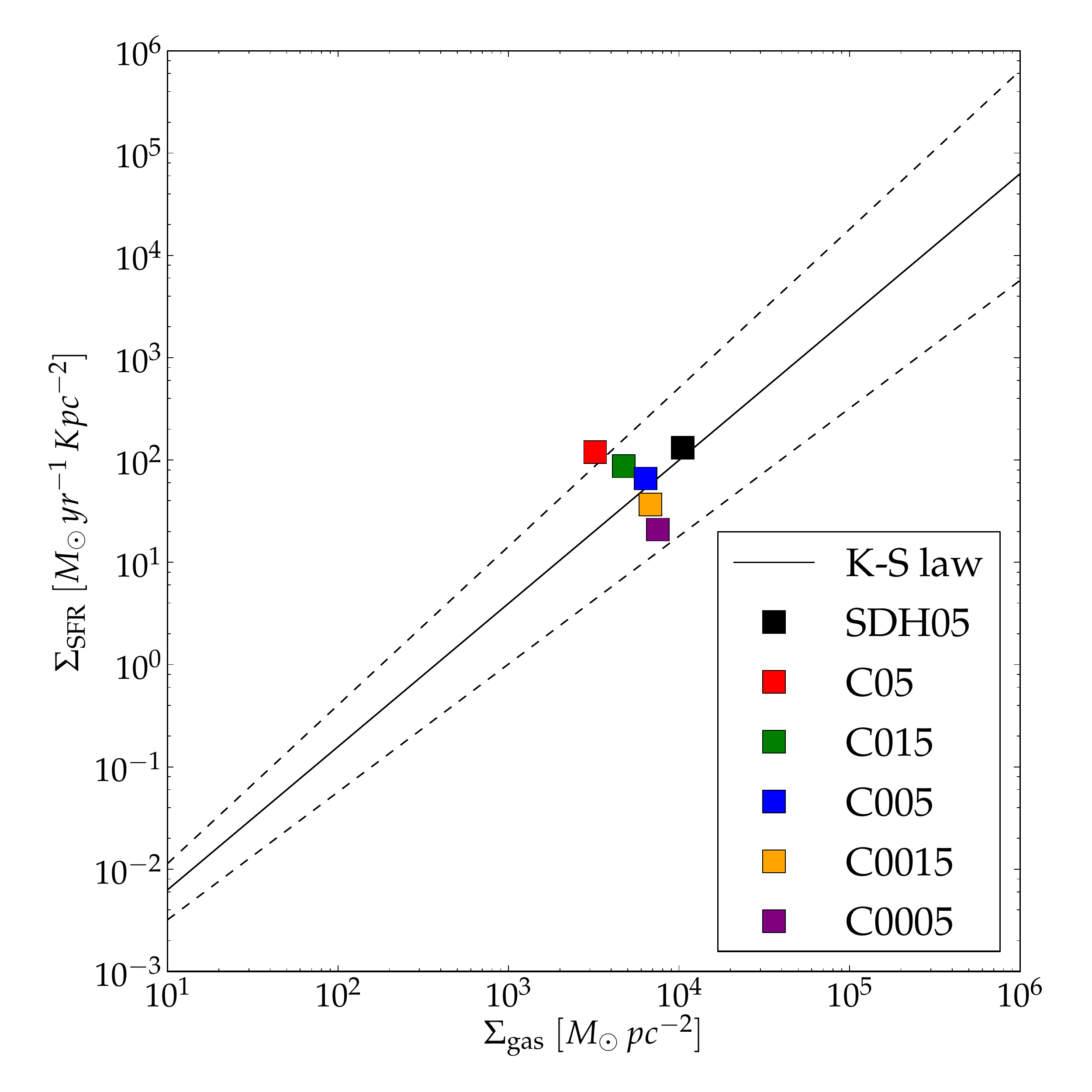}
    \caption{
      Correlation between disk-averaged SFR per unit area and
        average gas surface density.
        The continuous line corresponds to the best fit of the correlation for 61
        normal spiral galaxies and 36 infrared-selected star-bust galaxies obtained
        by (Kennicutt 1998).
        The dashed lines delimit the scatter of the observational relation obtained
        by (Kennicutt 1998).
    }
    \label{fig1}
\end{figure}

\begin{figure}
    \centering
    \includegraphics[width=0.5\textwidth]{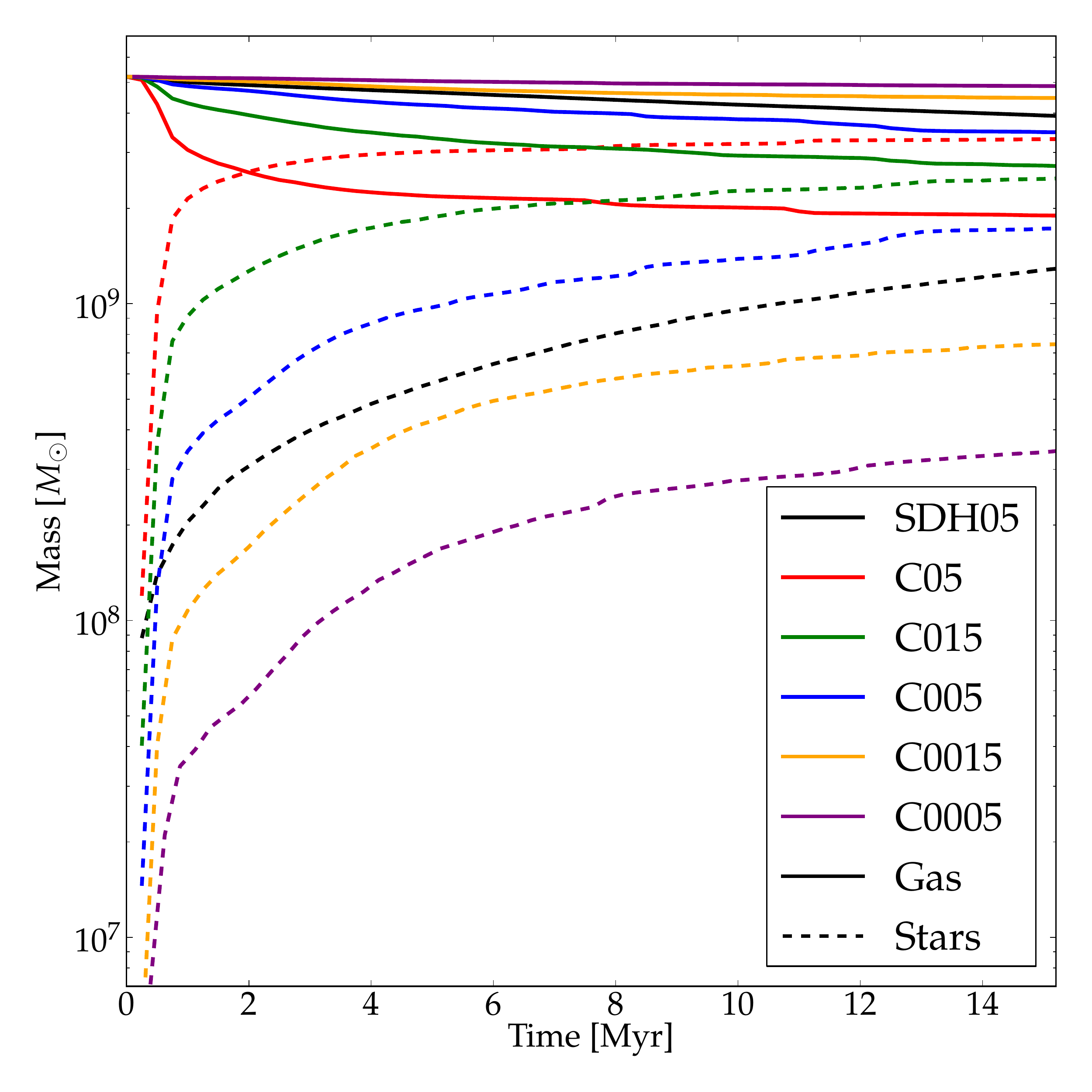}
    \caption{Time evolution of the mass of gas (continuous curves) and stars (dashed curves) for six
        simulations.
	The black lines corresponds to a simulation using {\tt{Gadget-3}}'s hybrid multiphase model for star formation. 
	The colored continuous lines correspond to simulations using our implementation of star formation, 
	feedback and cooling for five different values of the star formation efficiency $C_\star$ that appears in equation 
	\ref{probability}.       
    }
    \label{fig2}
\end{figure}

In order to resolve the formation of a clumpy multiphase medium, 
we assume a high star formation density threshold $n_{\rm min}=10^4\, {\rm cm^{-3}}$, 
and a low star formation temperature threshold $T_{\rm max}=10^3$ K (Stinson et al. 2006; Ceverino \& Klypin 2009). 
We set the mass of the star particles as half of the original mass of the gas particles. 
Only the parameter $C_{\star}$ is varied from between simulations to obtain different star formations rates. 
However, we restrict our selection of $C_{\star}$ to values that reproduce  an average star 
formation rate that is consistent with the empirical Kennicutt-Schmidt relation (Kennicutt 1998).
In figure \ref{fig1}, we show the surface gas density and star formation rate (SFR) obtained for 
different values of $C_{\star}$, and compare this to the observed relation.  
The values of the parameter $C_{\star}$ that we use are $0.005$, $0.015$, $0.05$, $0.15$, and $0.5$ 
(in table \ref{table1} these are the runs C0005, C0015, C005, C015 and C05
respectively). We also plot \ref{fig2} the time evolution of the gas mass of the disk
and the mass on new stars for the five different values of $C_{\star}$.

In order to see how the resolution of the gravitational forces affects the simulations, 
we ran ten simulations using lower gravitational softening for the black holes.  
In five of these simulations we set $\epsilon_{\rm BH}\,=\,0.04$ pc
and in the other five we set $\epsilon_{\rm BH}\,=\,0.004$ pc.
We label these runs with the suffixes ``\_$\epsilon$.04'' and ``\_$\epsilon$.004'', respectively.

We also ran two additional simulations to compare orbital evolution using our star formation recipe to orbital evolution with different recipes. 
In the first additional simulation, we used the hybrid multiphase model for star formation implemented in \texttt{Gadget-3} (Springel \& Hernquist 2003;
Springel et al. 2005a). 
This model assumes that the star formation is self-regulated and is parameterized only by the star formation timescale $t_{0}^*$. 
We use the typical value for this parameter, $t_{0}^*\,=\,$2.1, 
which provides a good fit to the Kennicutt law (see figure \ref{fig1}).  
The other simulation (run E05) uses more idealized gas physics.  
There is no star formation and the gas follows an adiabatic equation of state. 
This simulation corresponds to the \texttt{run A} of (Escala et al. 2005).
 
All the simulations that we run are summarized in table \ref{table1}.

\begin{table}
    \centering
    \caption{List of simulations and their parameters}
    \scriptsize
    \begin{tabular}{lllrrr}
    \hline
    {\bf Label} &
    {\bf Code} &
    {\bf SF} &
    {\bf $T_{\rm f}$ [K]} &
    {\bf $C_{\star}$} &
    {\bf $\epsilon_{\rm BH}$ [pc]} \\ \hline

    C05      & G3 Mod.      & yes & 25     & 0.5   & 4     \\
    C015     & G3 Mod.      & yes & 25     & 0.15  & 4     \\
    C005     & G3 Mod.      & yes & 25     & 0.05  & 4     \\
    C0015    & G3 Mod.      & yes & 25     & 0.015 & 4     \\
    C0005    & G3 Mod.      & yes & 25     & 0.005 & 4     \\
    \\
    C05\_$\epsilon$.04   & G3 Mod.      & yes & 25     & 0.5   & 0.04  \\
    C015\_$\epsilon$.04  & G3 Mod.      & yes & 25     & 0.15  & 0.04  \\
    C005\_$\epsilon$.04  & G3 Mod.      & yes & 25     & 0.05  & 0.04  \\
    C0015\_$\epsilon$.04 & G3 Mod.      & yes & 25     & 0.015 & 0.04  \\
    C0005\_$\epsilon$.04 & G3 Mod.      & yes & 25     & 0.005 & 0.04  \\
    \\
    C05\_$\epsilon$.004   & G3 Mod.      & yes & 25     & 0.5   & 0.004 \\
    C015\_$\epsilon$.004  & G3 Mod.      & yes & 25     & 0.15  & 0.004 \\
    C005\_$\epsilon$.004  & G3 Mod.      & yes & 25     & 0.05  & 0.004 \\
    C0015\_$\epsilon$.004 & G3 Mod.      & yes & 25     & 0.015 & 0.004 \\
    C0005\_$\epsilon$.004 & G3 Mod.      & yes & 25     & 0.005 & 0.004 \\
    SDH05    & G3           & yes & $10^4$ & 0.05  & 4     \\
    E05      & G1           & no  & -      & -     & 4     \\
    \hline
    \end{tabular}
    \label{table1}
\end{table}

\newpage

\section{Evolution of the SMBHs separation}
\subsection{The evolution in our simulations}

\begin{figure}
    \centering
    \includegraphics[width=0.5\textwidth]{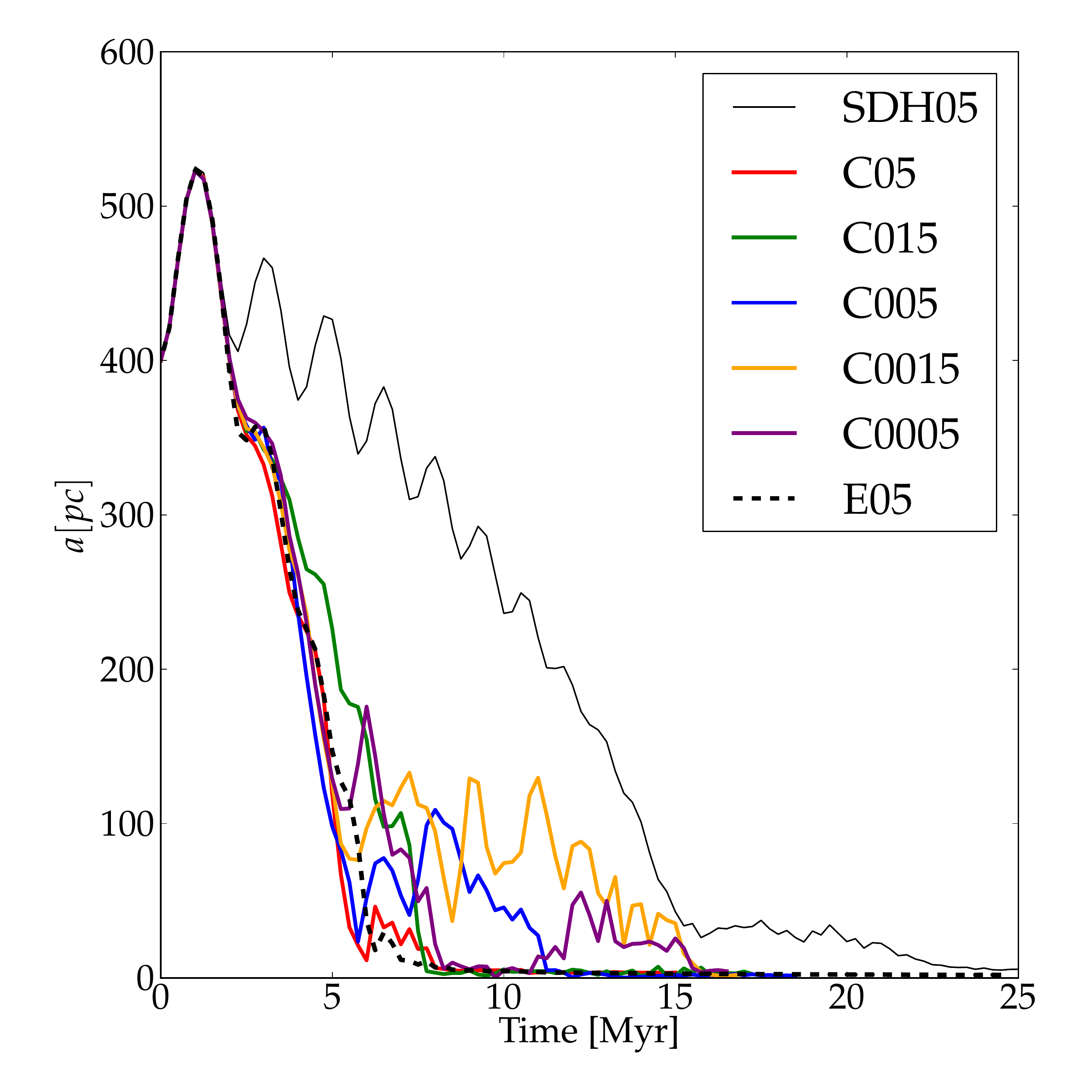}
    \caption{Time evolution of the SMBHs' separation for seven different
        simulations.
	The black continuous line corresponds to a simulation using {\tt{Gadget-3}}'s hybrid multiphase model for star formation. 
	The black dashed line corresponds to the \texttt{run A} of the idealized simulations of Escala et al. (2005).  
	The colored continuous lines correspond to simulations using our implementation of star formation, 
	feedback and cooling for five different values of the star formation efficiency $C_\star$ that appears in equation 
	\ref{probability}.       
    }
    \label{fig3}
\end{figure}

In figure \ref{fig3} we show the time evolution of the SMBHs' separation $a$ for seven simulations. 
Five of these runs use our prescription for star formation, feedback and cooling. 
The values of the star formation efficiency $C_{\star}$ (see equation \ref{probability}) used in these runs are 
$0.005$, $0.015$, $0.05$, $0.15$ and $0.5$, which in figure \ref{fig3} correspond to the purple, 
yellow, blue, green, and red continuous lines respectively. 
The black continuous line corresponds to a simulation using {\tt{Gadget-3}}'s hybrid multiphase 
model for star formation (SDH05).  
The black dashed line corresponds to \texttt{run A} of Escala et al. (2005) (E05), which uses 
idealized gas physics and does not consider star formation.

From figure \ref{fig3} we see that the time it takes for the SMBHs to reach a 
separation comparable to the gravitational softening is in the range of 7 to 25 {\Myr}.  
The fastest migration time corresponds to the SMBHs in the E05 run (black dashed line) 
and the slowest migration corresponds to the SMBHs in the SDH05 run (black continuous line). 
In the simulations with our prescriptions, the SMBHs reach
this separation at a time between these two limits.

\begin{figure}
    \centering
    \includegraphics[width=0.5\textwidth]{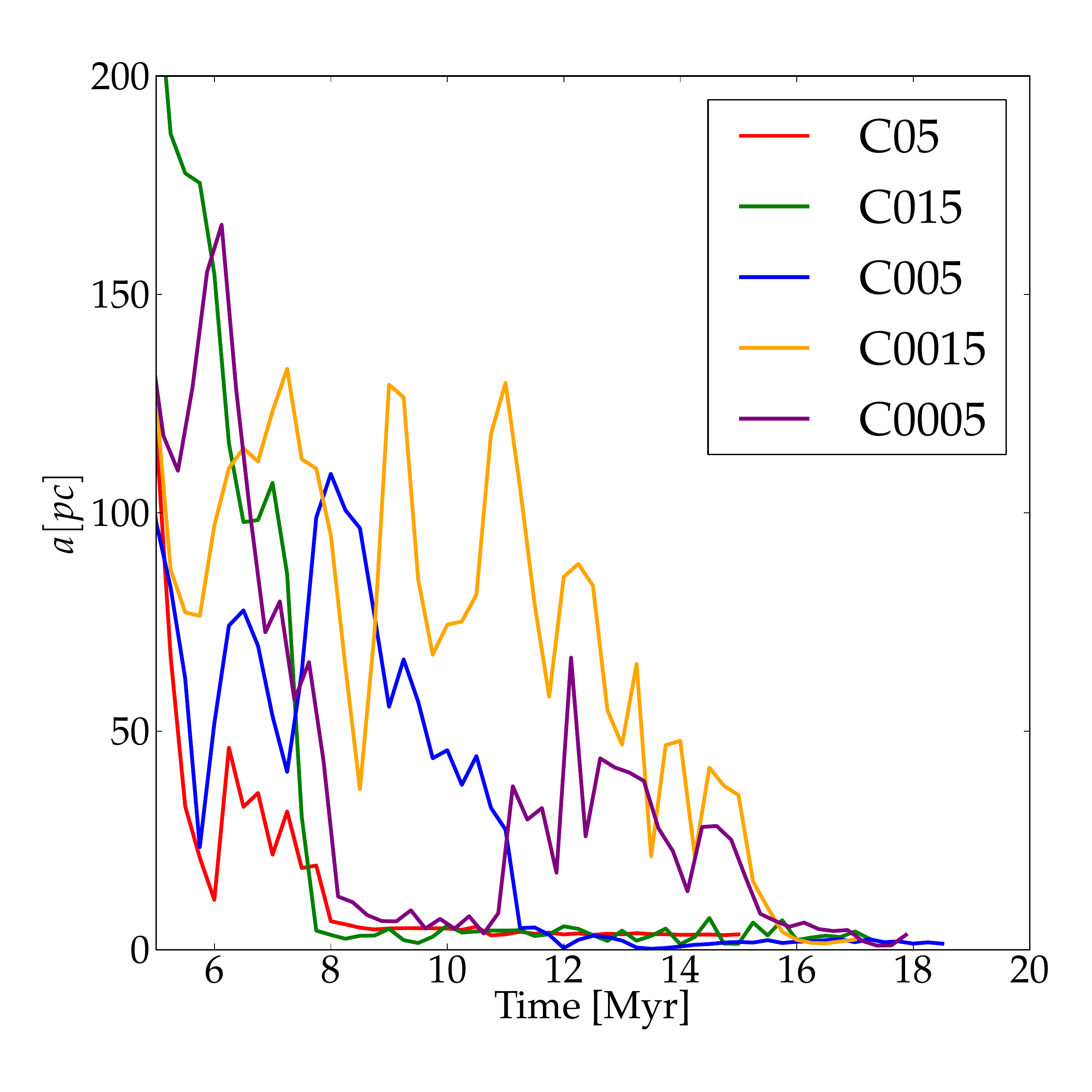}
    \caption{Zoom of figure 2. In this figure it can be seen that the time it takes for the SMBHs to stabilize
	at a separation comparable to the gravitational softening is shorter for higher values of $C_{\star}$.
    }
    \label{fig4}
\end{figure}

In figure \ref{fig4} we show a zoom of figure \ref{fig3} for times larger than 5 {\Myr}. 
Here we see that the time it takes for the SMBHs to stabilize at a separation
comparable with the gravitational softening is shorter for higher values of $C_{\star}$. 
These times are $8$, $7.9$, $11.6$, $16$, and $17.2$ {\Myr} for runs C0005, C0015, C005, 
C015, and C05 respectively (see table \ref{table1}).
Also we found that in the two simulations with highest star formation efficiency, there 
are no large fluctuations in the SMBHs separation after $t=$ 8 {\Myr}.

From figures \ref{fig3} and \ref{fig4}, we concluded that orbital decay of the SMBH pair 
occurs over a timescale at most $\sim2.4$ times longer in simulations using our recipes than 
in simulations using more idealized gas physics (E05).

It is important to note that in our simulations the orbital decay timescale varies by a factor of $\sim2.1$ 
while the star formation efficiency extends over two orders of magnitude.  
So the orbital decay timescale in our simulations has a weak dependence on star formation efficiency.

The effect produced by $C_{\star}$ on the orbital decay of the SMBH pair 
should be expected because the orbital evolution of the pair depends on the torques produced by the gas, 
and the structure of the gas is sculpted by the star formation, cooling, and SNII heating.  
In the next section we explore this issue, computing the gas distribution of the CND for different star formation efficiencies.

\subsection{Effect of the gas distribution on the orbital decay}

\begin{figure}
    \centering
\includegraphics[width=0.5\textwidth]{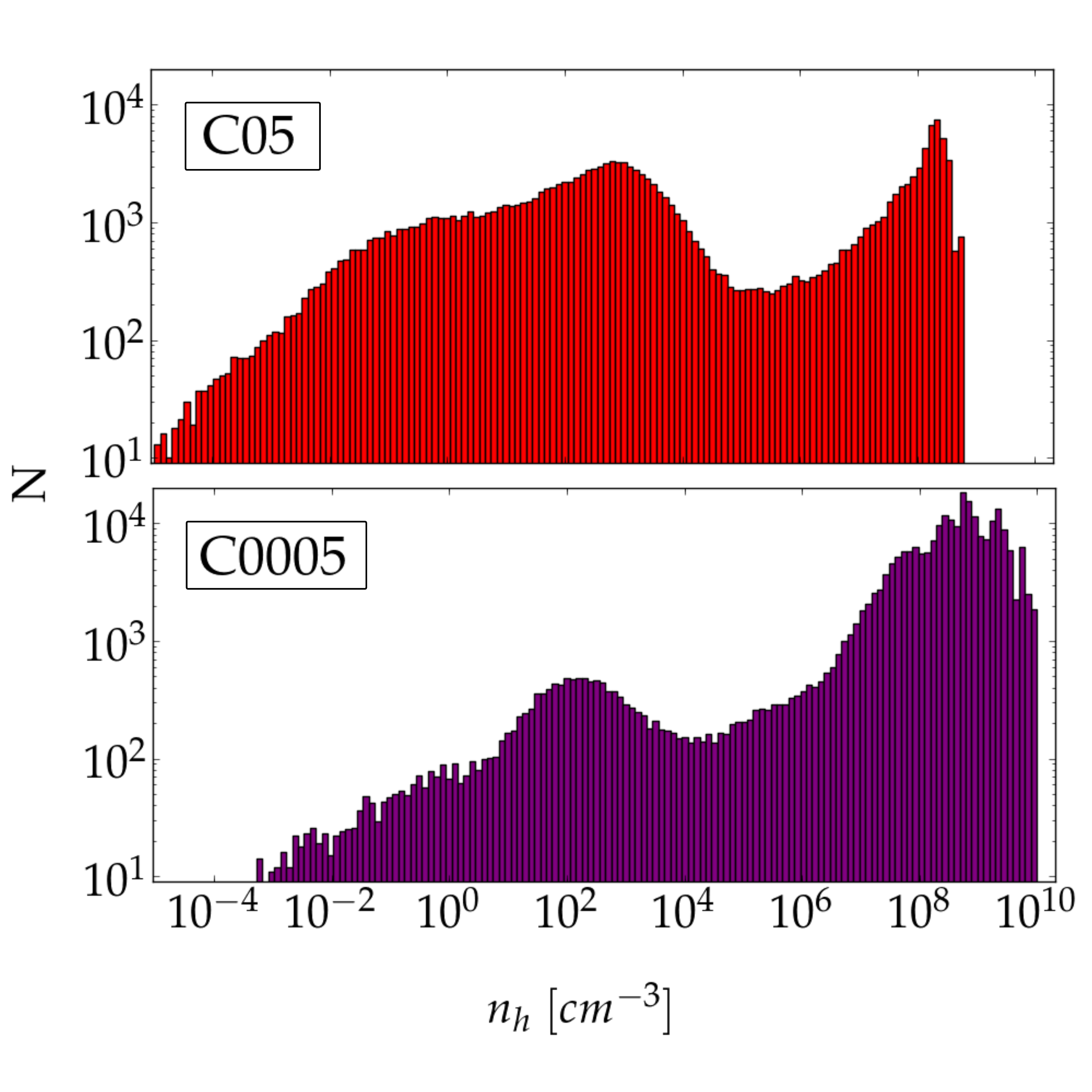}
\caption{Gas density histogram at $t =$ 10 {\Myr} for two different values of
	$C_{\star}$. Note that for higher values of $C_{\star}$
	the gas shows a more prominent low density phase. Also note that
	the maximum density of the gas is lower for higher values of $C_{\star}$. }
\label{fig5}
\end{figure}

In figure \ref{fig5} we show the density distribution of the gas at $t =$ 10 {\Myr} 
for two different values of $C_{\star}$. 
From this figure, we find that for higher values of $C_{\star}$ a greater proportion 
of the gas has a low density and the maximum gas density is lower.  
This is because with more vigorous star formation, high density gas is more efficiently 
converted into stars.  
With a greater number of stars, there is a greater number of supernovae explosions, 
which heat the cold dense gas around the stars and drive the formation of a hot diffuse medium.

\begin{figure}
    \centering
    \includegraphics[width=0.5\textwidth]{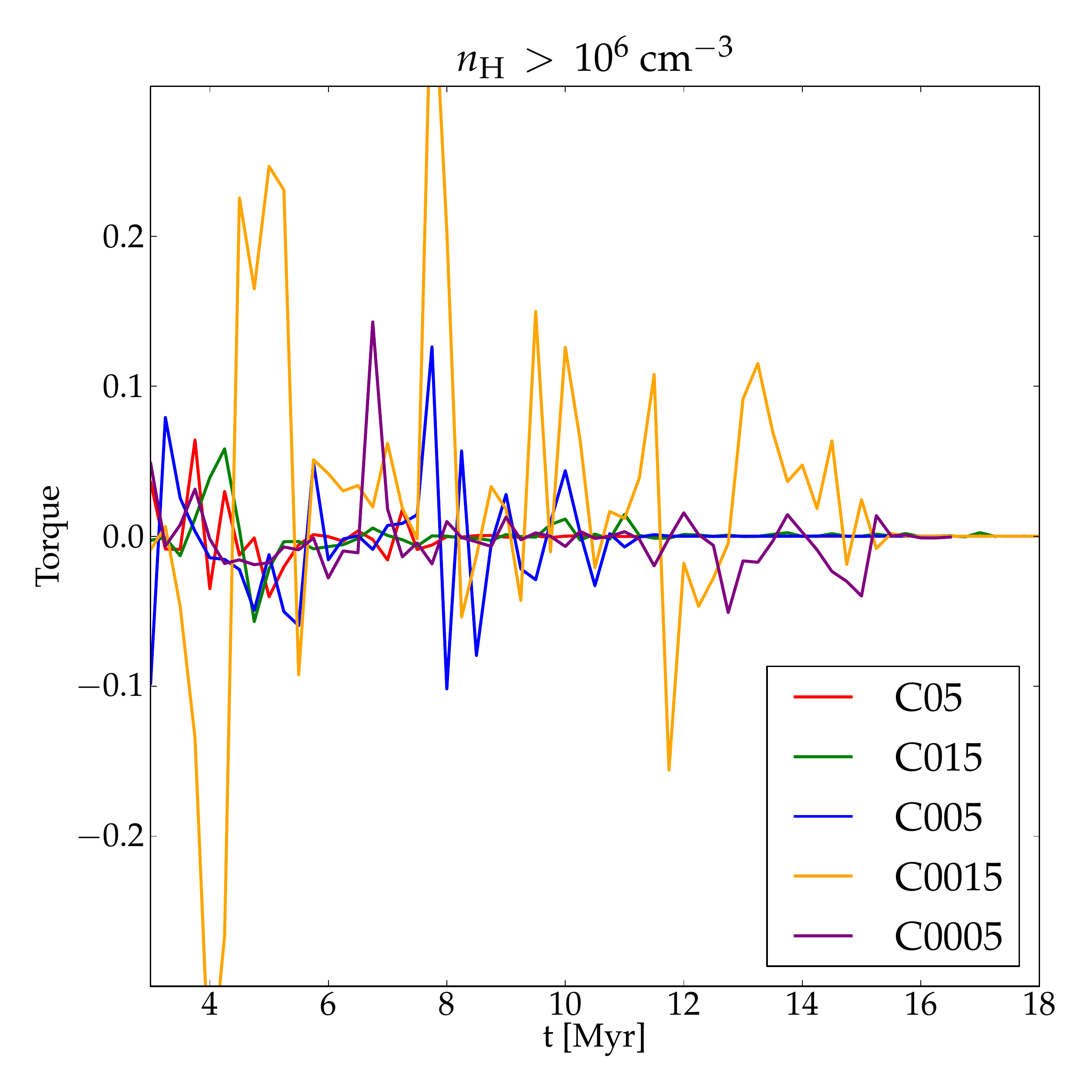}
    \caption{Torque experienced by the SMBHs due to gas with density above $10^6$ cm$^{-3}$ for different values of $C_{\star}$. 
	From this figure we see that practically all the gravitational torque is produced
	by gas with density higher than $10^6$ cm$^{-3}$. }
    \label{fig6}
\end{figure}

\begin{figure}
    \centering
    \includegraphics[width=0.5\textwidth]{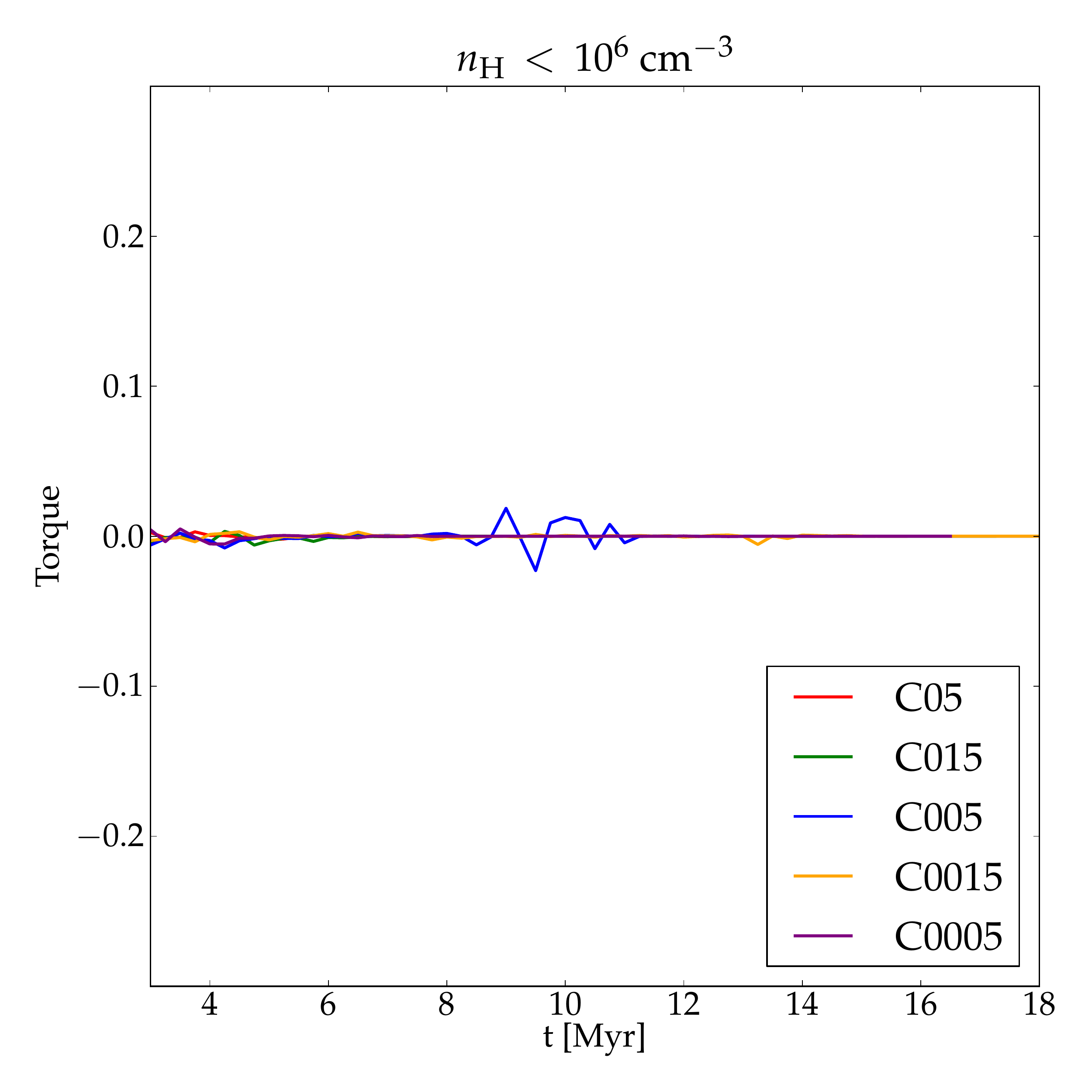}
    \caption{Torque experienced by the SMBHs due to gas with density less than $10^6$ cm$^{-3}$ for different values of $C_{\star}$. 
	We can see from this figure that the torque produced by gas with density less than $10^6$ cm$^{-3}$ is negligible. }
    \label{fig7}
\end{figure}

The torques the gas exerts on the SMBHs are density dependent, 
and we expect that most of the torque experienced by the SMBHs is due to high density gas.  
From figure \ref{fig5}, we would expect that in simulations with smaller values of $C_{\star}$, 
where the portion of high density gas is greater, 
the torque of the gas on the SMBHs is more intense and the in-spiral of the SMBHs faster.  
However, figure \ref{fig6} shows that the torque produced by gas with a density greater 
than $10^6$ cm$^{−3}$ fluctuates repeatedly about zero, indicating that the high density gas 
does not always extract angular momentum from the SMBHs, but also deposits angular momentum into the SMBHs. 
This is consistent with the large fluctuations that we observe on the separation of the SMBHs in figure \ref{fig4}. 
For completeness, we plot in figure \ref{fig7} the torque produced by gas less dense than $10^6$ cm$^{−3}$.  
It is clear from this figure that the effect of the low density gas on the orbits of the SMBHs is negligible.

\begin{figure}
    \centering
    \includegraphics[width=0.45\textwidth]{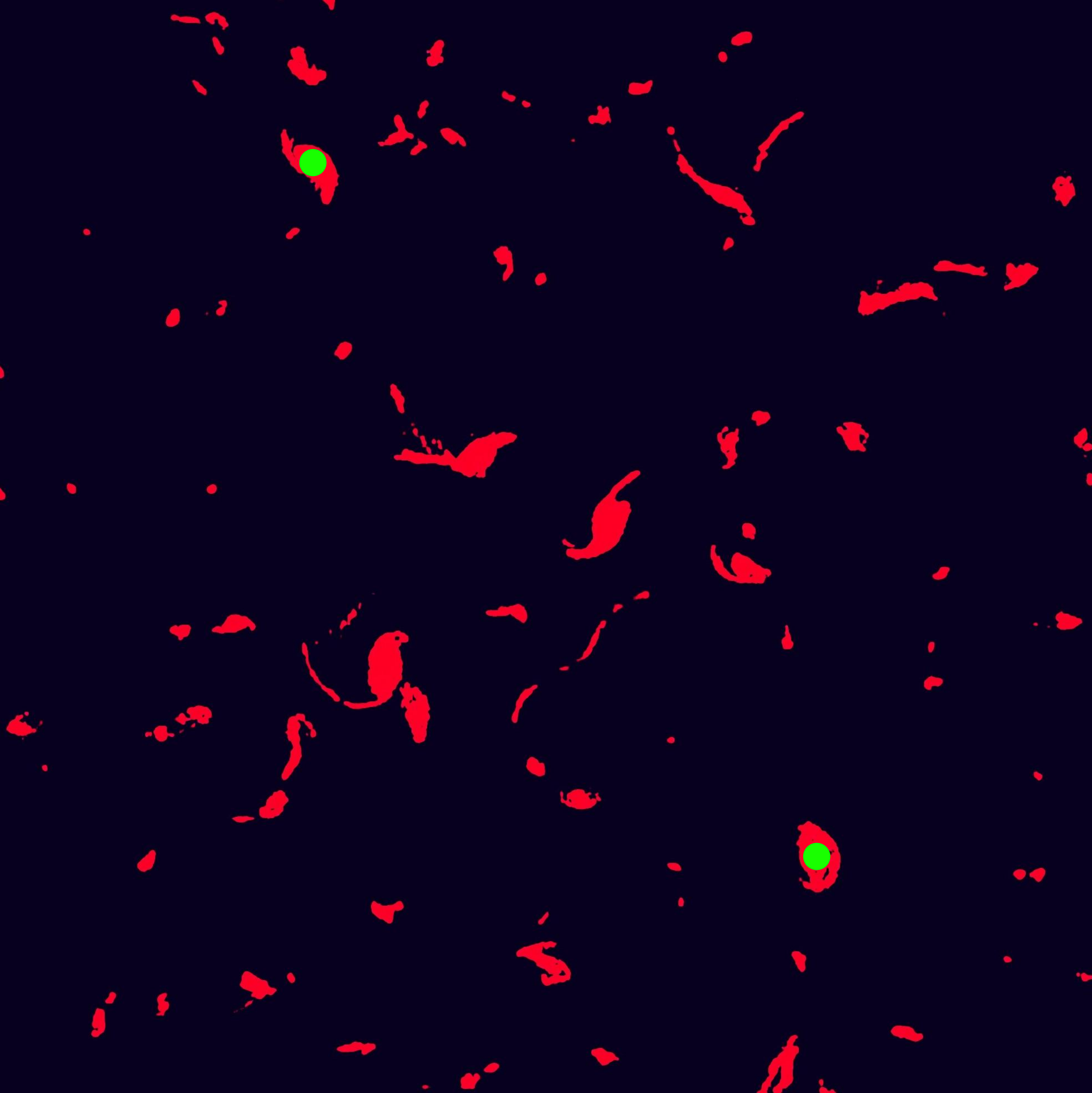}
    \caption{Distribution of gas in the plane of the disk at $t =\ 10$ {\Myr} for
        C0005.
        We only show the gas with density greater than $10^6$ cm$^{-3}$.
        The green circles are the two black holes.}
    \label{fig8}
\end{figure}

\begin{figure}
    \centering
    \includegraphics[width=0.45\textwidth]{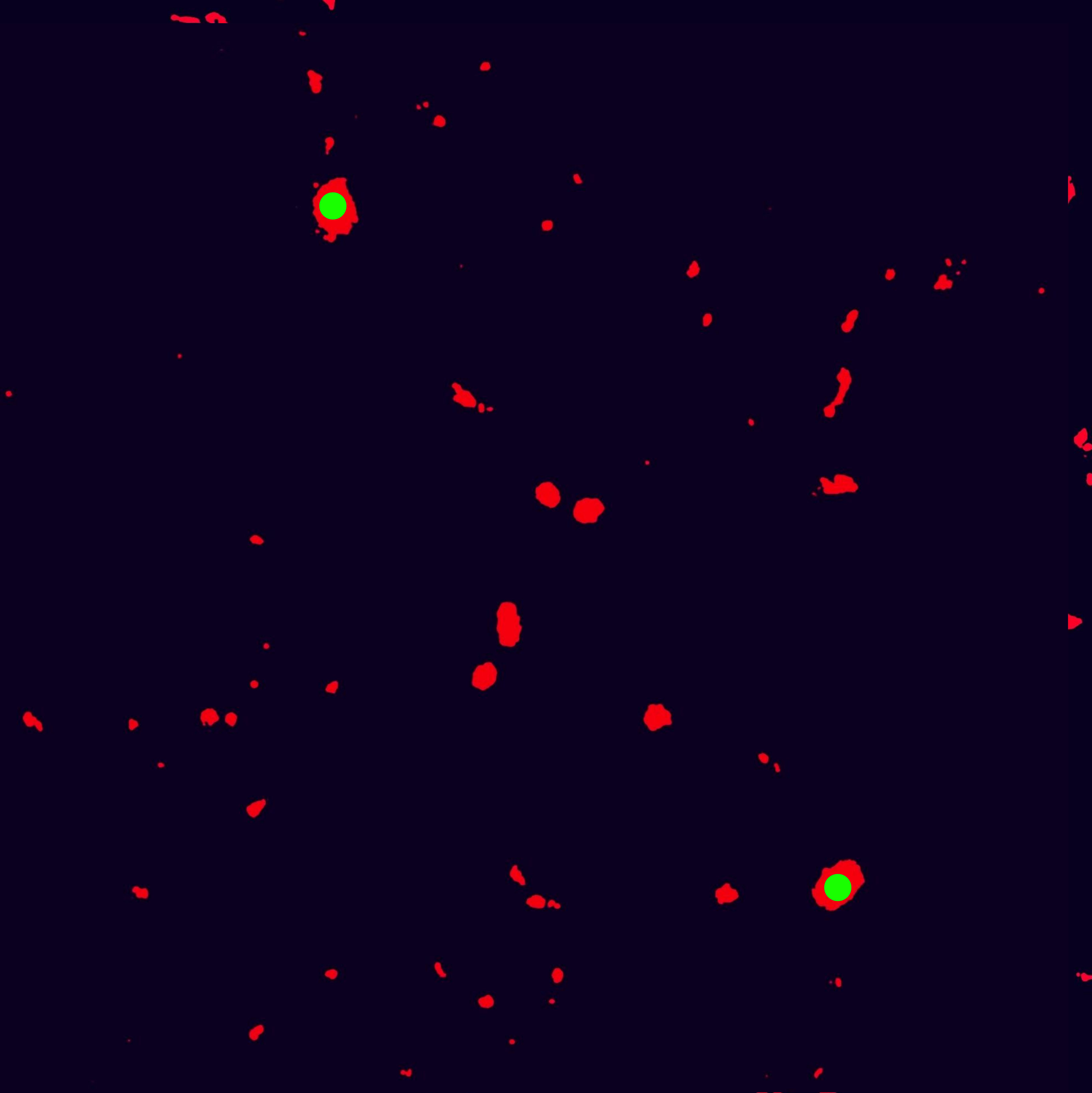}
    \caption{Distribution of gas in the plane of the disk at $t =\ 10$ {\Myr} for
        C05. We only show the gas with density greater than
        $10^6$ cm$^{-3}$.  The green circles are the two black holes.}
    \label{fig9}
\end{figure}

The fluctuations of the torque exerted by the high density gas on the SMBHs may be the 
result of the spatial distribution of this gas in the disk. 
To explore this, we show the distribution of gas with densities greater than $10^6$ cm$^{-3}$ in the plane of the disk at 
$t =$ 10 {\Myr} for run C0005 in figure \ref{fig8} and for run C05 in figure \ref{fig9}.  
From these figures, we find that gas with density greater than
$10^6$ cm$^{-3}$ is mainly concentrated in clumps, two of which surround the two black holes. 
In this clumpy CND, the SMBHs have encounters with these high density gaseous clumps, and depending
on the characteristics of the encounter, they gain or lose angular momentum.  
This is reflected in the fluctuations of the gravitational torque produced by the high density gas on the SMBHs (figure \ref{fig6}).

From figures \ref{fig8} and \ref{fig9}, 
it seems that the main difference between run C0005 and run C05 is the number of clumps formed.  
This difference may be the cause of the different numbers of fluctuations in the SMBHs' separation seen in these runs, 
and the slightly different migration timescale of the SMBHs.  
Simulations with lower SFR yield a greater number of gaseous clumps, 
so it is more probable for the SMBHs to closely interact with one or more clumps in these simulations.

\begin{table}
    \label{table2}
    \centering
    \caption{Gas clumps number and densities.}
    \begin{tabular}{lrrrrrr}
        \hline
        $C_{\star}$     &
        $N_{\rm cl,t1}$ &
        $N_{\rm cl,t2}$ &
        $N_{\rm cl,t3}$ &
        $\Delta n_{\rm cl}\, [\text{cm}^{-3}]$ &
        $\bar{ n_{\rm cl}}\, [\text{cm}^{-3}]$ \\ \hline

        0.005 & 86 & 44 & 38 & $10^6-10^{8.7}$ & $10^{7.6}$ \\
        0.015 & 84 & 39 & 25 & $10^6-10^{9.3}$ & $10^{7.9}$ \\
        0.05  & 87 & 32 & 19 & $10^6-10^{8.6}$ & $10^{7.7}$ \\
        0.15  & 70 & 27 & 15 & $10^6-10^{8.8}$ & $10^{7.9}$ \\
        0.5   & 37 & 16 &  9 & $10^6-10^{8.0}$ & $10^{7.1}$ \\
        \hline
    \end{tabular}
    \begin{center}
        $N_{\rm cl,t1}$ corresponds to the number of clumps at $t1=$5 {\Myr},
        $N_{\rm cl,t2}$ corresponds to the number of clumps at $t2=$10 {\Myr}, and
        $N_{\rm cl,t3}$ corresponds to the number of clumps at $t3=$15 {\Myr}.
        $\Delta n_{\rm cl}$ corresponds to the range of densities of the clumps
        and $\bar{n_{\rm cl}}$ to the mean density of the clumps.
\end{center}
\end{table}

In order to determine if the number of clumps depends on and $C_{\star}$, 
we compute the number of clumps ($N_{\rm cl}$) as the number of groups of gas particles
that are gravitational bound with central density greater than $10^6$ cm$^{-3}$.  
We compute $N_{\rm cl}$ at three different times-- 5, 10, and 15 {\Myr}-- for each value of $C_{\star}$ explored.  
We also compute the mean density and the range of densities of these clumps. 
We summarize this information in table 2.

From table 2 we found that the number of clumps decreases as $C_{\star}$ increases.  
This is consistent with figure \ref{fig5}, 
where we found that higher SFR correspond to a lower proportion of high density gas.  
Therefore, in simulations with lower $C_{\star}$, 
the SMBHs are prone to interact with a higher number of gaseous clumps.  
This is reflected in the slight increase in fluctuations of both the separation and the 
gravitational torque experienced by the SMBHs that we observe in simulations with higher SFR.

Having explained in this section why the orbital decay timescale changes with $C_{\star}$ in simulations 
using our recipe, in the next section we analyze why the orbital decay of the SMBHs in these simulations 
is faster than in simulation SDH05 and typically slower than in simulation E05.
 
\section{Gas physics and its effect on orbital decay}

Before the SMBHs form a binary, the gas extracts angular momentum from the SMBHs
through dynamical friction, which leads to the in-spiral of the SMBHs toward the center of the disk.
As the intensity of the dynamical friction is proportional to the density of the gas,
in a disk where the gas density is higher, the orbital decay of the SMBHs is faster 
(Chandrasekhar 1943; Ostriker 1999; Kim \& Kim 2007).

\begin{figure}
    \centering
    \includegraphics[width=0.54\textwidth]{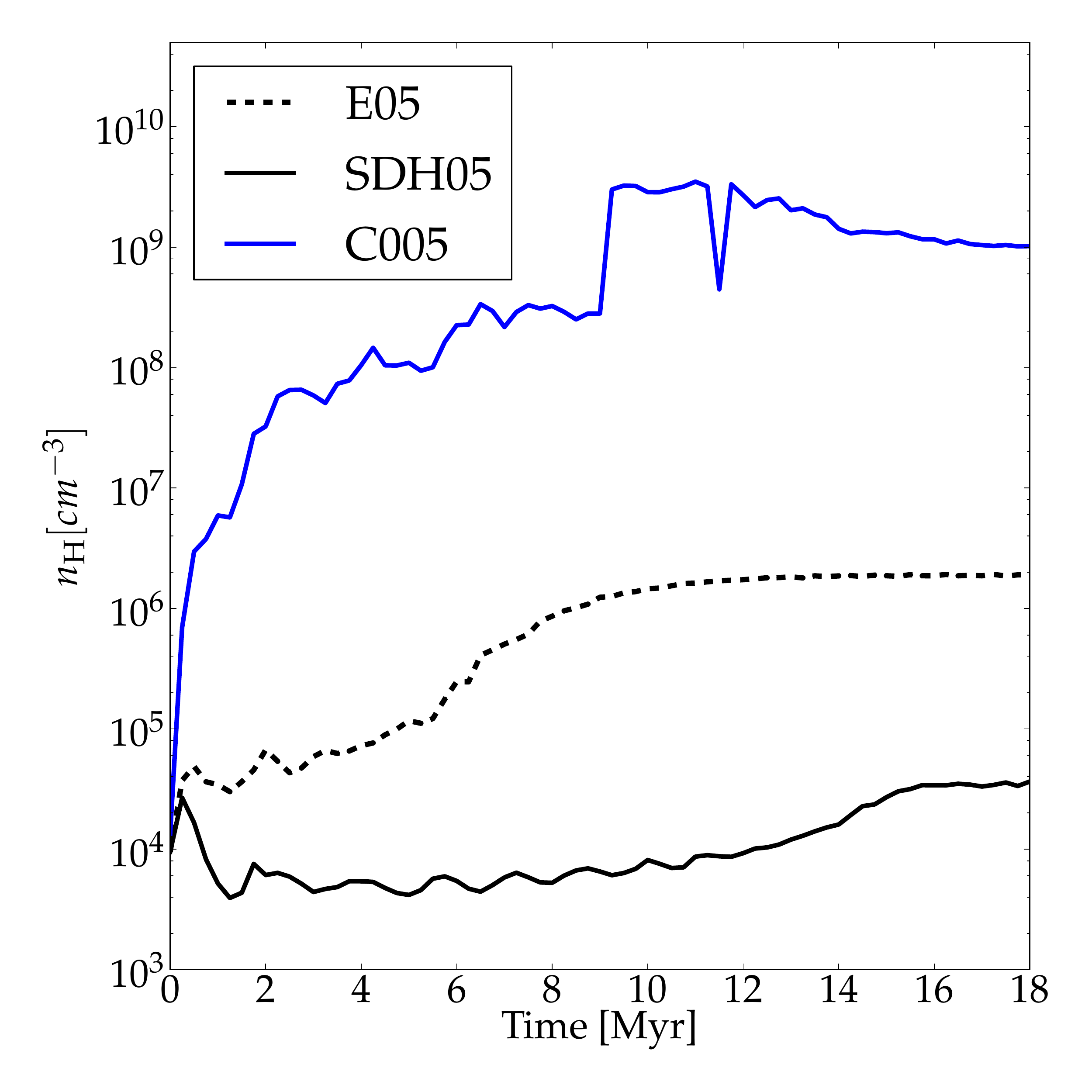}
    \caption{Mean density of the gas around the SMBHs as a function of time for simulations E05, SDH05, and C0005. 
	The mean is taken over all gas particles within twice the gravitational influence radius of the SMBH.}
    \label{gasPhysicsDens}
\end{figure}

In figure \ref{gasPhysicsDens} we show the mean density of the gas around the SMBHs 
for the simulations E05, SDH05, and C005. 
The mean is taken over all gas particles within twice the gravitational 
influence radius (bound radius) of the SMBH: $R_{\rm bound}=2GM_{\rm BH}/(c_{\rm s}^2+v_{\rm rel}^2)$,
where $c_{\rm s}$ is the sounds speed of the gas and $v_{\rm rel}$ is its 
velocity relative to the SMBH.  
We choose this region around each SMBH because most of the torque experienced 
by the SMBH comes from gas particles in this region.  

As seen in figure \ref{gasPhysicsDens}, our recipes yield far greater gas 
density than the recipes used in simulation SDH05.  
This is because our recipes produce a greater density threshold for star 
formation than the threshold of simulation SDH05.  
Additionally, the temperature floor for star formation is much lower using our 
recipes (25 K in ours, as compared to $10^4$ K in SDH05).  
As a result, the orbital decay of the SMBHs in simulations using our recipes 
is faster than in simulation SDH05.

Figure \ref{gasPhysicsDens} also shows that the density of the gas in 
simulations with our recipes is much greater than in simulation E05.
However, after $t \sim 6$ {\Myr}, in some of the simulations with our recipes,
the migration of the SMBHs slows.  
This is because in the simulations with our recipes, the CND is fragmented 
in a few tens of high density gaseous clumps and these gaseous clumps 
erratically perturb the orbits of the SMBHs.  
So even though in these simulations the density is higher than in simulation E05, 
the orbital decay of the SMBHs is, in some cases, delayed due to the interaction 
of the SMBHs with these gaseous clumps. 
The CND of simulation E05 is smoother than the CND in simulations with our recipes, 
and there are no high density gaseous clumps in it to perturb the SMBHs orbits.

In the next section we investigate whether these perturbations are likely to affect 
the evolution of SMBHs in real CNDs by comparing properties of the clumps in our 
simulations to observations.

\section{ Gaseous clumps}
\subsection{ Density of the gaseous clumps. }

As seen in section \S3.2, the stochastic fluctuations of the SMBHs' separation 
comes from gravitational interaction between the black holes and high density gas 
clumps ($n_{\rm h}>10^6$cm$^{-3}$).  
However, these fluctuations are unlikely to occur in real CNDs because intensity 
of the gravitational torques is density dependent, and the densities of the gas 
clumps in our simulations are higher than the densities of gas clumps in observed CNDs.  
For example, if we compare our clumps' densities with the density of molecular 
clouds in isolated galaxies (Mathis 1990; Oka et al. 2001; Struve \& Conway 2010) 
or ultra luminous infrared galaxies (Downes \& Solomon 1998; Schulz et al. 2007), 
which are typically $\sim 10^2-10^5$cm$^{-3}$, we find that the density of the 
clumps in our simulations is at least two orders of magnitude greater.
Inside the molecular clouds, there are star forming regions (cold or hot cores) 
with densities comparable to the mean density of the clumps in our simulations 
($<n_{\rm h}>\,\sim10^7$cm$^{-3}$).  
However, they are one or two orders of magnitude less dense than roughly half 
of the gas clumps in our simulations ($n_{\rm h}\,\sim\, 10^{6}-10^{9.3}$cm$^{-3}$).  
Even though one of the gas clumps that perturbs the SMBH orbits may have a 
density comparable with the density of these cores, the mass of these cores 
is on the order of $\sim 10^3\, M_{\odot}$ (Garay et al. 2004; Mu\~noz et al. 2007), 
which is at least two orders of magnitude smaller than the mass of any of the clumps 
in our simulations.

The clumps in our simulations have these extremely high densities because our recipes 
do not account for all of the physical processes that sculpt the clumps.  
For example, in our simulations we use a cooling that assumes that the gas is optically 
thin while the clumps in our simulations are optically thick ($\tau >> 1$).  
Also, there is a large amount of energy that we are not considering in our simulations, 
such as the energy of photons emitted by the stars and by the black hole’s accretion disk.  
The effects of turbulence, which is typically damped in SPH codes, are also omitted.  
All these physical processes will prevent the further collapse of the clumps, 
promoting the formation of clumps with lower (and hence more realistic) densities.  

\subsection{ Force resolution and the SMBH-clump interaction. }

The high densities of the gaseous clumps can affect the orbits of the SMBHs in a spurious way.  
The gravitational pull at the edge of these clumps can be greater than the gravitational 
pull at the ``edge'' of the SMBHs in our simulations, where we define the edge of an SMBH 
to be its gravitational softening. 
We define the effective mass density of our black holes ($\rho_{\rm BH}$) to be the mass of the black hole ($M_{\rm BH}$) 
enclosed within a sphere of radius equal to the gravitational softening of the black holes ($\epsilon_{\rm BH}$), i.e.
$\rho_{\rm BH}\,=\,3\,M_{\rm BH}/(4\pi \epsilon_{\rm BH}^3)$.  
We can compare $\rho_{\rm BH}$ with the mass density of the gaseous clumps ($\rho_{\rm cl}\,=\,m_{\rm H}\,n_{\rm cl}$) 
to determine if, in our simulations, the clumps are more compact than the black holes. 
For a gravitational softening of 4 pc, we find that $\rho_{\rm BH}\,=\,$3.8$m_{\rm H}\times10^6$cm$^{-3}$.  
Hence, in our simulations the density of the gaseous clumps is greater than the 
effective mass density of the black holes. 
This implies that the gravitational pull at the edge of the black holes 
($F_{\rm BH}(\epsilon)\,\propto\,\epsilon\rho_{\rm BH}$) is typically smaller than the 
gravitational pull at the edge of the clumps ($F_{\rm cl}(R_{\rm cl})\,\propto\,R_{\rm cl}\,\rho_{\rm cl}\,=\,4-5\,\epsilon\,\rho_{\rm cl}$). 
Indeed, considering a gravitational softening of 4 pc, and the minimum and maximum 
densities of the gaseous clumps in our simulations, $F_{\rm BH}(\epsilon)/F_{\rm cl}(R_{\rm cl})\sim$ 0.04 - 10. 
Therefore, a black hole sees the clumps as point masses, and it is not able to 
disrupt them when it has a close encounter with one of them, within a distance $\lesssim 4$ pc. 
Instead, in our simulations a close encounter within this distance will result in the scattering of the black hole.

\begin{figure*}
    \centering
    \includegraphics[height=3.6cm]{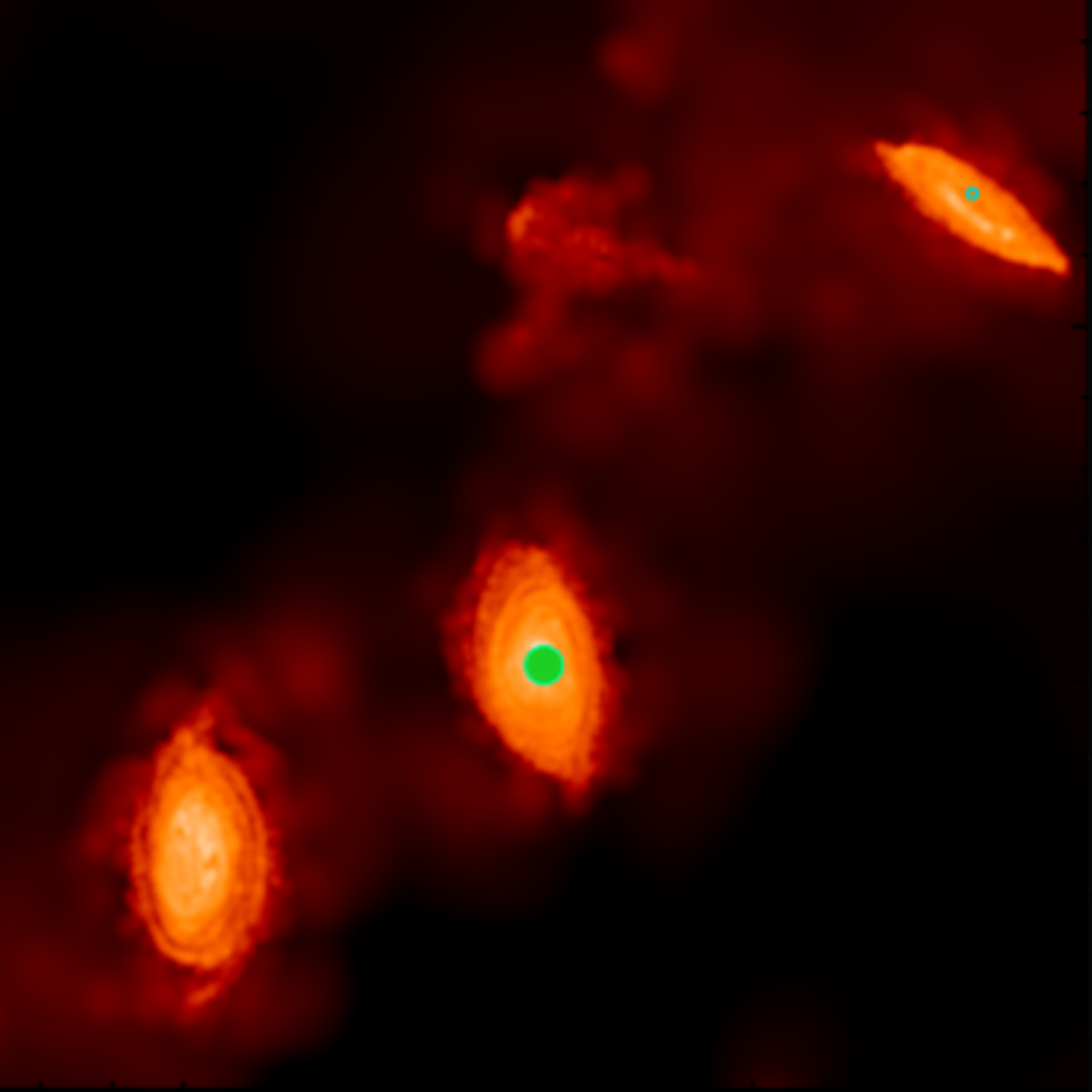}
    \includegraphics[height=3.6cm]{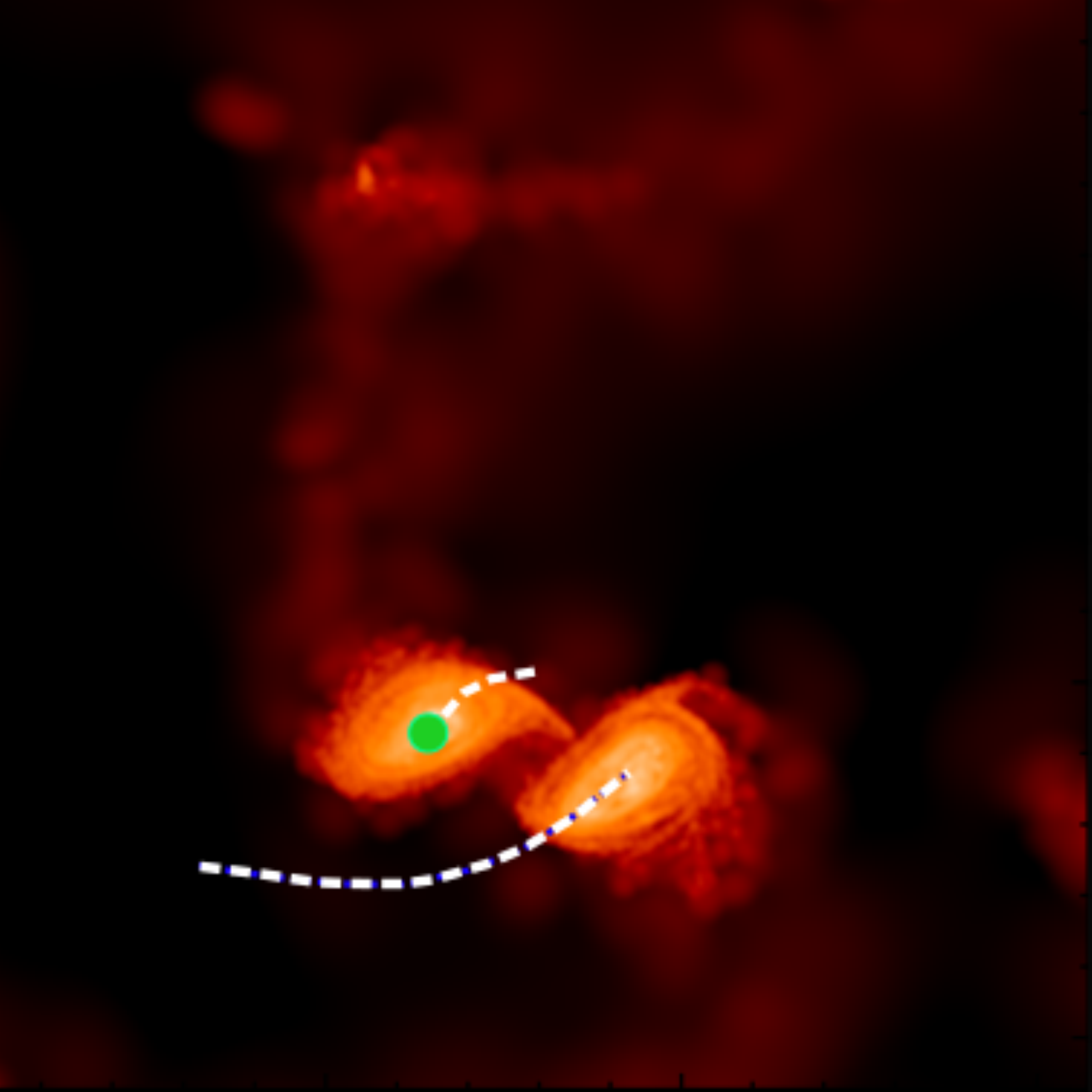}
    \includegraphics[height=3.6cm]{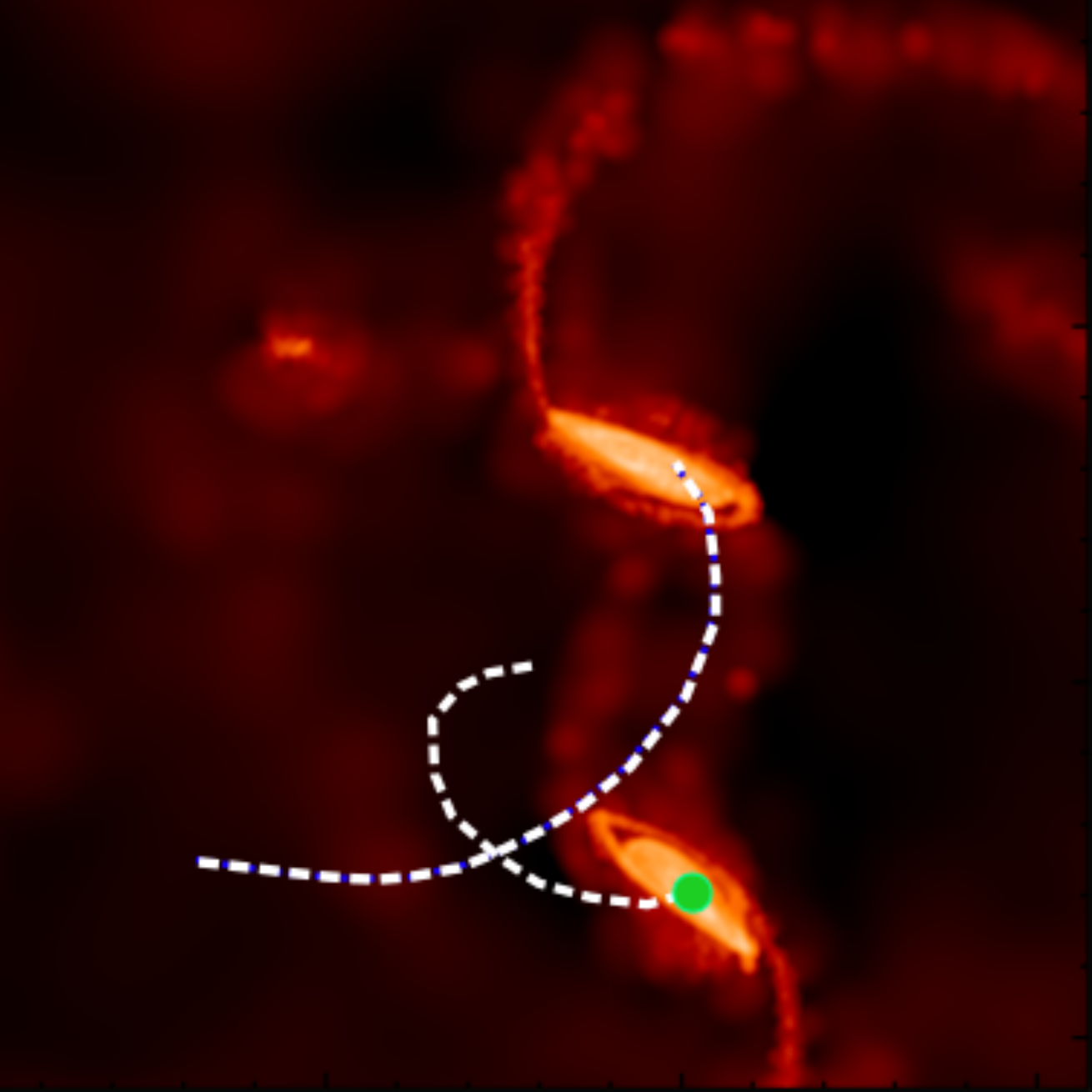}
    \includegraphics[height=3.6cm]{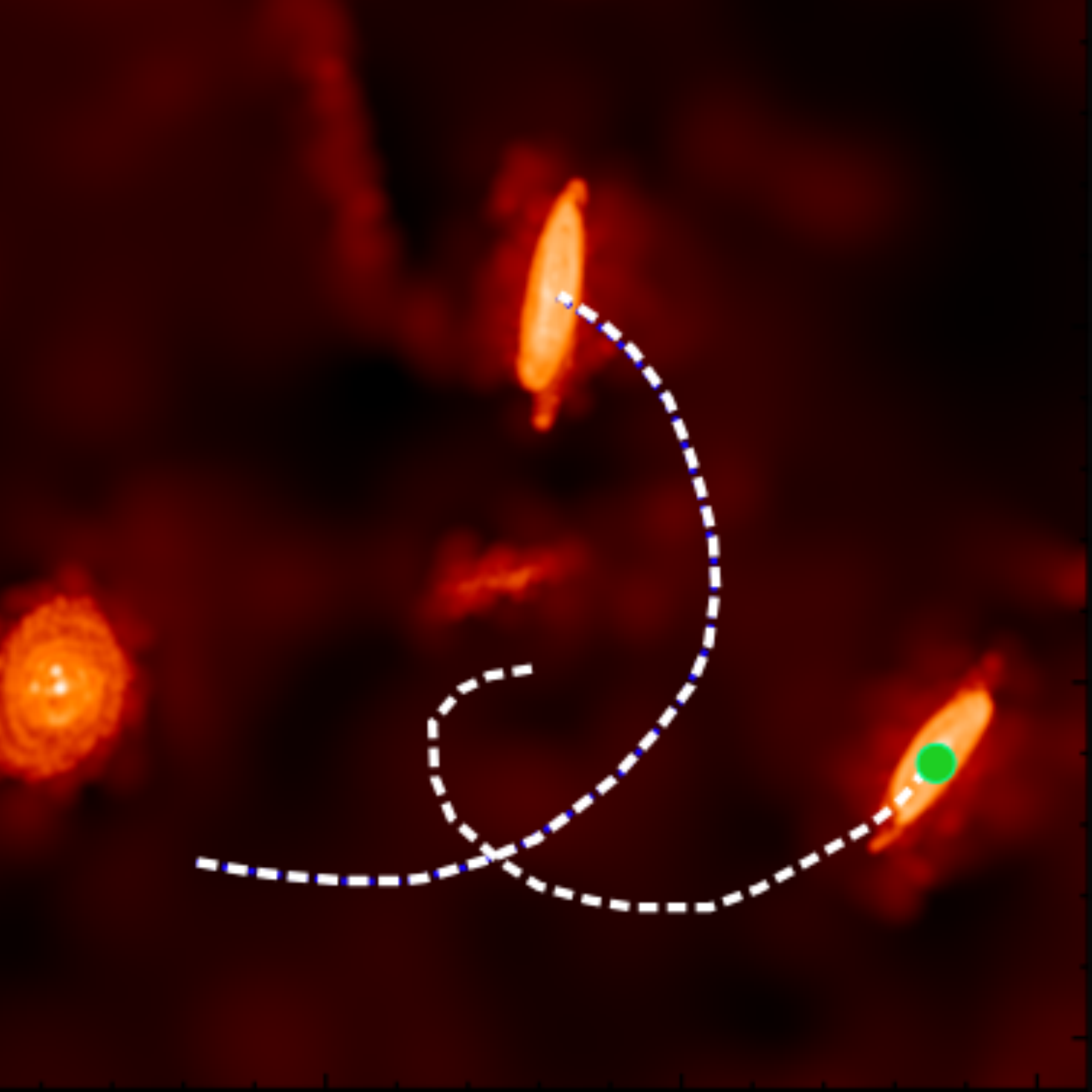}\\
    \includegraphics[height=3.6cm]{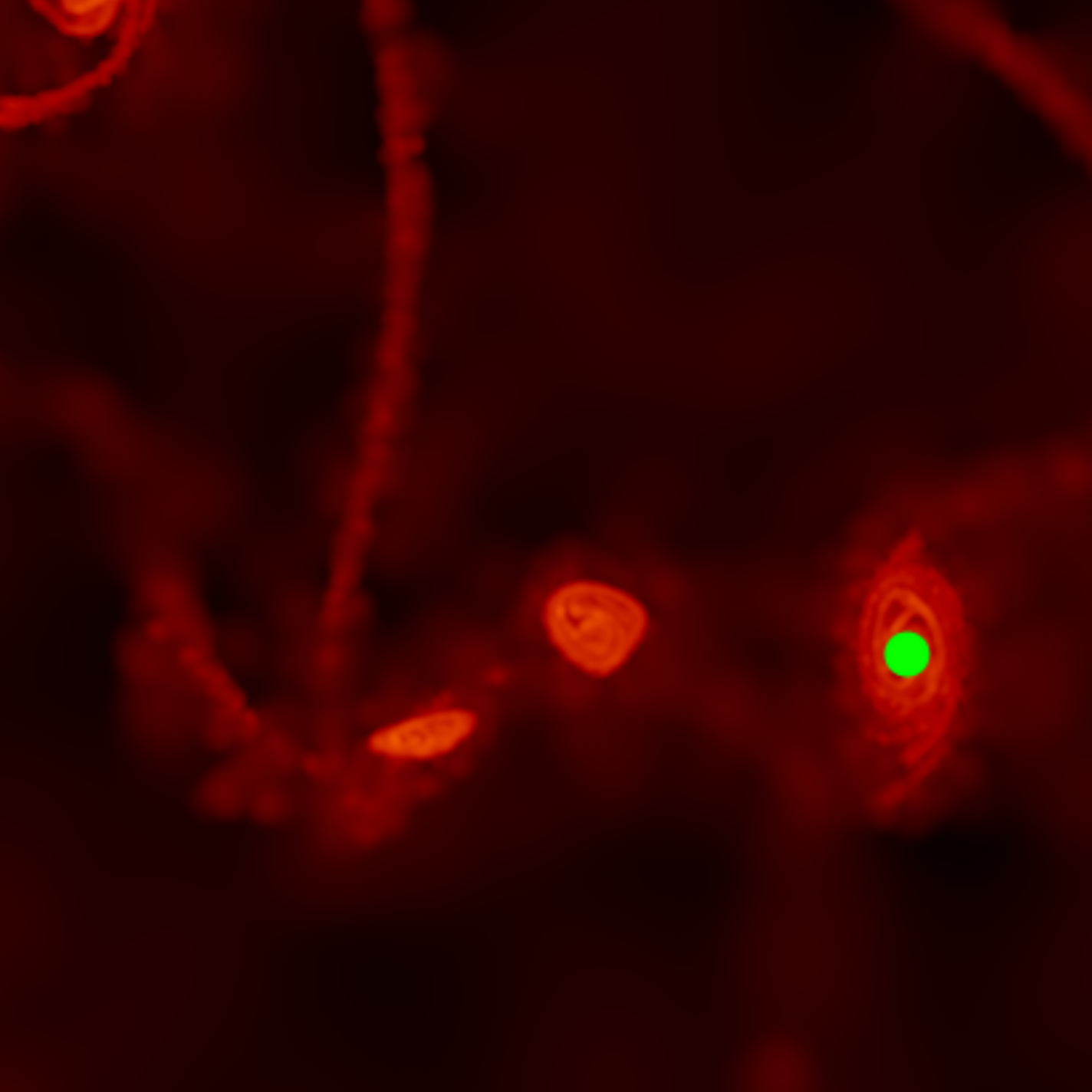}
    \includegraphics[height=3.6cm]{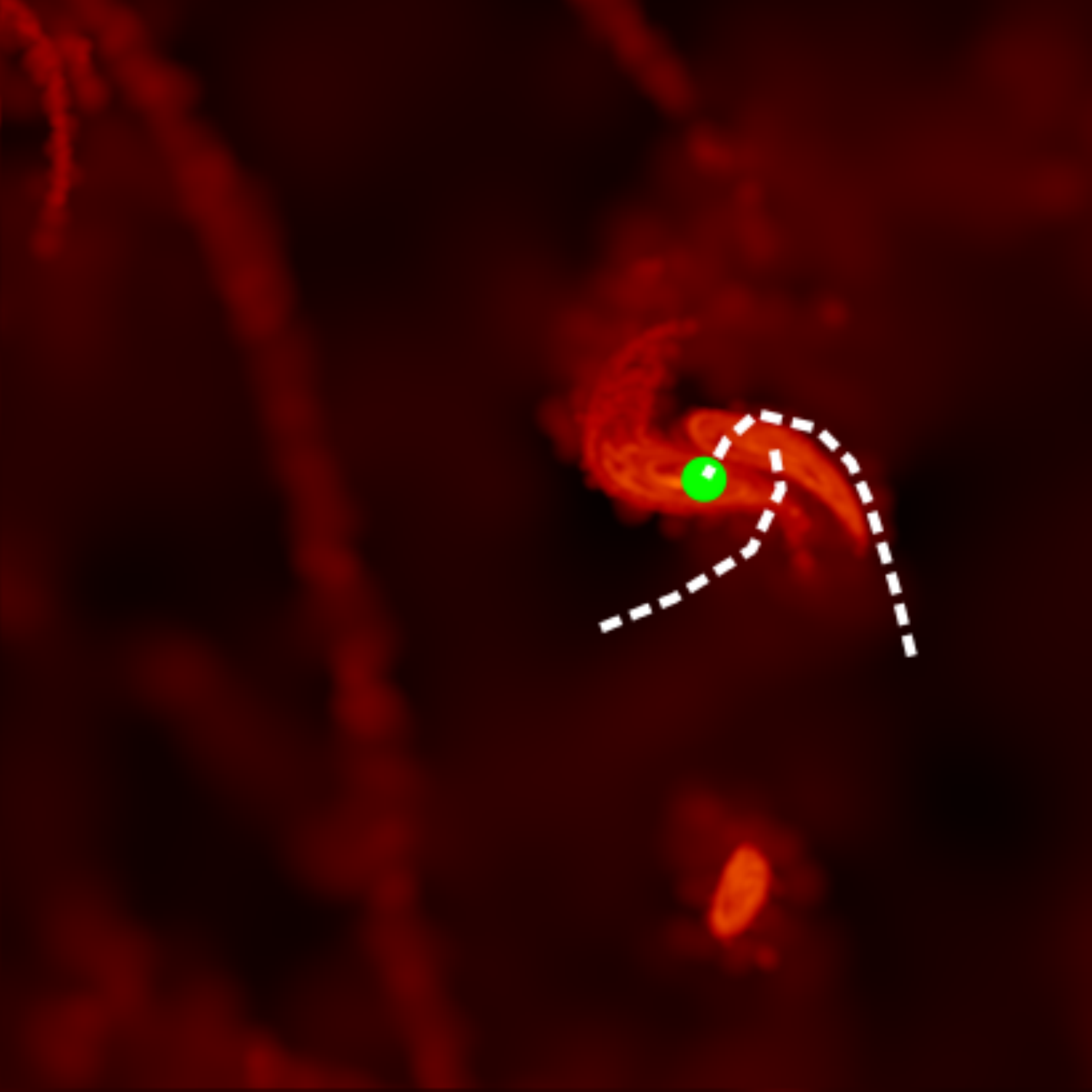}
    \includegraphics[height=3.6cm]{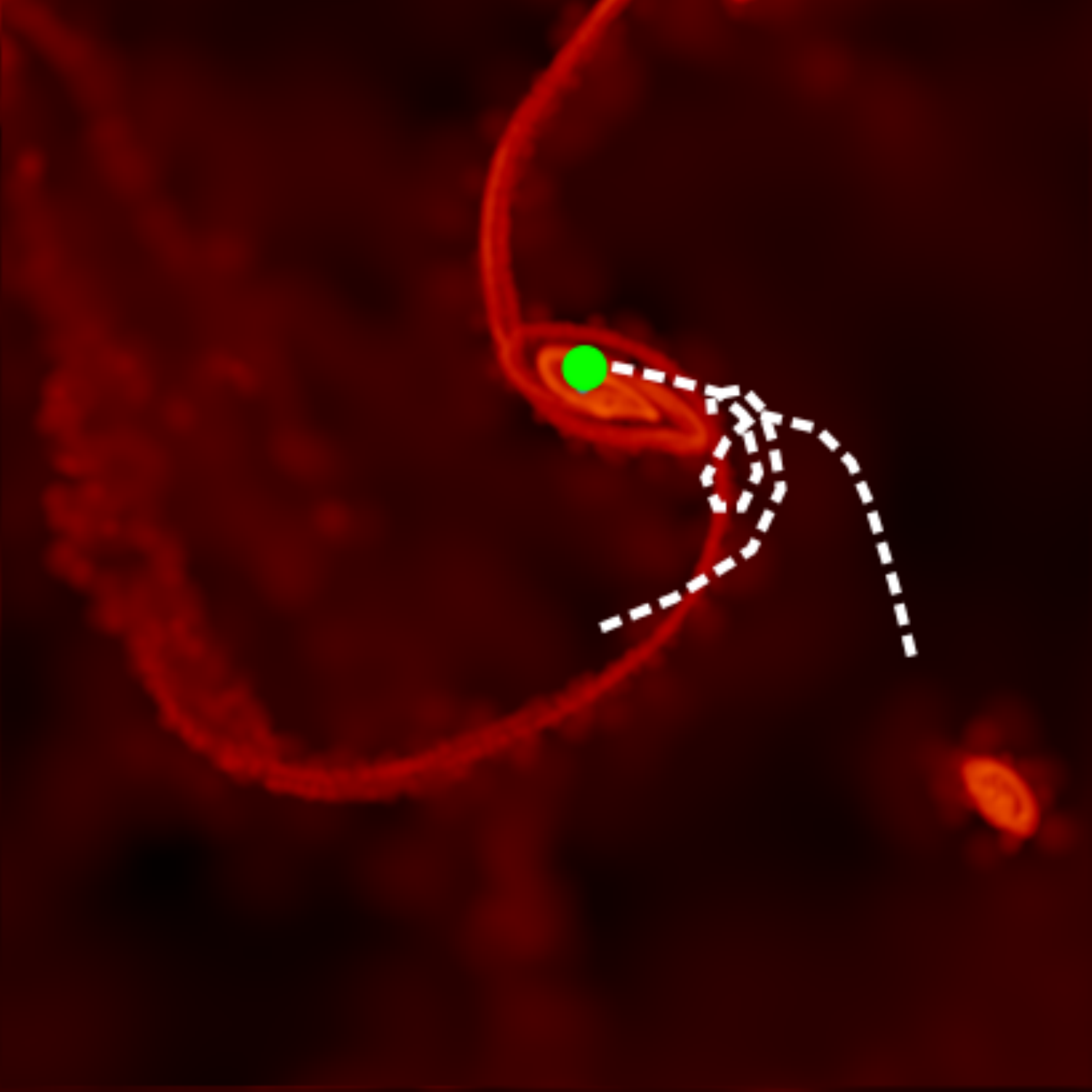}
    \includegraphics[height=3.6cm]{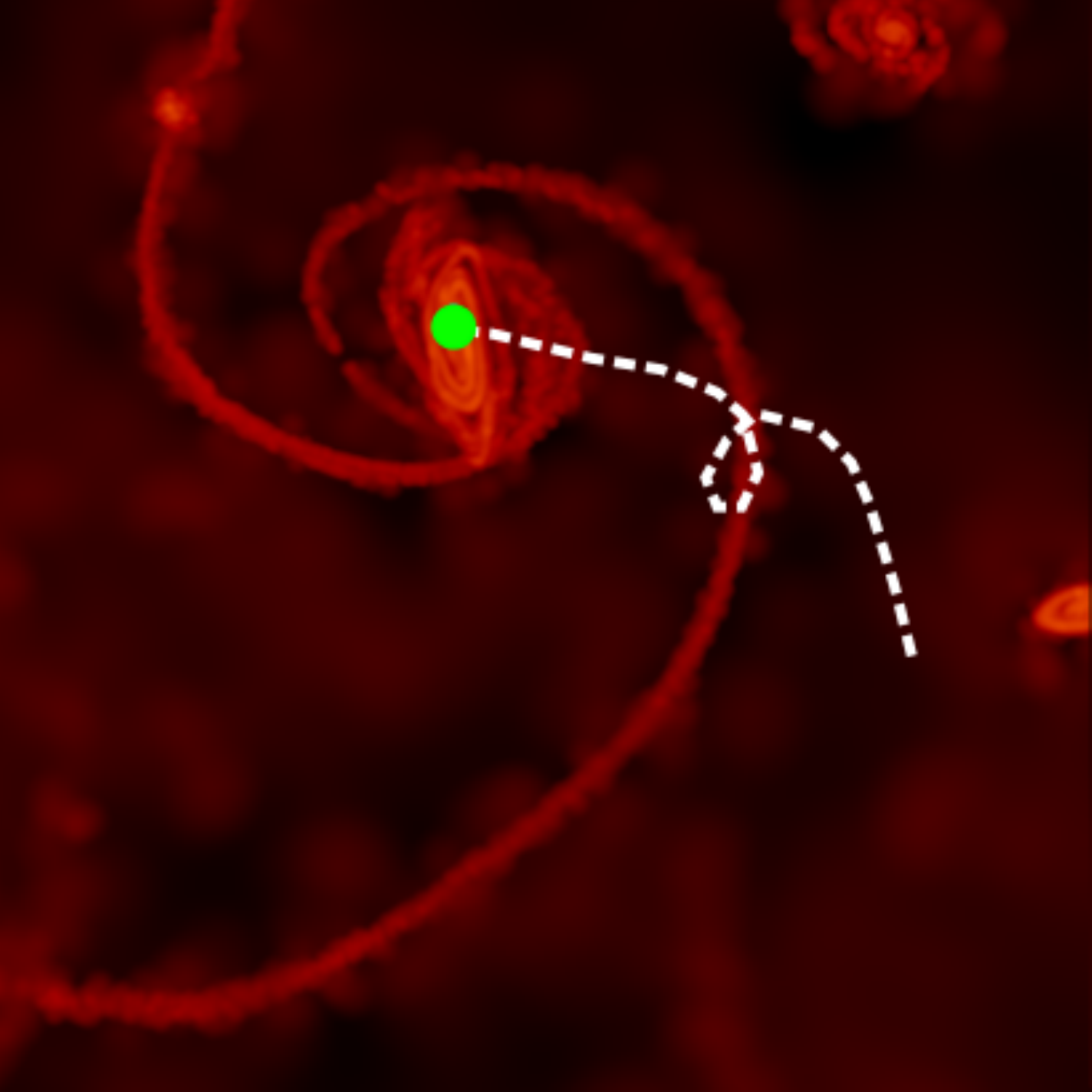}\\
    \includegraphics[height=3.6cm]{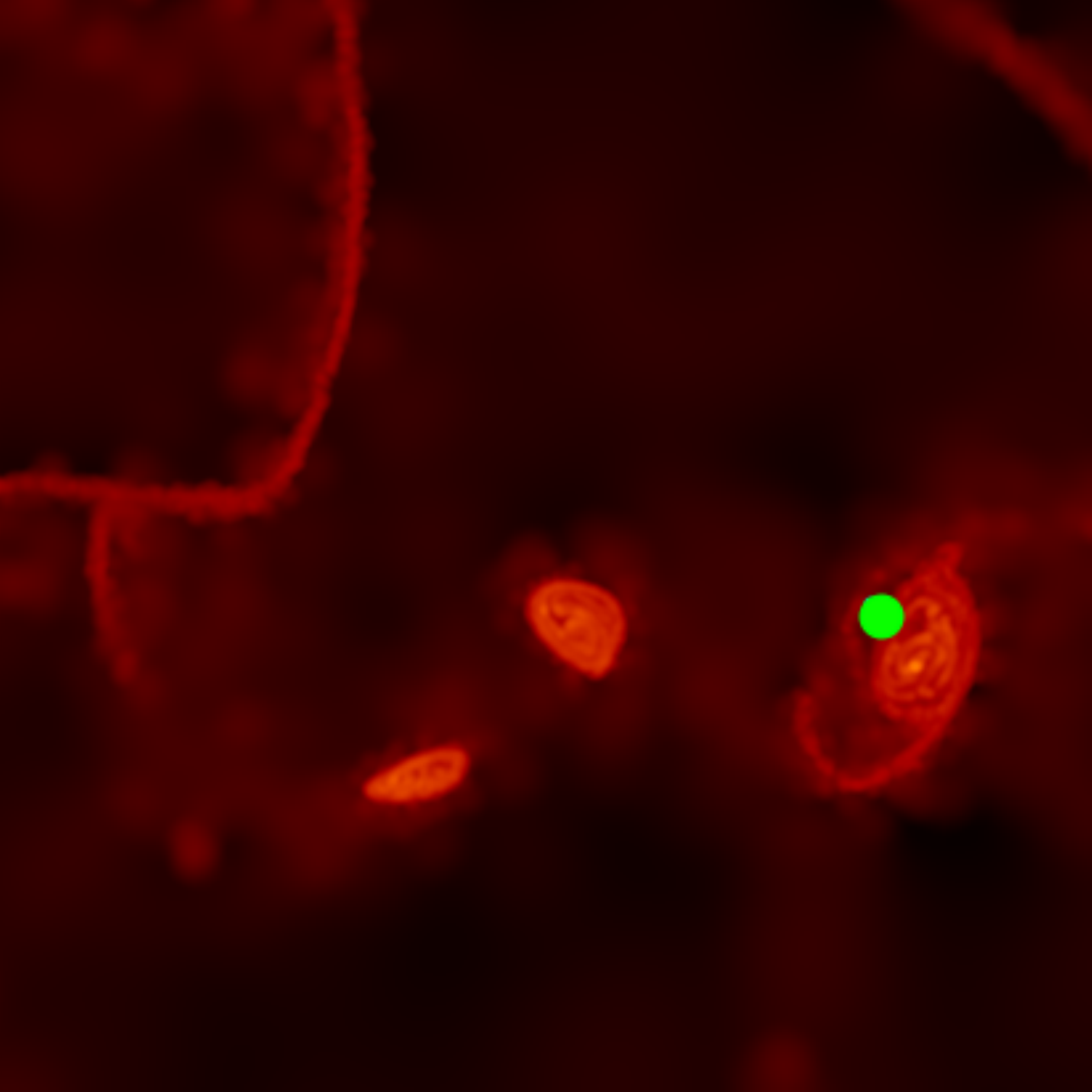}
    \includegraphics[height=3.6cm]{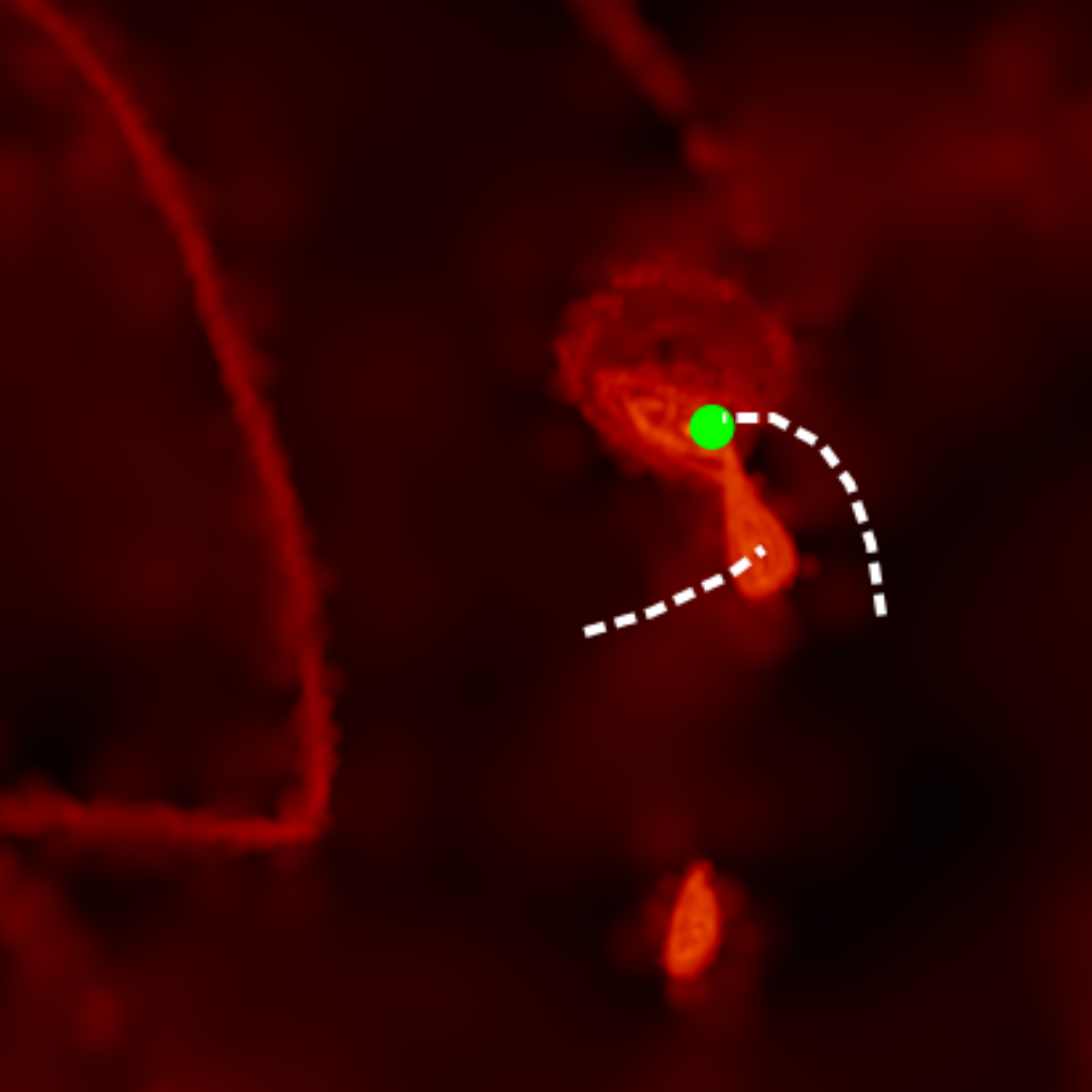}
    \includegraphics[height=3.6cm]{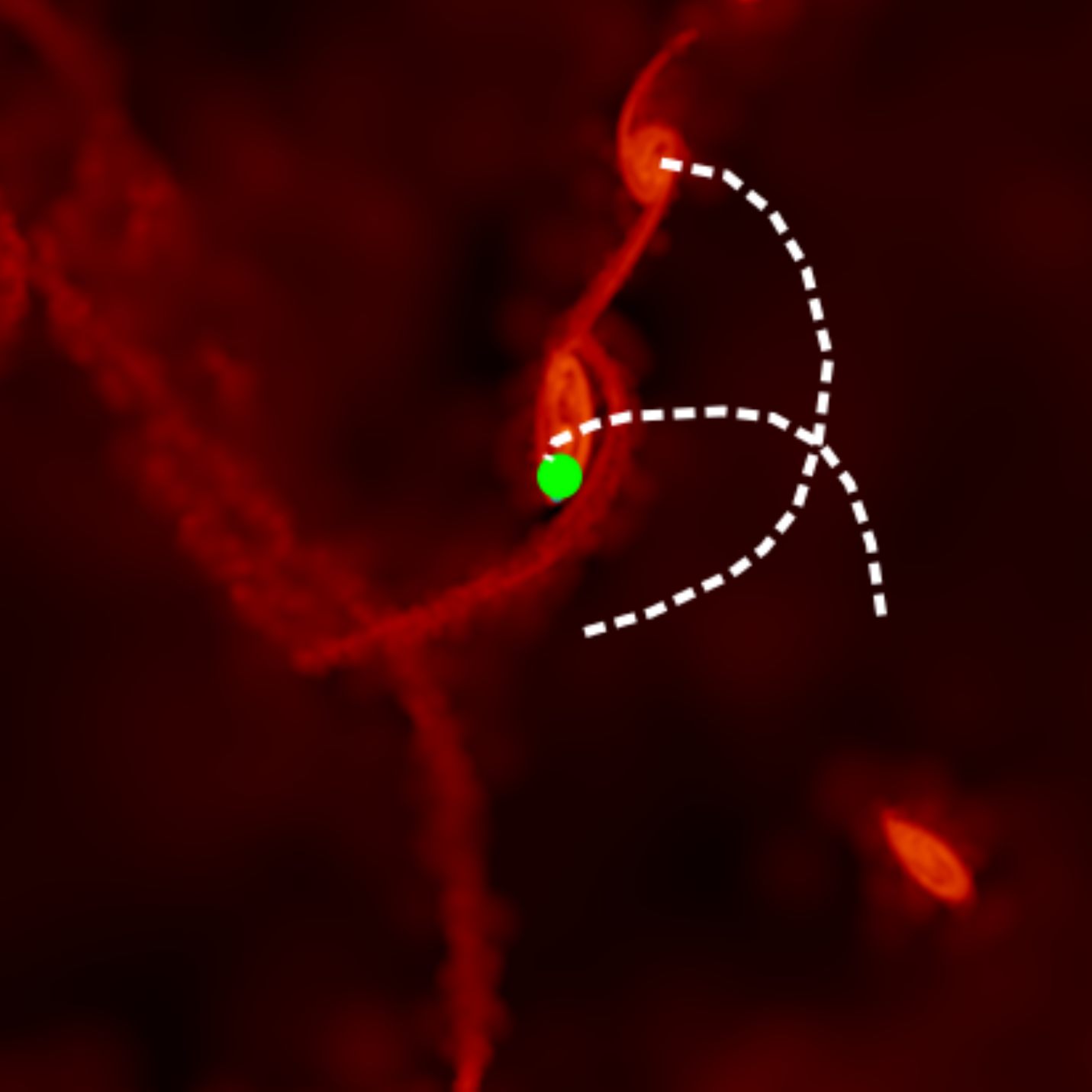}
    \includegraphics[height=3.6cm]{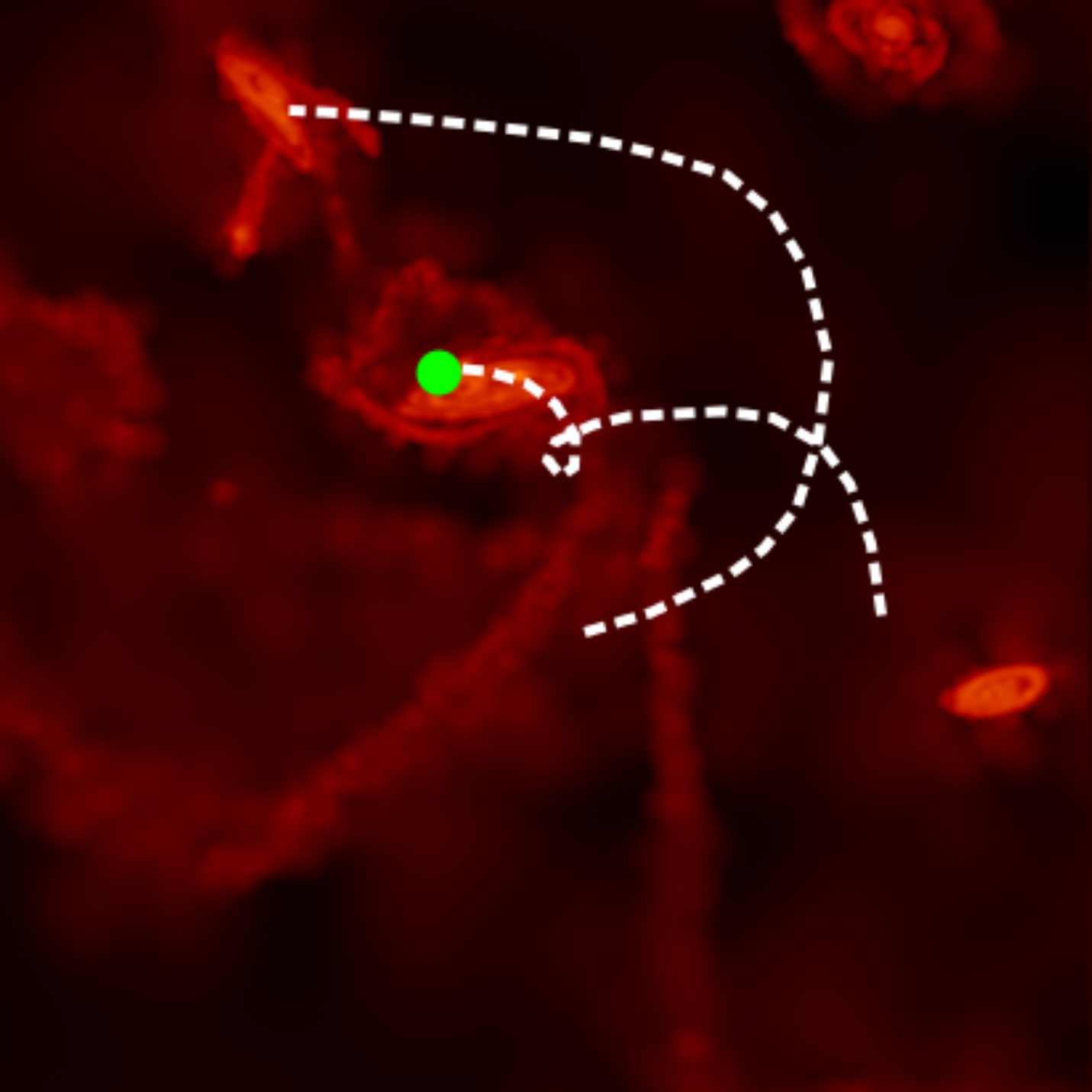}\\
    \caption{Distribution of gas in the plane of the disk. 
        Time of the snapshots increases from left to right. 
	The figures show the interaction between one black hole and a 
        clump for three different simulations (upper, middle, and lower rows).
        The upper row corresponds to the run C0015, where $\epsilon_{\rm BH} =4$ pc;
        the middle row correspond to the run C0015\_$\epsilon$.004, where
        $\epsilon_{\rm BH} =0.004$ pc; and the lower row correspond to a run
        with $\epsilon_{\rm BH} =4$ pc restarted from simulation C0015\_$\epsilon$.004
        in order to follow the same close encounter that we show in the middle row
        but with a higher gravitational softening.
        In all the panels the SMBH is represented as a green filled circle.
        In the upper and lower row we see that, where $\epsilon_{\rm BH} =4$
        and thus the density of the black hole is comparable with the density of
        the gas, the close encounter produces a slingshot effect on the black hole.
        In contrast, we see in the middle row that, when
        $\epsilon_{\rm BH} =0.004$, as the effective density of the black hole is
        greater than the density of the clumps, the clump is disrupted in the close
        encounter with the black hole. }
    \label{fig11}
\end{figure*}

If we decrease $\epsilon_{\rm BH}$, we would expect the higher resolution of the black hole's
 gravitational force to allow the black hole to disrupt a clump in a close encounter, 
where the minimum distance is $\lesssim$ 4 pc. 
To illustrate how decreasing $\epsilon_{\rm BH}$ affects the outcome of close encounters, 
we show how an encounter proceeds for different values of $\epsilon_{\rm BH}$ in the bottom 
two rows of figure \ref{fig11}.  
Snapshots that are farther to the right in this figure correspond to later times.  
The first row of this figure shows a different encounter-- one that illustrates
 scattering of the SMBH.  
In the middle row (run C0015\_$\epsilon$.004), we see that the SMBH's 
gravitational potential disrupts the clump, and the SMBH does not scatter significantly.  
On the other hand, in the lower row (run C0015) where $\epsilon_{\rm BH}$ is larger, 
the black hole is not able to disrupt the clump.  
The orbit of the SMBH is perturbed and the gaseous clump is scattered.

If these close encounters, where the minimum distance between the black hole and the clump is $\lesssim$ 4 pc, 
are the primary source of large fluctuations in the separation of the black holes, 
then we expect decreasing $\epsilon_{\rm BH}$ sufficiently would cause these fluctuations to disappear.  
More precisely, we expect the large fluctuations to disappear once $F_{\rm BH}(\epsilon)/F_{\rm cl}(\epsilon)>1$ 
for all clumps in the simulation.

\begin{figure}
    \centering
    \includegraphics[width=0.45\textwidth]{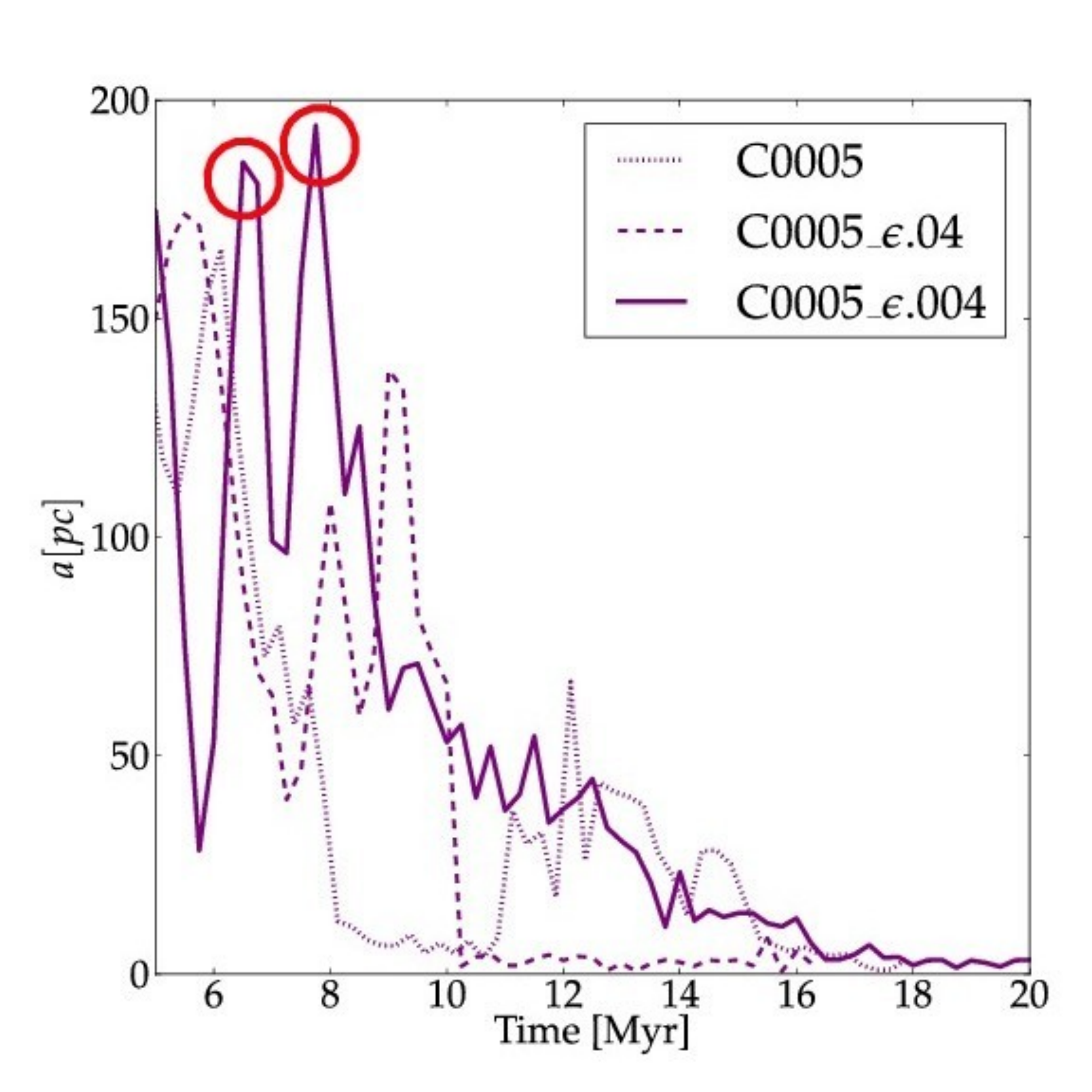}
    \caption{
        Evolution of the SMBHs' separation for simulations with different
        values of the gravitational softening of the black holes.
        The simulations correspond to the run C0005 (dotted purple line),
        the run C0005\_$\epsilon$.04 (dashed purple line), and the run C0005\_$\epsilon$.004
        (continuous purple line).
        In all three simulations, the value of the star formation efficiency
        is $C_{\star}=0.005$, and thus the number of clumps and their range of densities are the same in each run.}
    \label{fig12}
\end{figure}

\begin{figure}
    \centering
    \includegraphics[width=0.45\textwidth]{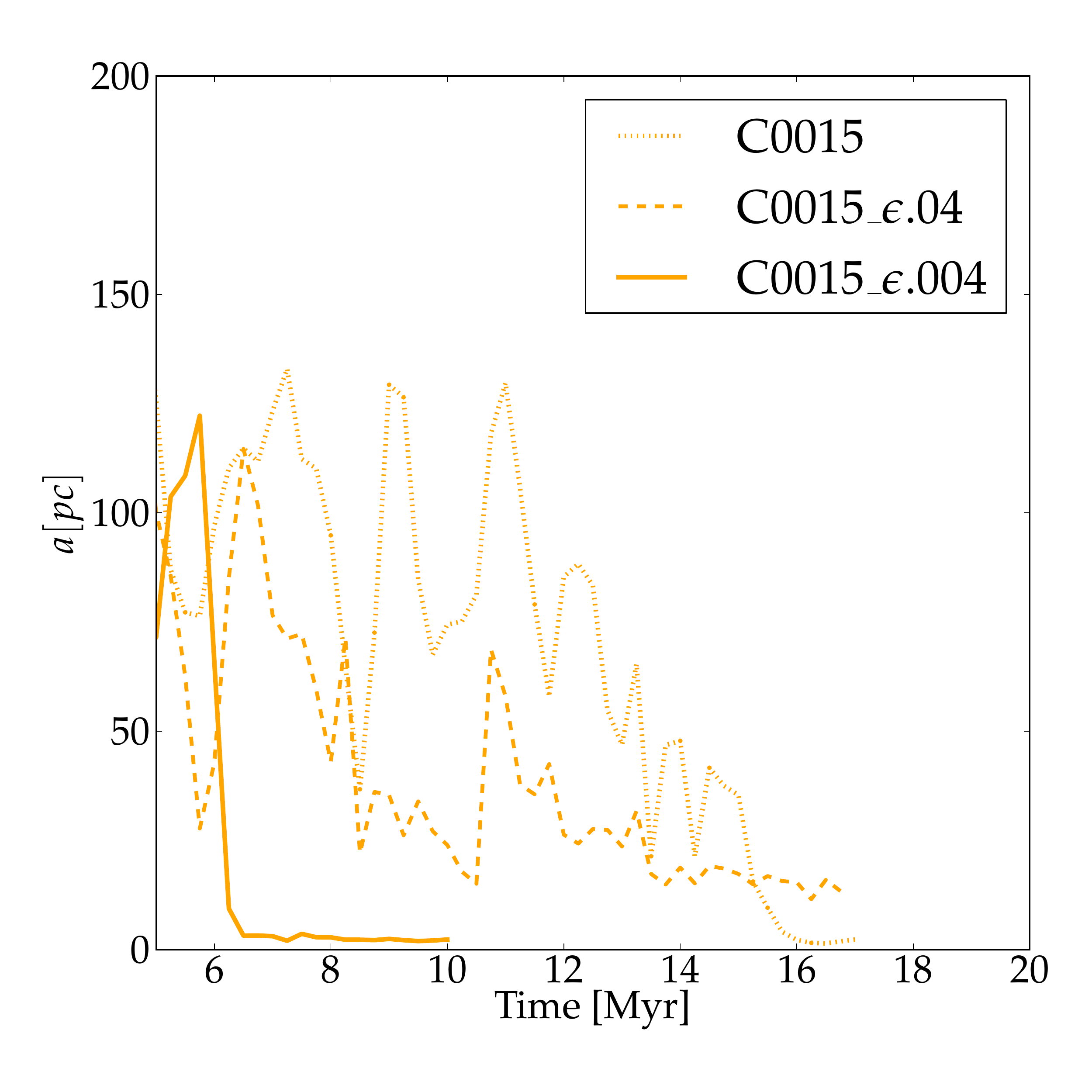}
    \caption{ Same as figure \ref{fig12} for  $C_{\star}=0.015$.}
    \label{fig13}
\end{figure}

\begin{figure}
    \centering
    \includegraphics[width=0.45\textwidth]{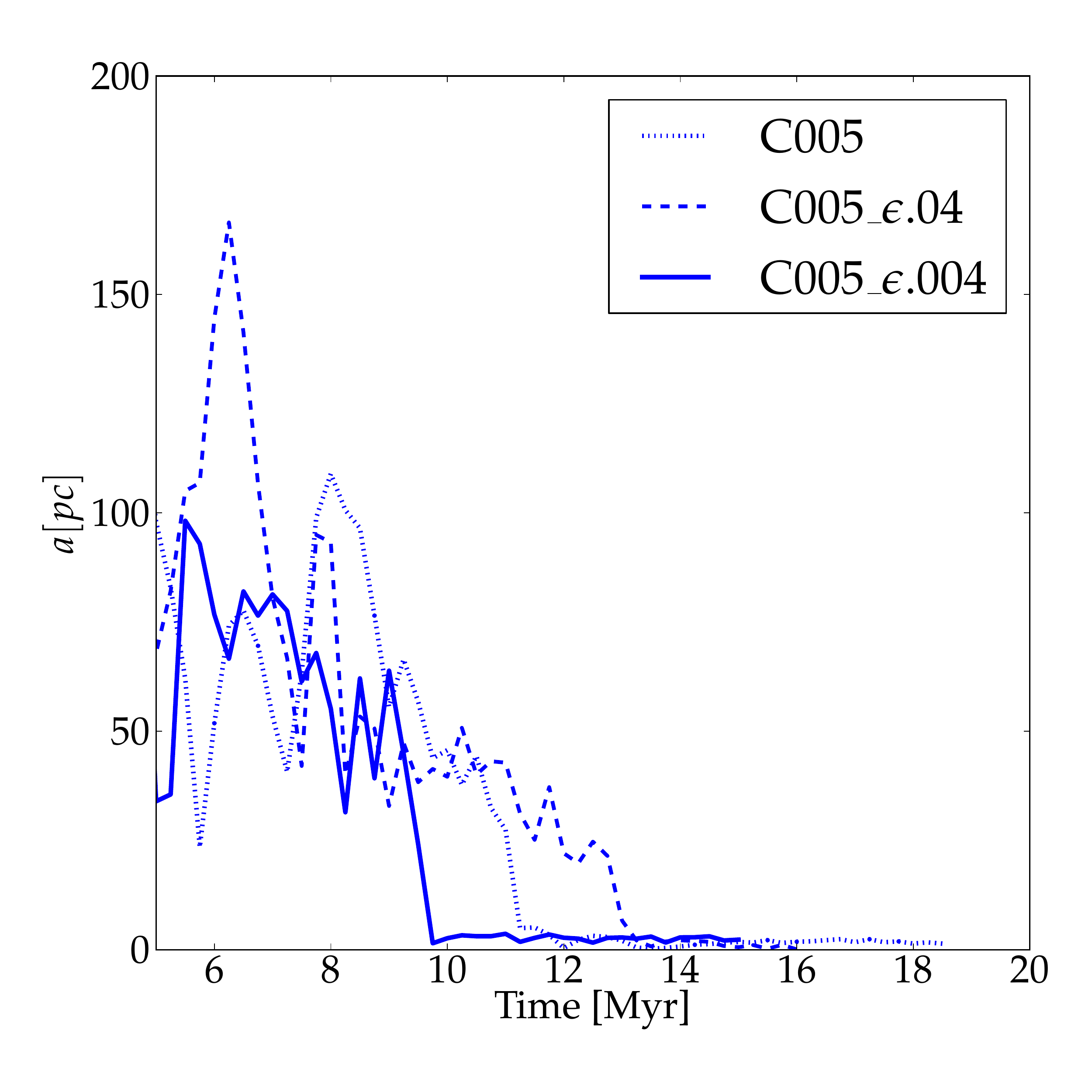}
    \caption{ Same as figure \ref{fig12} for  $C_{\star}=0.05$.}
    \label{fig14}
\end{figure}

\begin{figure}
    \centering
    \includegraphics[width=0.45\textwidth]{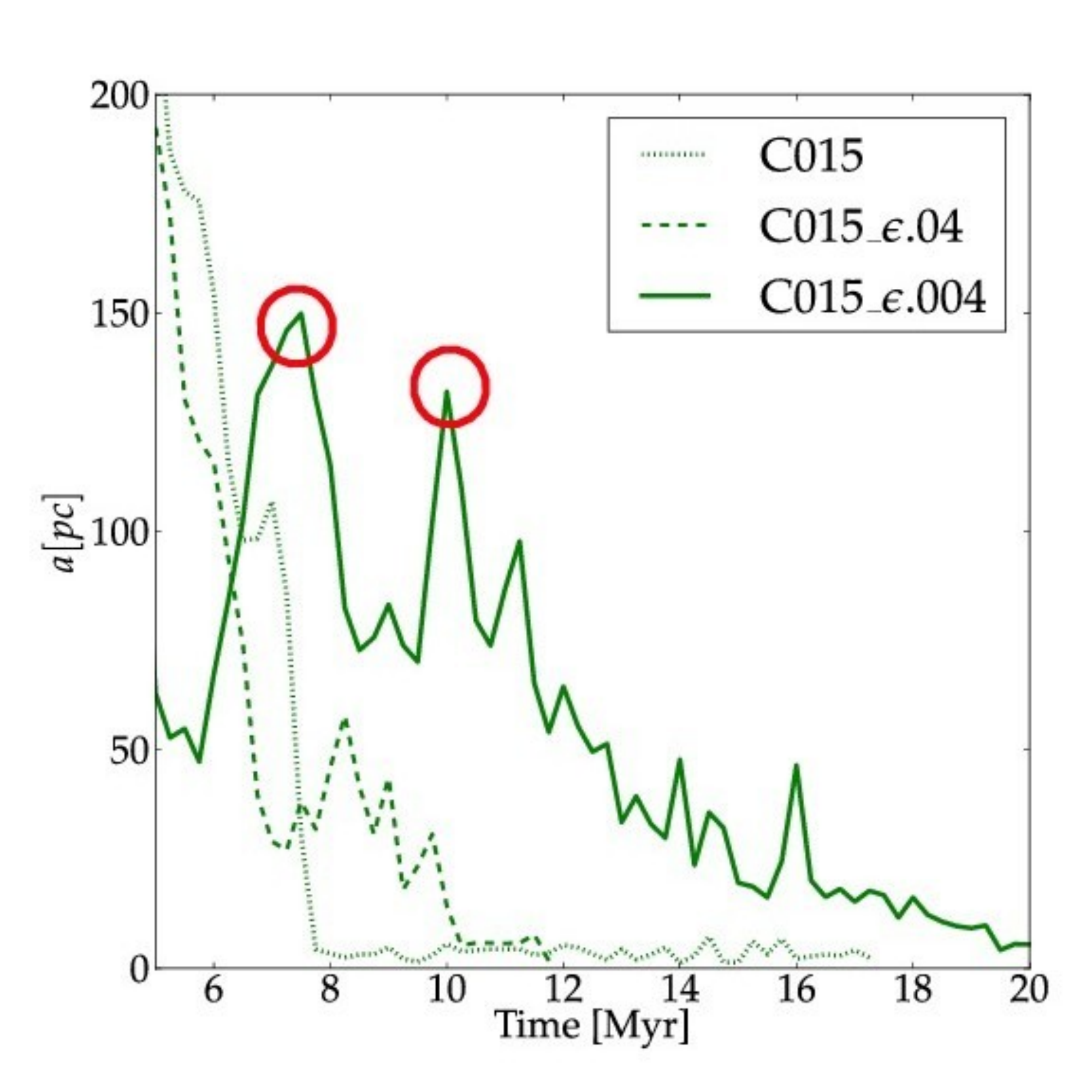}
    \caption{Same as figure \ref{fig12} for  $C_{\star}=0.15$.}
    \label{fig15}
\end{figure}

\begin{figure}
    \centering
    \includegraphics[width=0.45\textwidth]{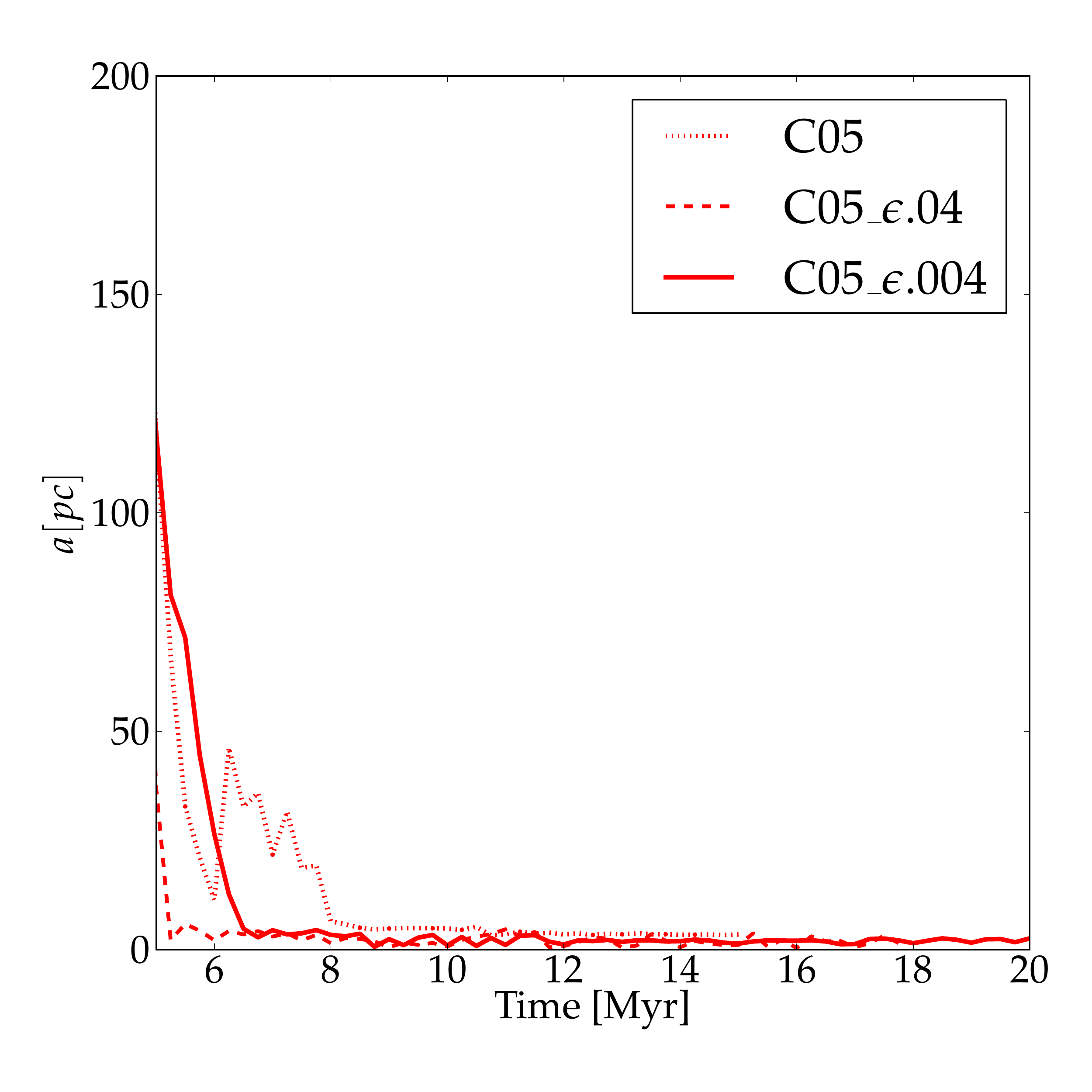}
    \caption{Same as figure \ref{fig12} for  $C_{\star}=0.5$.}
    \label{fig16}
\end{figure}

In figures \ref{fig12}, \ref{fig13}, \ref{fig14}, \ref{fig15}, and \ref{fig16}
we show the evolution of the SMBHs' separation for different values of
 $\epsilon_{\rm BH}$. 
The smallest value that we choose for  $\epsilon_{\rm BH}$ is still much greater than
 the Schwarzschild radius of the black holes in our simulations 
($R_{\rm sch}\,=\,4.78\times10^{-6}$ pc).
We note that, as we change $\epsilon_{\rm BH}$, the orbits of the black holes change.
So, in simulations with the same SFR but different $\epsilon_{\rm BH}$,
the black holes don't have close encounters with the same gaseous clumps.
We found that in three of our simulations (runs C0015\_$\epsilon$.004,
C005\_$\epsilon$.004, and C05\_$\epsilon$.004), making $\epsilon_{\rm BH}$ smaller
 causes the large fluctuations to disappear. 
As expected, the better resolution of the gravitational force of the black holes in these 
simulations allows them to disrupt the clumps in all close encounters.
In contrast, we found that in the runs C0005\_$\epsilon$.004 and C015\_$\epsilon$.004, some
large fluctuations in the separation of the pair occurred (see peaks enclosed by red circles 
in figures \ref{fig12} and \ref{fig15}) that are even larger than the fluctuations observed 
in the simulations with the same SFR, but larger $\epsilon_{\rm BH}$ (runs C0005 and C015). 
We analyzed each of these large fluctuations to determine why they are present even when 
the black hole's gravitational force is resolved down to scales of $0.004$ pc, where 
$F_{\rm BH}(\epsilon_{\rm BH})$ is larger.

\begin{figure*}
    \centering
    \includegraphics[height=3.6cm]{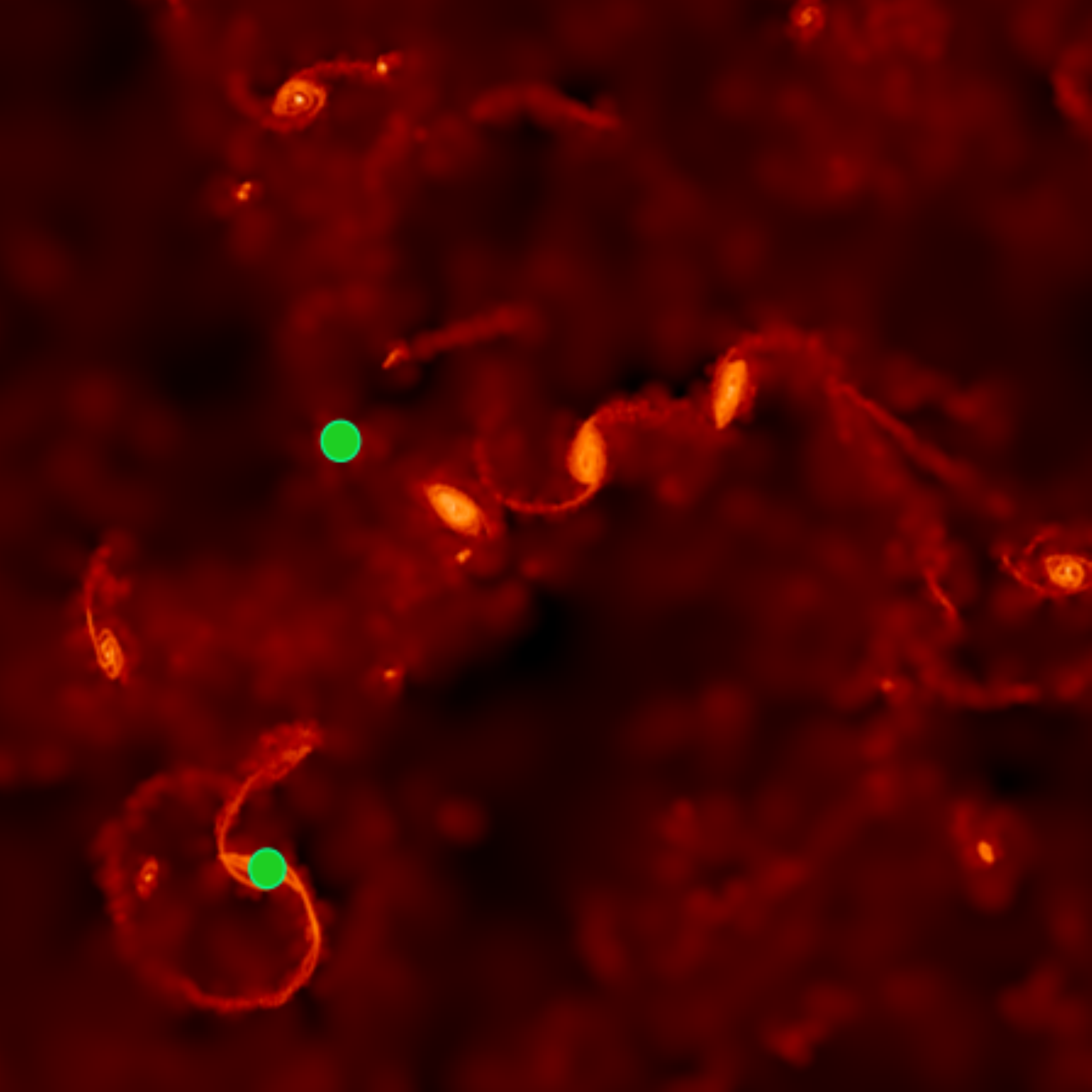}
    \includegraphics[height=3.6cm]{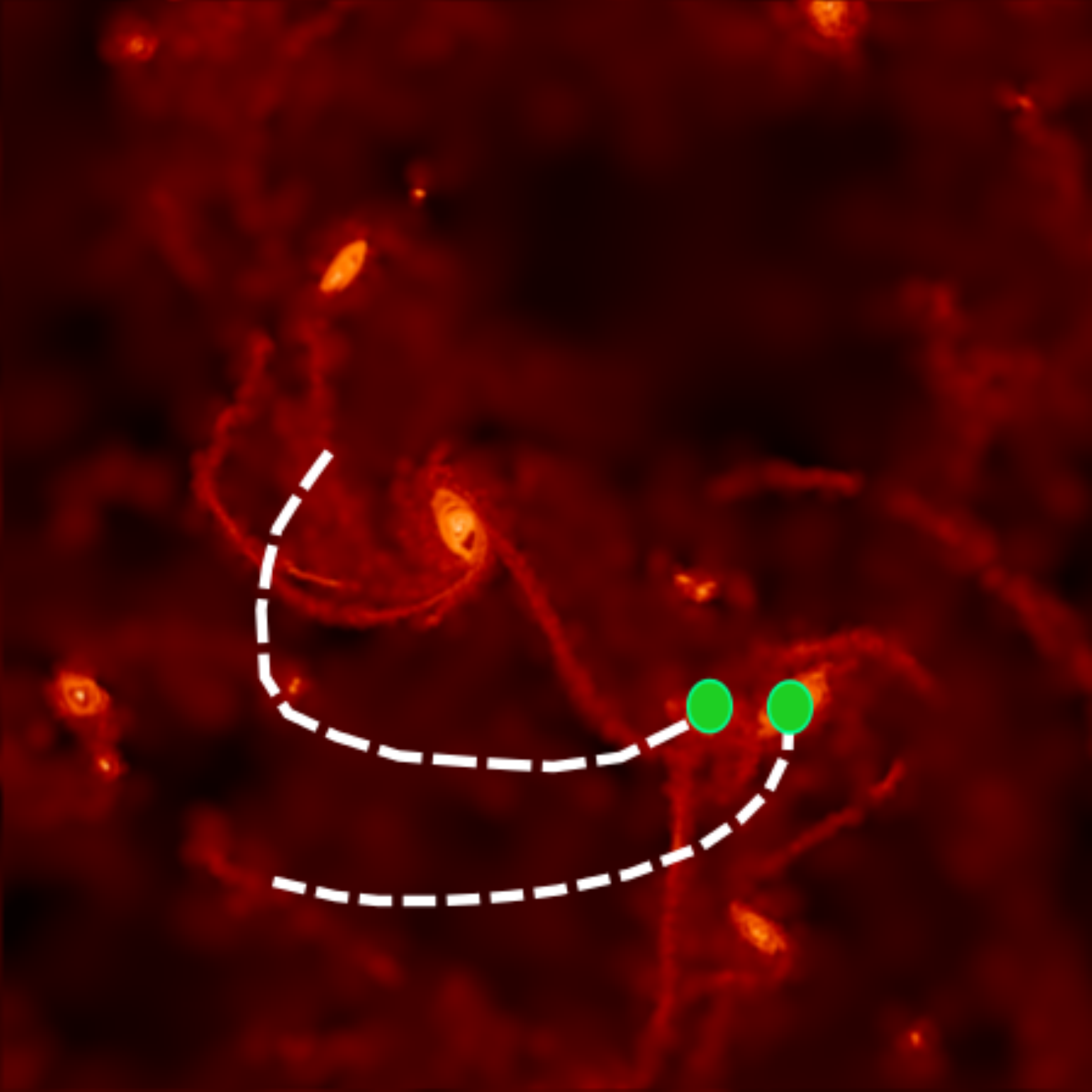}
    \includegraphics[height=3.6cm]{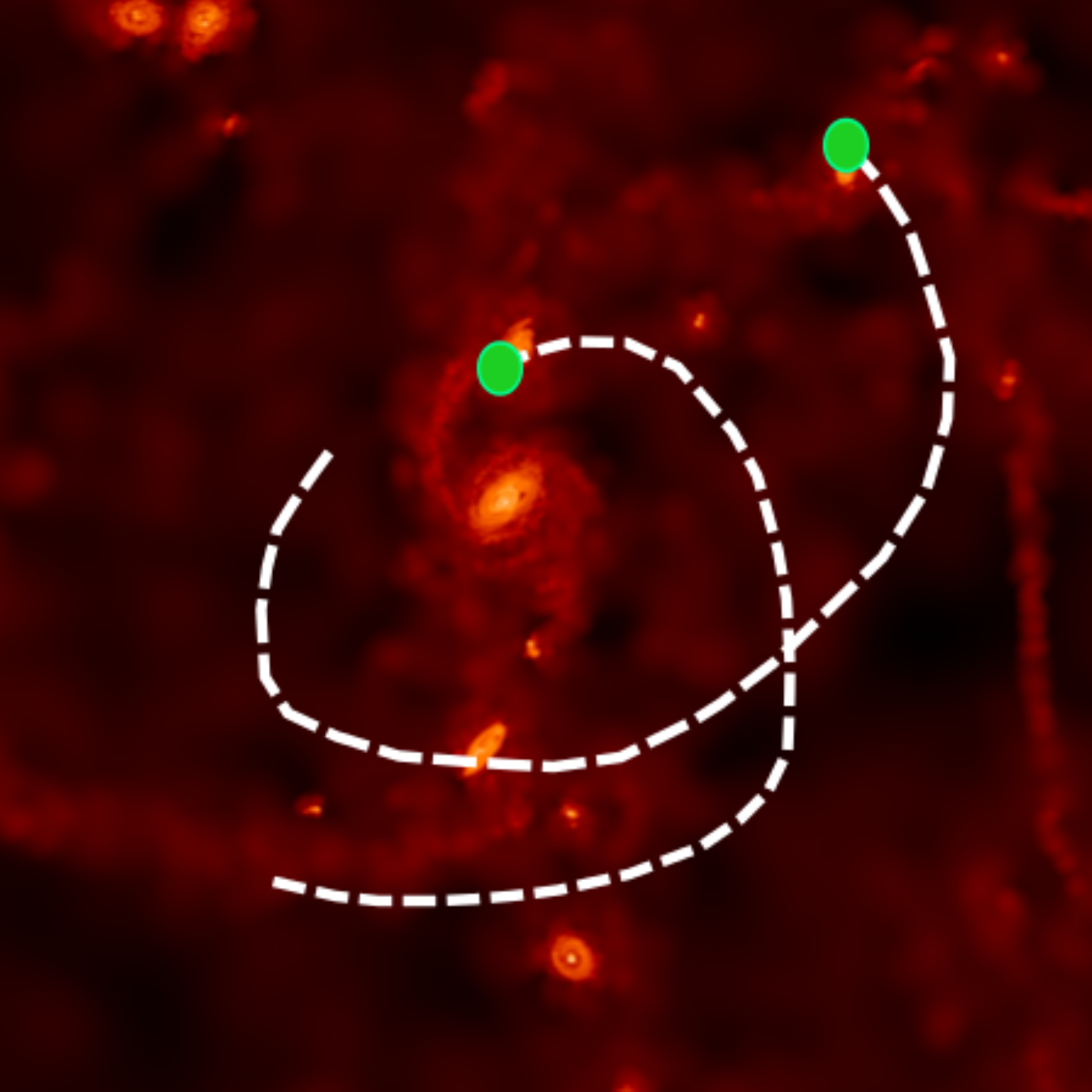}
    \includegraphics[height=3.6cm]{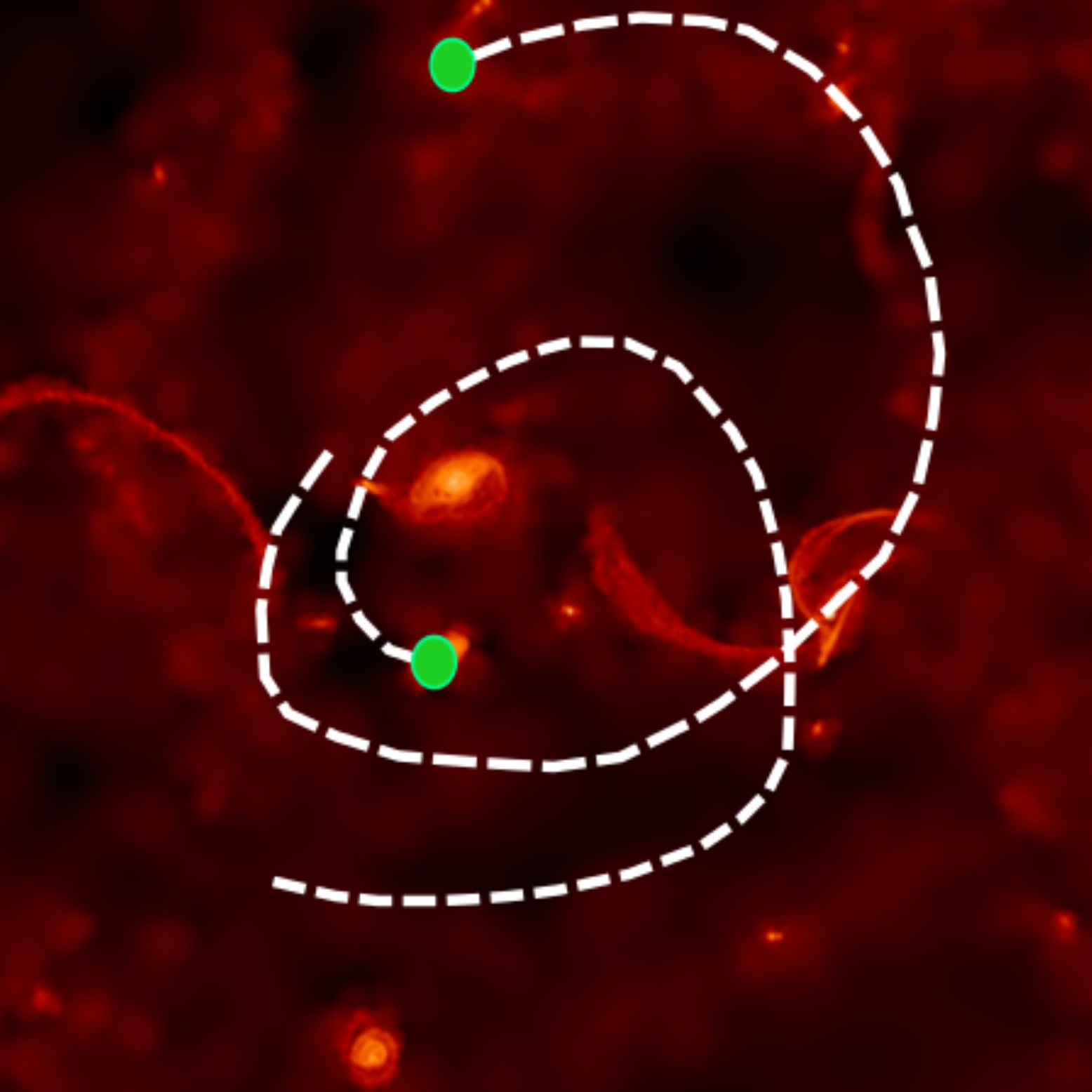}\\
    \includegraphics[height=3.6cm]{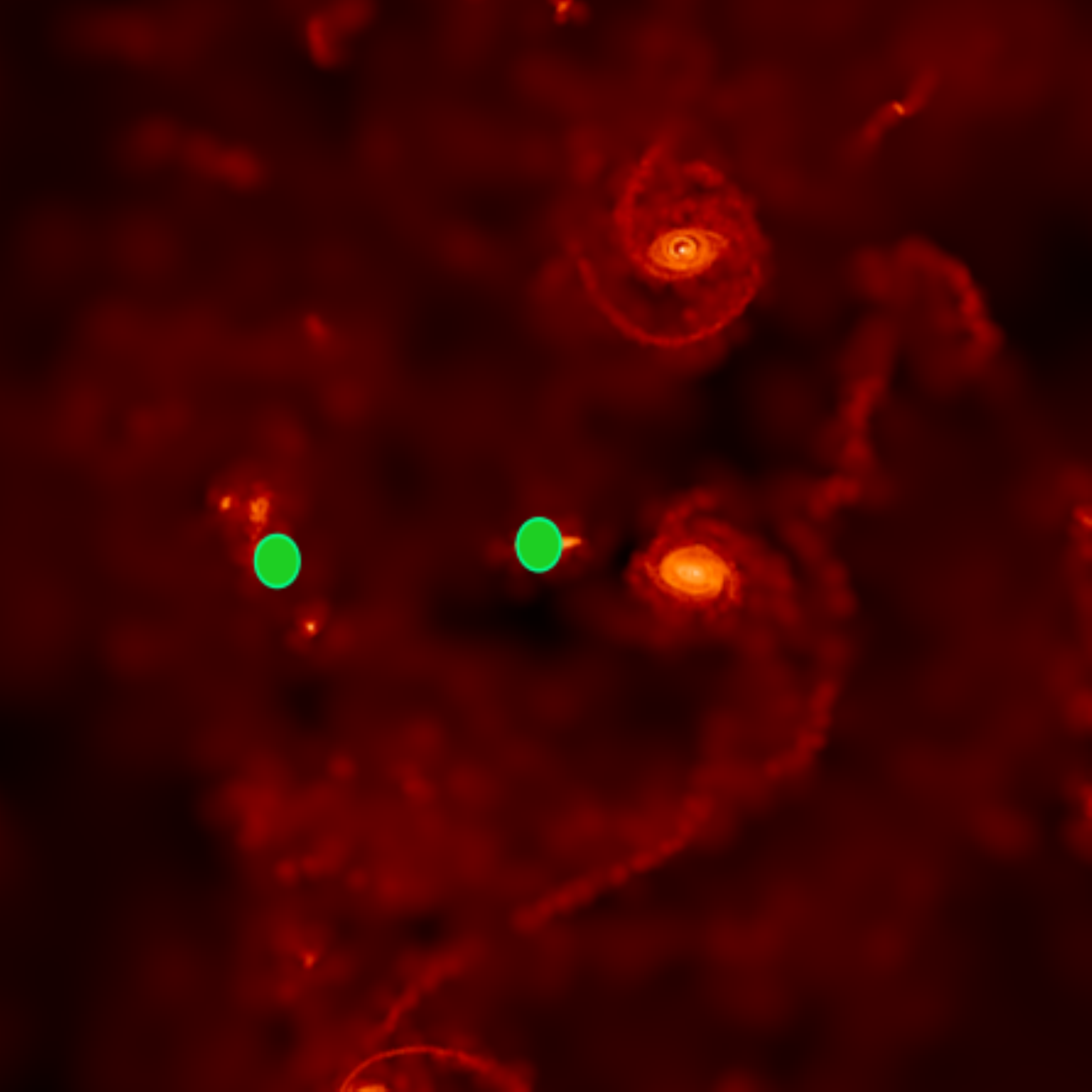}
    \includegraphics[height=3.6cm]{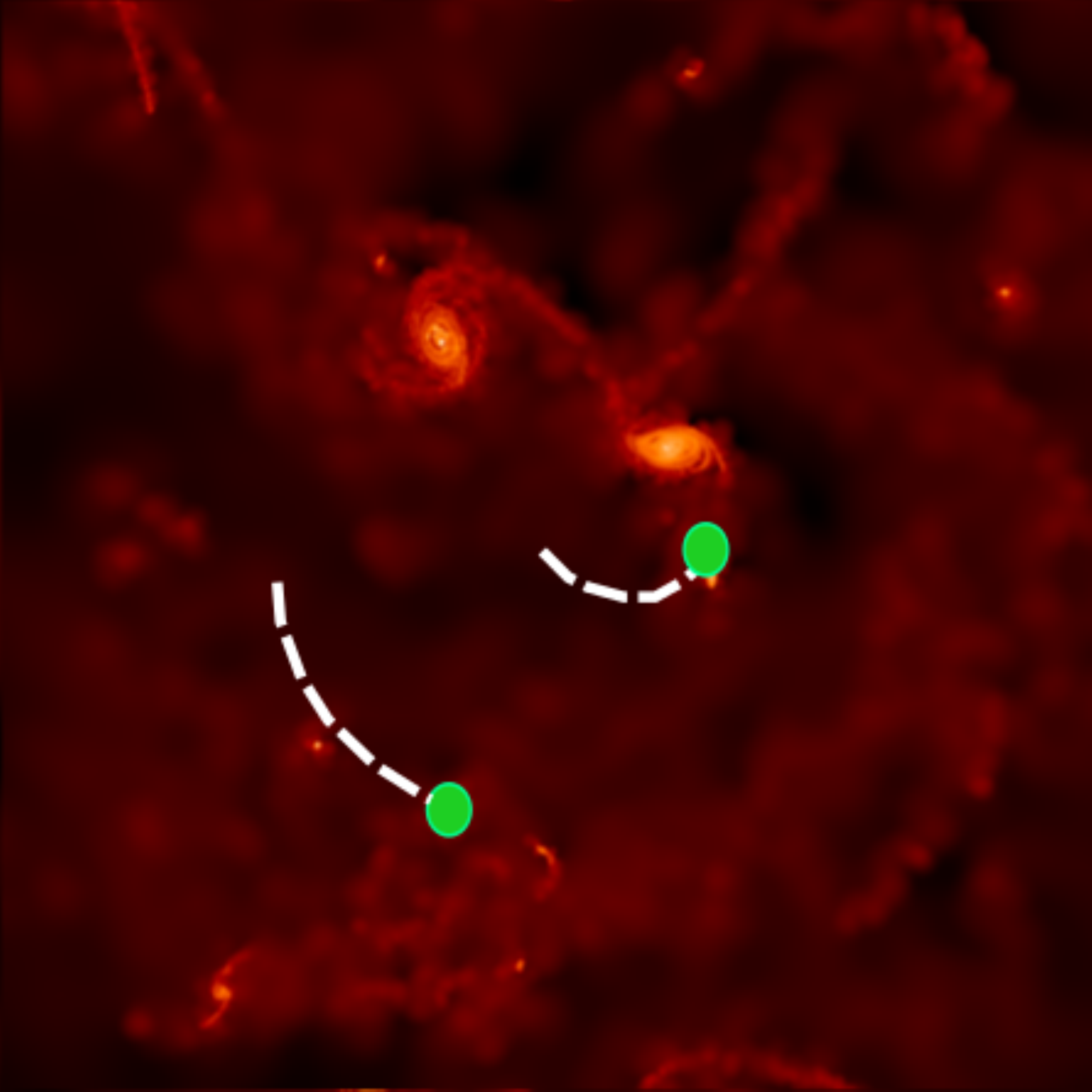}
    \includegraphics[height=3.6cm]{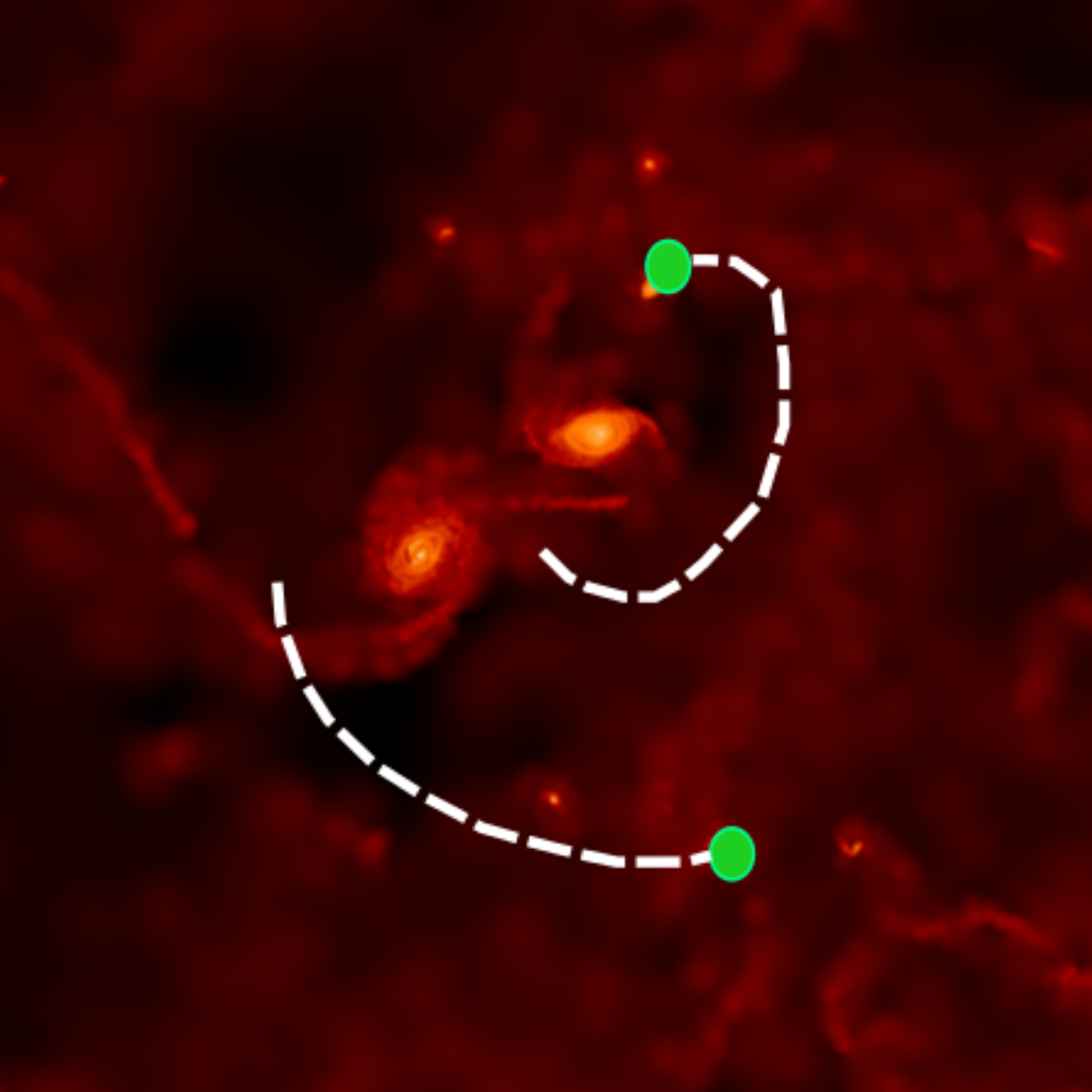}
    \includegraphics[height=3.6cm]{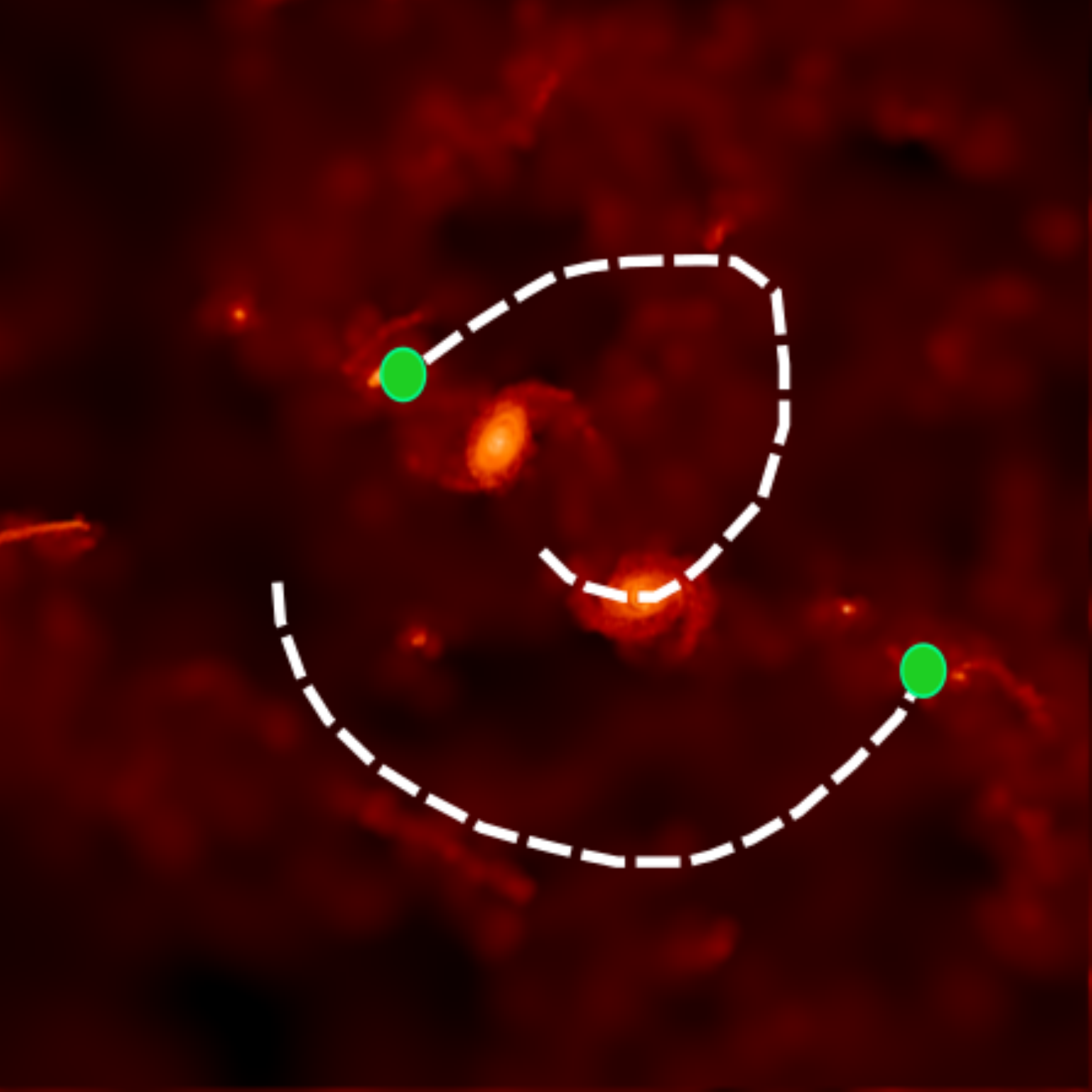}\\
    \caption{Distribution of gas in the plane of the disk, from left to right the
         time of the snapshots increases. The figures show the orbit of the black holes
         in the run C0005\_$\epsilon$.004. 
         The upper row shows the evolution during the first large fluctuation in the SMBHs' separation (first red circle in figure \ref{fig12}).
         The lower row shows the evolution during the second large fluctuation on the SMBHs' separation (second red circle in figure \ref{fig12}).
         Here we can see that the two large fluctuations are the result of the changing separation of the black holes 
	 as they orbit a common center, near the center of very dense and massive gas clump, with different orbital radii.
	 The fluctuations are not the result of close encounters with high density gas clumps.}
    \label{fig17}
\end{figure*}

In the case of run C0005\_$\epsilon$.004, we identify two large fluctuations (enclosed by red
circles in figure \ref{fig12}). In figure \ref{fig17}, we show the evolution of the orbits of the black holes 
during these two large fluctuations. The upper row of
 figure \ref{fig17} shows the orbits of the black holes during the first of
these large fluctuations, and the lower row of figure\ref{fig17} shows the second of these large fluctuations. 
We found that these two large fluctuations occurred because
the black holes follow orbits of different radius and eccentricity, but with respect to nearly
the same center, which is located close to the densest and most massive gas clump of the
simulation ($n_{\rm cl}\,=\,7.4\times10^8$cm$^{-3}$  and $M_{\rm cl}\,=\,8.8\times 10^8$ $\Msun$).
Therefore, the large fluctuations in run C0005\_$\epsilon$.004 are not the result of close encounters with high density clumps.
Instead they are the result of the movement of the black holes which, at the moment of these fluctuations, are not bound together.

\begin{center}
\begin{figure*}
    \centering
    \includegraphics[height=3.6cm]{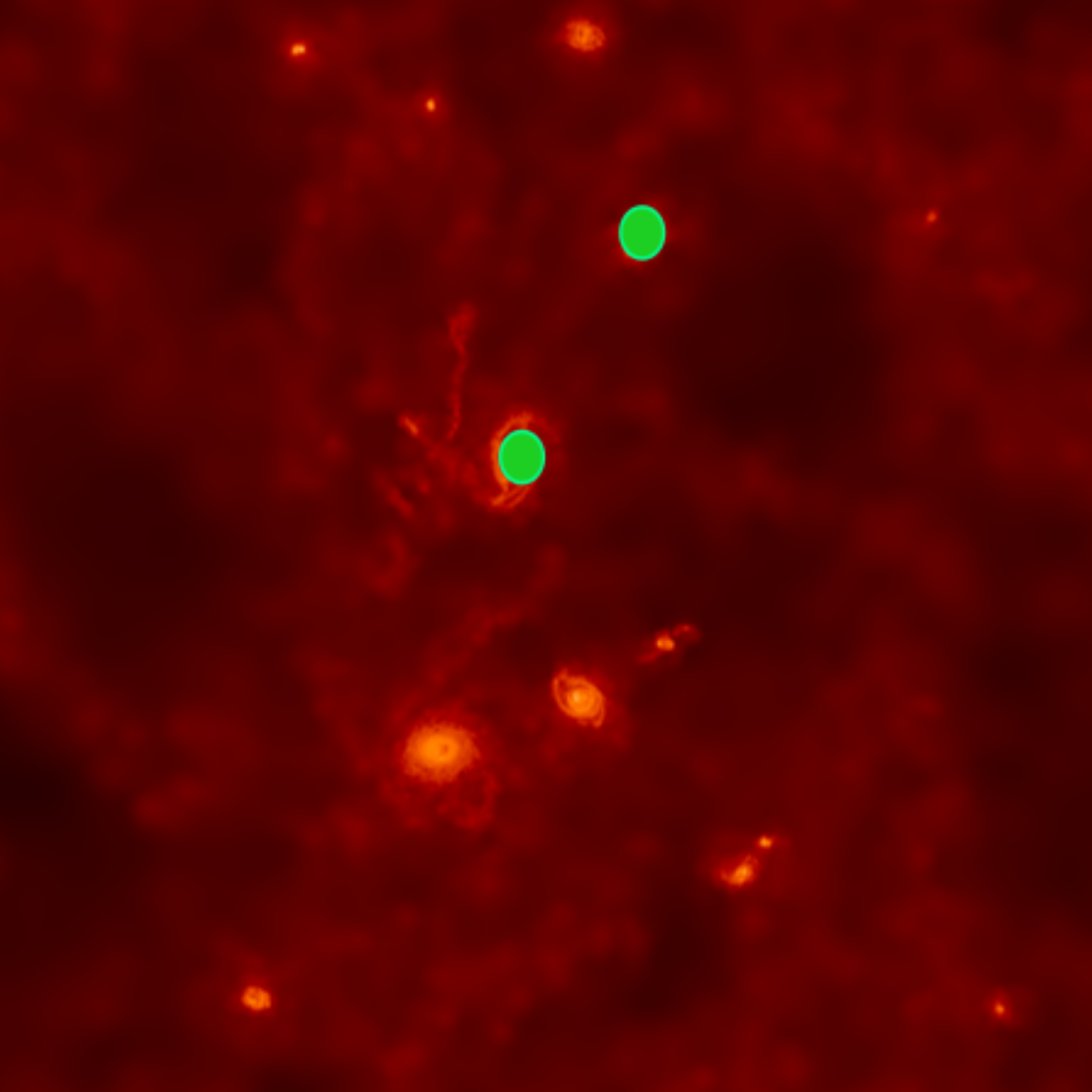}
    \includegraphics[height=3.6cm]{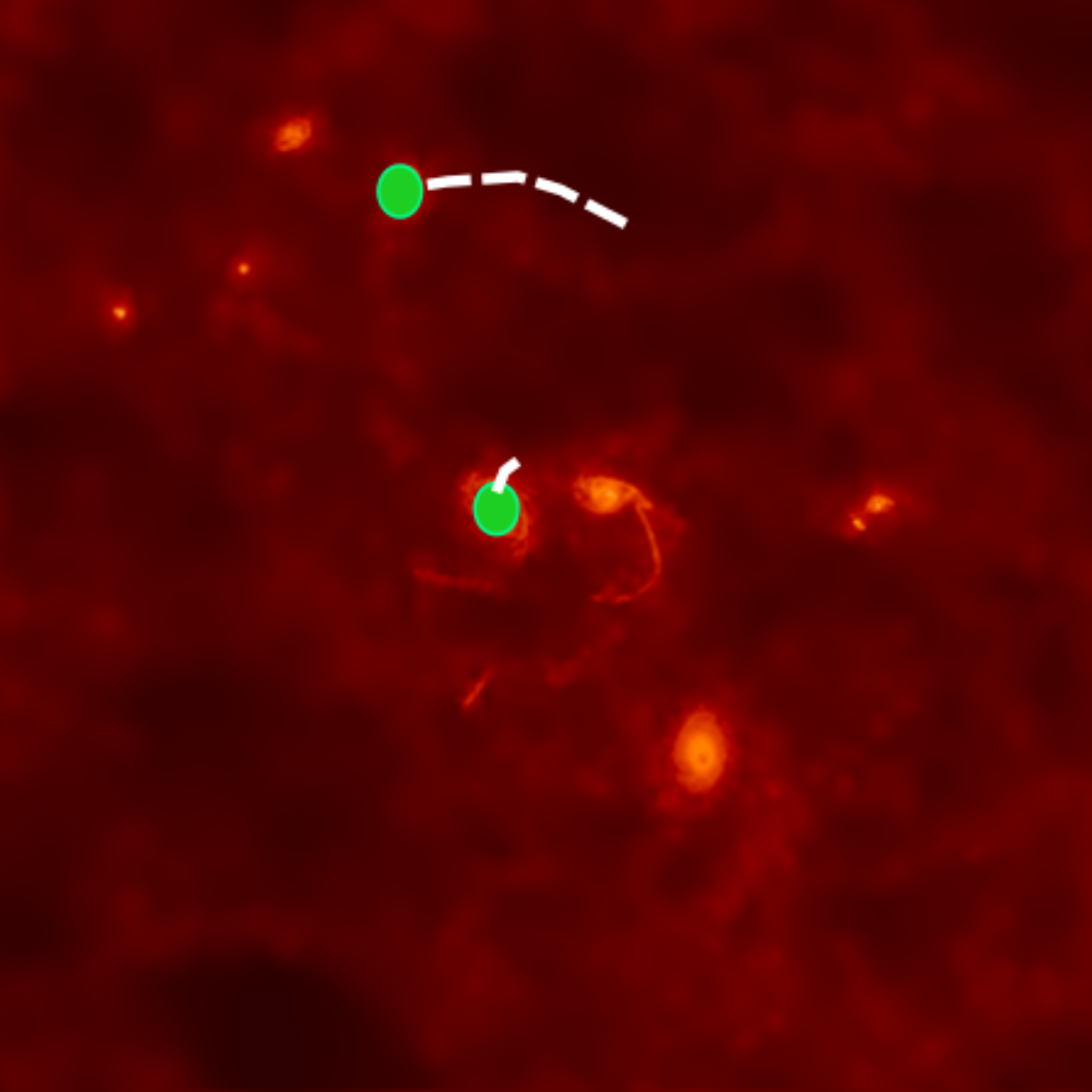}
    \includegraphics[height=3.6cm]{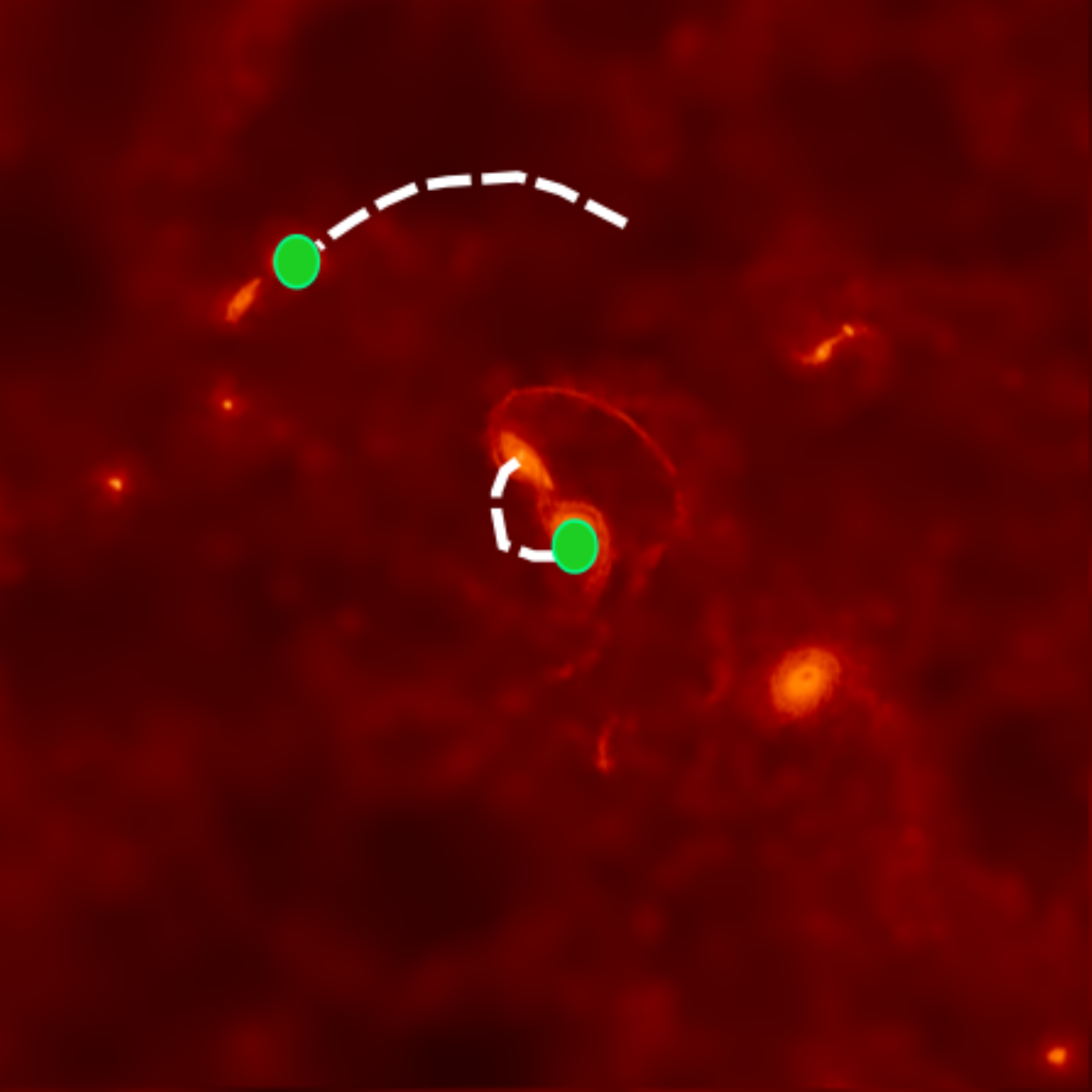}
    \includegraphics[height=3.6cm]{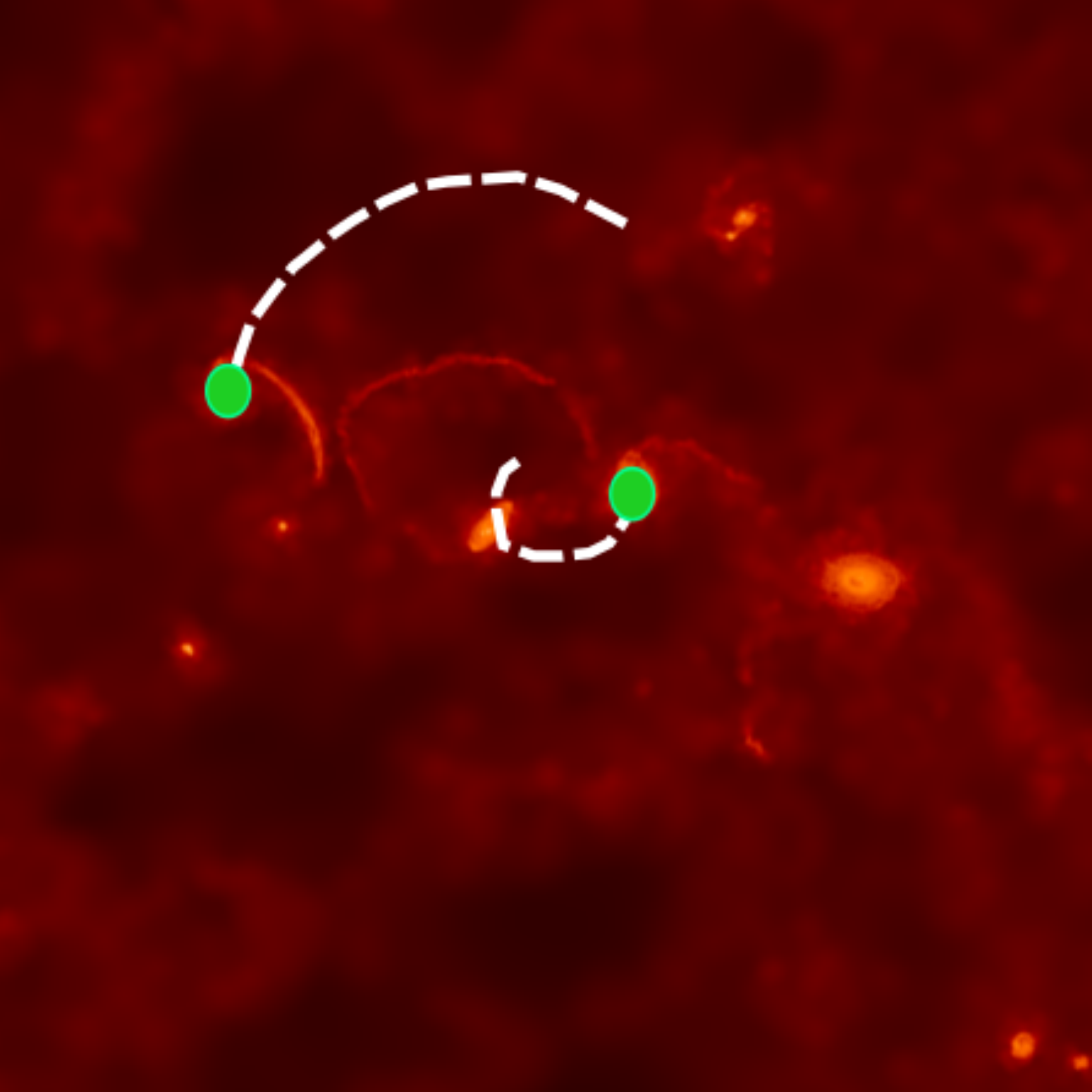}\\
    \includegraphics[height=3.6cm]{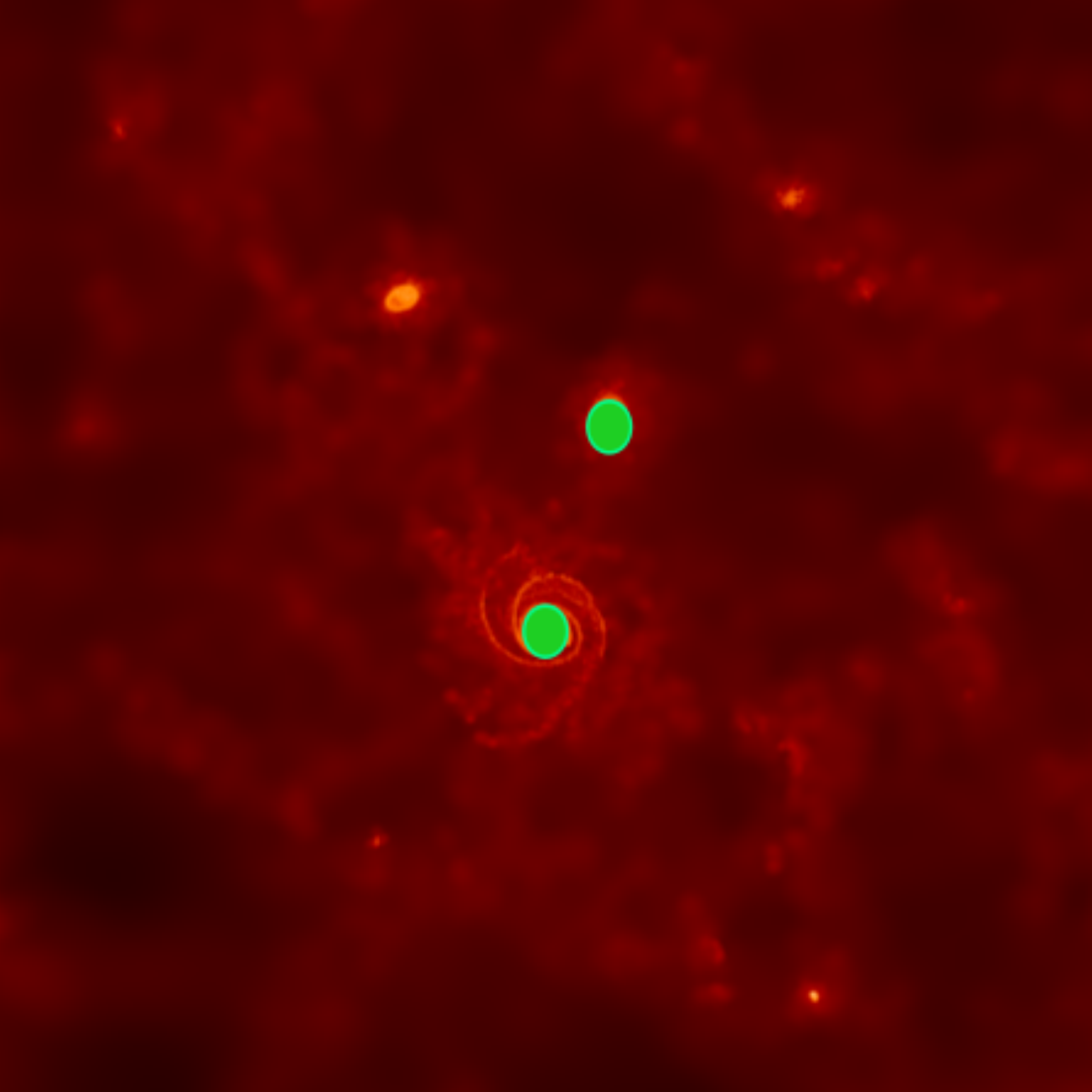}
    \includegraphics[height=3.6cm]{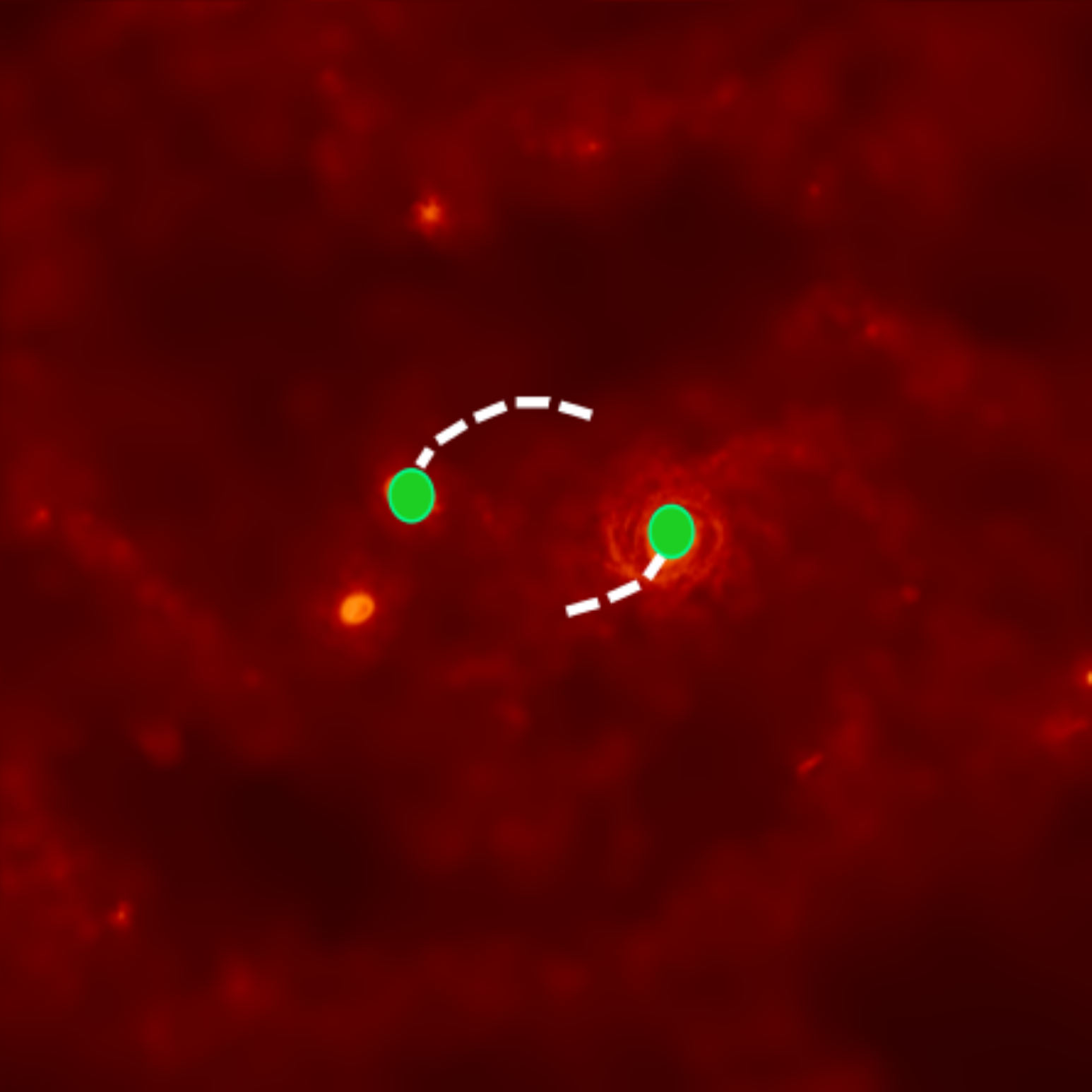}
    \includegraphics[height=3.6cm]{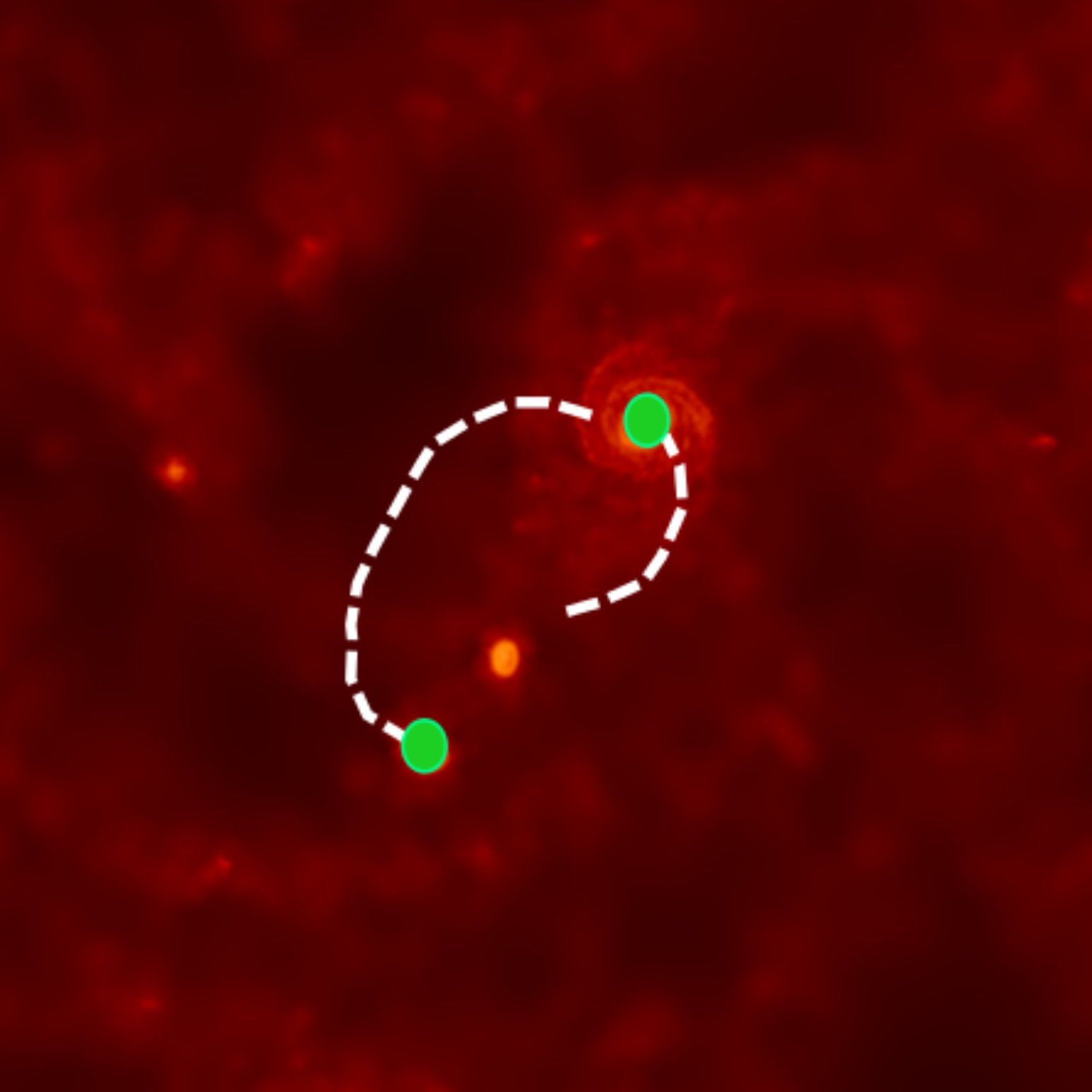}
    \includegraphics[height=3.6cm]{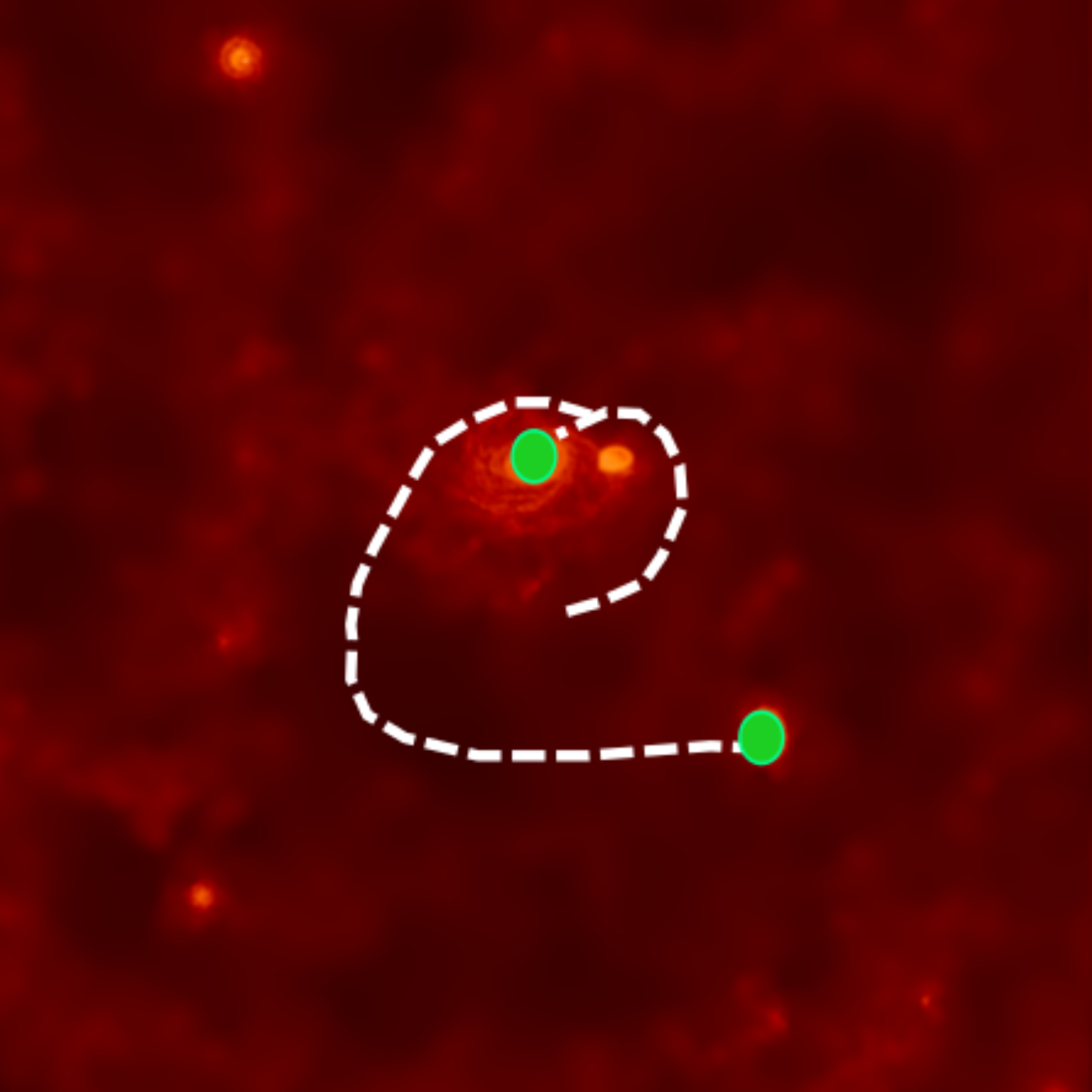}\\
    \caption{Distribution of gas in the plane of the disk, from left to right the
         time of the snapshots increases. The figures show the orbit of the black holes
         in the run C015\_$\epsilon$.004.
	 The upper row shows the evolution during the
         first large fluctuation on the SMBHs' separation (first red circle in figure \ref{fig15}).
         The lower row shows the evolution at the moment of the
         second large fluctuation on the SMBHs' separation (second red circle in figure \ref{fig15}).}
    \label{fig18}
\end{figure*}
\end{center}

Run C015\_$\epsilon$.004 also has two large fluctuations in the separation of 
the black holes (enclosed by red circles in figure \ref{fig15}).
We show the evolution of the orbit of the black holes during these two large 
fluctuations in figure \ref{fig18}. From the upper row of figure \ref{fig18}, we found
that the first large fluctuation is the result of scattering of one of the black holes due 
to gravitational interaction with a high density gas clump.
This gas clump had in turn been ejected from a past scattering with another high density gas clump.
In the lower row of figure \ref{fig18}, we see that the second large fluctuation is
the result of scattering with the same gas clump responsible of the first large fluctuation.
Both of these close encounters have a minimum distance which is greater than 10 pc, and therefore they 
are not artificially produced by low resolution of the gravitational force of the black holes. 
However, the clump that produces these two large fluctuations in the separation of the black holes 
is the most dense and massive clump in run C015\_$\epsilon$.004 ($3.5\times10^8$cm$^{-3}$ and $4\times 10^8$ $\Msun$).
So this clump is the most extreme clump in our simulation and hence, the two scatterings produced
by it are highly unlikely to happen in a real CND. We note that shortly after the last
scattering ($\sim$ 2 {\Myr}), this clump is disrupted by one of the black holes, and after
this disruption the orbital decay of the SMBH pair continues relatively smoothly.

Intense, fluctuating gravitational torques experienced by SMBH pairs in a 
clumpy medium have also been observed in numerical simulations by other authors
(Escala et al. 2004; Fiacconi et al. 2013; Roskar et al. 2014).
To compare our results with the ones obtained by Fiacconi et al. (2013), we estimate
the effective density of the black holes and the clumps of their simulations.
Using figure 3 of Fiacconi et al. (2013), we estimate that the density of their
clumps range between $0.8\times10^6$cm$^{-3}$ and $13\times10^6$cm$^{-3}$.
This estimate is a lower bound for the density, because we assume that 
the vertical size of their clumps is comparable to the thickness of their disk.
If we compare this density range with the effective density of their black holes
(which is $\rho_{\rm BH}\,=\,7$.$9\,m_{\rm H}\,\times10^7$cm$^{-3}$), we find that 
$\rho_{\rm BH}$ is typically greater than the density of their clumps and therefore 
their simulations don't suffer of a bad resolved black holes force.
However, from our estimation, we found that the density of these gas clumps 
is greater, or equal, than the density of the densest observed molecular clouds.
This means that, in these simulations the stochastic gravitational torques 
experienced by the black holes due to the gravitational interaction with the 
densest gas clumps can be overestimated.

We note that, although in our simulations the maximum density of the gaseous 
clumps is one or two order of magnitude greater, the delay on SMBH orbital 
decay is comparable with the obtained by Fiacconi et al. (2013) for SMBHs that 
are initially in a circular orbit. 

In Fiacconi et al. (2013) they found that the SMBH orbital decay 
is slower for SMBHs that are initially in a eccentric orbit. 
This dependence between the orbital decay timescale and the initial eccentricity 
may be caused by an indirect retarding effect on the SMBH orbital decay
which is stronger for SMBHs in an initially eccentric orbit.
This indirect effect is caused by the spiral arms produced by the 
high density gaseous clumps, they exert gravitational torques 
on the SMBHs and perturb the hydrodynamical wake of the SMBHs. 
We expect to found a similar behaviour in our simulations, 
however, as our simulations have star formation included,
the clumps' spiral arms may be transformed into stars and their 
indirect effect on the SMBH orbital decay may be smaller than the 
observed in simulations without star formation. 
Therefore, it is not clear if in our simulations a SMBH that is 
initially in an eccentric orbit will have a slower orbital 
decay than a SMBH that is initially in a circular orbit.

\section{Conclusions}

We studied the evolution of a pair of SMBHs embedded in a star forming CND 
to explore how different SFR and gas physics can result in 
different timescales for the orbital decay of the pair.
For this purpose, we ran twenty three N-body/SPH simulations. 
In eighteen of these simulations we used a modified version of the code {\tt{Gadget-3}}
 in which we implemented recipes for star formation, cooling and supernovae 
explosions that resemble the ones used by other authors in the study of SMBH pairs
 (Callegari et al. 2011; Van Wassenhove et al. 2012; Roskar et al. 2014).
The other two simulations used different gas physics. In one of these 
simulations we used the hybrid multiphase
model for star formation implemented in {\tt{Gadget-3}} Springel \& Hernquist 2003; Springel et al. 2005a),
where it is assumed that star formation is self regulated (simulation SDH05). 
In the other one, the gas thermodynamics is simpler and there is no star formation (simulation E05).

Before the SMBH pair forms a binary, their orbital decay is driven 
by dynamical friction coming from their gaseous environment.
As this dynamical friction is density dependent, we expect the 
orbital decay of the SMBHs to occur more quickly in simulations where 
the gas density of the CND is higher. 
The simulations we ran with our recipes resulted in a higher mean 
gas density around the SMBHs than simulation SDH05, and accordingly 
we found the SMBHs' orbital decay to be faster using our recipes.
However, in simulations with our recipes the orbital decay is sometimes slower than
in simulation E05, even though the gas density in these simulations is higher 
than in simulation E05.
This happen because in these simulations, the CND is fragmented in a few tens of high density
($\sim 10^6-10^9$ cm$^{-3}$) gaseous clumps, which erratically perturb the orbits of the SMBHs,
delaying their orbital decay.

The density of these gaseous clumps are extremely high compared
to the observed density of molecular clouds in isolated galaxies
(Mathis 1990; Struve \& Conway 2010) or ULIRGs (Downes \& Solomon 1998; Schulz et al. 2007),
which is typically on the order of $\sim 10^2-10^5$cm$^{-3}$.
The reason for this discrepancy is that we don't include all the different
physical process that sculpt the properties of the gas clumps.
For example, to properly model the formation of gas clumps in the
CND, we need to consider a more realistic cooling function which computes
the effects of photons trapped inside the optically thick clumps, or
follow the radiative transfer of these photons inside the clumps.
Also, we have to include heating due to the stars and the accretion disks
of the SMBHs, which can have an important effect on the temperatures and
densities of the gas clumps. As the gas clumps are regions of vigorous star
formation, we also expect that they will be highly turbulent. 
This turbulence, which would sustain the gas clumps against gravitational collapse, 
is also a missing ingredient in our simulations since it is damped in SPH codes
like {\tt{Gadget-3}}.

We also found that in simulations with our recipes,
the density of the gas clumps is comparable to or greater than the effective
mass density of the SMBHs, which we define as $\rho_{\rm BH}\,=\,3\,M_{\rm BH}/(4\pi \epsilon_{\rm BH}^3)$
(for $\epsilon_{\rm BH}\,=\,4$ pc, this density is $\rho_{\rm BH}\,=\,3.8\,m_{\rm H}\times10^6$cm$^{-3}$).
We ran simulations where the gravitational softening of the SMBHs is smaller,
and hence their effective mass density higher.
With these simulations we show that the outcome of a close encounter, within
a distance comparable with the gravitational softening, depends on the
force resolution of the SMBHs.
If the force resolution (i.e. gravitational softening) is such that the effective density
of the SMBHs is smaller than the density of the gas clumps, a close encounter
results in the scattering of the SMBH. On the other hand, if the force resolution
is such that the effective density of the SMBHs is higher than the density of the gas clumps,
a close encounter results in the tidal disruption of the gas clump and the SMBH
orbit is less affected.

Recently, it has been argued that the orbital evolution of a pair of SMBHs in a clumpy CND has
a stochastic behavior, due to the gravitational interaction of the SMBHs with high density
gaseous clumps (Fiacconi et al. 2013; Roskar et al. 2014).
However, in the simulations of these studies, as in our simulations, the densities of the 
denser gaseous clumps are higher than the densities of the observed, denser molecular clouds 
(Mathis 1990; Oka {\it et al} 2001; Struve \& Conway 2010). 
Thus, in these simulations the effect of the denser gaseous clumps on the orbital evolution 
of the SMBHs, which is density dependent, can be overestimated. 
So the delay produced on the orbital decay of the SMBHs by these gaseous clumps 
in real CNDs can be even smaller than the observed in simulations. 

We emphasize that even though in our simulations the orbit-perturbing gravitational torques
produced by these gaseous clumps are overestimated, the migration timescale is still at most 
a factor three longer than the migration timescale in simulations with simpler gas physics 
(Escala et al. 2005; Dotti et al. 2006). 
Even with this overestimation, gravitational interaction with the gas with the SMBHs 
produces a migration timescale which is much shorter than the typical migration timescale 
due to gravitational interaction with a background of stars.  
Interaction with gas yields a migration timescale on the order of $10^7$ yr, while migration
 timescale is on the order of one to ten Gyr in the case of a triaxial distribution of stars 
(Berczik et al. 2006; Khan et al. 2011).

In our simulations we assume that the orbits of the SMBHs are totally embedded 
in the CND, a scenario that is expected for massive CNDs. As we see in our simulations
the SMBHs can experience strong scatterings due to the gravitational interaction 
with high density clumps and we may expect that a similar type of scattering can 
affect the orbits of the SMBHs before they reach the CND, making their way into the CND 
more difficult.
For example, a SMBH can have a close encounter with a globular cluster, however, 
the densest globular clusters have densities on the order of $10^4$ cm$^{-3}$ 
and therefore, as the intensity of the scattering its density dependent,  
their effect on the orbits of the SMBHs will be much smaller than the observed
in our simulations.

Another scenario where we may expect that a SMBH experience an intense scattering 
is in a multiple-merger of galaxies, where multiples SMBHs can coexist in a 
compact region.
However, multiples SMBHs will coexist in a compact region only if the time interval 
between the mergers is smaller than the timescale in which two SMBHs coalesce.
If this is the case, the multiple interaction of SMBHs can slow down the 
SMBH orbital decay in an analogue way as the high density gaseous clumps 
in our simulations retard the orbital decay of the SMBH pair. 
For this scenario we also expect that, in a 3-body interaction some of the SMBHs 
will be ejected from the central region of the multiple-merger.
These kicked SMBHs may have interesting observational consequence for multiple-merger
systems like the formation of an AGN spatially offset form its host galaxy,
a signature which is usually associated to SMBHs ejected by gravitational
wave recoil kicks (Schnittman et al. 2008, Blecha et al. 2012). 
However, to determine the relevance of this scenario a detail 
statistic of the timescale between mergers in multiple-merger systems and the
dynamics of the gas and stars on these systems is needed.

From our numerical study and previous studies of binaries embedded in gaseous circumbinary disks  
(Escala et al. 2005; Dotti et al. 2006; Cuadra et al. 2009), we conclude that from the moment 
the SMBHs are separated by hundreds of parsecs until shortly after they form a binary, SFR 
has a much smaller effect on their orbits than cavity formation in the CND does.
(This cavity formation is the result of inefficient viscous dissipation of angular momentum extracted 
from the binary (del Valle \& Escala 2012, 2014).
Indeed, we show that for a two order of magnitude of difference on the SFR the migration timescale only
change in a factor of two, and in comparison, the formation of a cavity in the CND can extend the migration 
timescale by two orders of magnitude.
However, since we are limited by the resolution of our simulations, we don't explore for a sufficiently
long period the effects of the star formation on the evolution of the SMBHs after they form a binary.
Amaro-Seoane et al. (2013) have made advances in this direction by studying the evolution of an SMBH binary which
resides inside the central cavity of a star-forming gaseous circumbinary disk. Their results indicate that
star formation slows the orbital decay of the binary. However, they do not explore how star formation affects 
the evolution of an SMBH binary when it is embedded in a circumbinary disk without a gap, a regime in which 
the gravitational torque of the gas on the binary is stronger.

As a future work, the effects of the star formation on the orbital evolution of an SMBH 
binary embedded in a circumbinary disk without a gap should be explored. 
We also need to model the gas around the binary more realistically. 
Ultimately, we would like to determine if the gas can extract sufficient angular momentum 
from the SMBH binary to drive its separation down to scales small enough for gravitational 
radiation to take over, ensuring coalescence of the SMBHs.\\[12pt]

{\it Acknowledgments}.
L del V's research was supported by CONICYT-Chile through Grant D-21090518 and a
Redes (12-0021) grant, DAS Universidad de Chile and proyecto anillo de ciencia 
y tecnología ACT1101.
A.E. acknowledges partial support from the Center of Excellence in Astrophysics
and Associated Technologies (PFB 06), FONDECYT Regular Grant 1130458.
J.C. acknowledges support from FONDECYT Regular Grant 1141175.
The simulations were performed using the HPC clusters Markarian (FONDECYT 11090216)
from the DAS, Geryon(2) (PFB 06, QUIMAL 130008 and Fondequip AIC-57 )  at the
Center for Astro-Engineering UC and Damiana from the AEI.
We used splash (Price 2007) for the SPH visualization in
Figures 7, 8, 9, 15 and 16.
We are grateful to V. Springel for allowing us to use the {\tt{Gadget-3}} code.

\section*{References}
\begin{itemize}
\item[] Amaro-Seoane P., Brem P., Cuadra J., 2013, \apj, {\bf 764}, 14
\item[] Barnes J. E., 2002, \mnras, {\bf 333}, 481
\item[] Barnes J. E., Hernquist L., 1996, \apj, {\bf 471}, 115
\item[] Berczik P., Merritt D., Spurzem R., Bischof H.-P., 2006, \apjl, {\bf 642}, L21
\item[] Blecha L., Civano F., Elvis M. \& Loeb A., 2012, \mnras {\bf 428}, 1341
\item[] Callegari S., Kazantzidis S., Mayer L., Colpi M., Bellovary
\item[] J. M., Quinn T., Wadsley J., 2011, \apj, {\bf 729}, 85
\item[] Ceverino D., Klypin A., 2009, \apj, {\bf 695}, 292
\item[] Chandrasekhar S., 1943, \apj, {\bf 97}, 255
\item[] Cuadra J., Armitage P. J., Alexander R. D., Begelman M. C., 2009, \mnras, {\bf 393}, 1423
\item[] del Valle L., Escala A., 2012, \apj, {\bf 761}, 31
\item[] del Valle L., Escala A., 2014, \apj, 780, 84
\item[] Dotti M., Colpi M., Haardt F., 2006, \mnras, {\bf 367}, 103
\item[] Downes D., Solomon P. M., 1998, \apj, {\bf 507}, 615
\item[] Escala A., Larson R. B., Coppi P. S., Mardones D., 2004, \apj, {\bf 607}, 765
\item[] Escala A., Larson R. B., Coppi P. S., Mardones D., 2005, \apj, 630, 152
\item[] Fiacconi D., Mayer L., Roˇkar R., Colpi M., 2013, \apjl, {\bf 777}, L14
\item[] Garay G. {\it et al.}, 2004, \apj, {\bf 610}, 313
\item[] AGerritsen J. P. E., Icke V., 1997, aap, {\bf 325}, 972
\item[] Gultekin {\it et al.}, 2009, \apj, {\bf 698}, 198
\item[] Hopkins P. F., Quataert E., 2010, \mnras, {\bf 407}, 1529
\item[] Katz N., Weinberg D. H., Hernquist L., 1996, {\it apjs}, {\bf 105}, 19
\item[] Kennicutt Jr. R. C., 1998, \apj, {\bf 498}, 541
\item[] Khan F. M., Just A., Merritt D., 2011, \apj, {\bf 732}, 89
\item[] Kim H., Kim W.-T., 2007, \apj, {\bf 665}, 432
\item[] Kormendy J., Ho L. C., 2013, araa, {\bf 51}, 511
\item[] Lodato G., Nayakshin S., King A. R., Pringle J. E., 2009, \mnras, {\bf 398}, 1392
\item[] Lupi A., Haardt F., Dotti M., 2015, \mnras, {\bf 446}, 1765
\item[] Magorrian J {\it et al.}, 1998, aj, {\bf 115}, 2285
\item[] Mathis J. S., 1990, araa, {\bf 28}, 37
\item[] Mayer L., Kazantzidis S., Escala A., Callegari S., 2010, \nat, {\bf 466}, 1082
\item[] Mayer L. {\it et al.},  2007, {\it Science}, {\bf 316}, 1874
\item[] Medling A. M., U V., Guedes J., Max C. E., Mayer L., Armus L., Holden B., Roˇkar R., Sanders D., 2014, \apj, {\bf 784}, 70
\item[] Mihos J. C., Hernquist L., 1996, \apj, {\bf 464}, 641
\item[] Miller G. E., Scalo J. M., 1979, apjs, {\bf 41}, 513
\item[] Monaghan J. J., 1992, araa, {\bf 30}, 543
\item[] Mu\~noz D. J. {\it et al.}, 2007, \apj, {\bf 668}, 906
\item[] Oka, T., Hasegawa, T., Sato, F., {\it et al.} 2001, \apj, {\bf 562}, 348
\item[] Ostriker E. C., 1999, \apj, {\bf 513}, 252
\item[] Price D. J., 2007, pasa, {\bf 24}, 159
\item[] Raiteri C. M., Villata M., Navarro J. F., 1996, aap, {\bf 315}, 105
\item[] Richstone D. {\it et al.},  1998, \nat, {\bf 395}, A14
\item[] Roskar, R., Fiacconi. D., Mayer, L., et al. 2014, MNRAS, 449, 494
\item[] Saitoh T. R. {\it et al.}, 2010, in Astronomical Society of the Pacific Conference Series, Vol. 423, Galaxy Wars: Stellar Populations and Star Formation in Interacting Galaxies, Smith B., Higdon J., Higdon S., Bastian N., eds., p. 185
\item[] Saitoh T. R. {\it et al.}, 2008, pasj, {\bf 60}, 667
\item[] Sanders D. B., Mirabel I. F., 1996, araa, {\bf 34}, 749
\item[] Schulz A. {\it et al.}, 2007, aap, {\bf 466}, 467
\item[] Schnittman, J.D. 2007, \apj, {\bf 667}, 133
\item[] Springel V., 2005, \mnras, {\bf 364}, 1105
\item[] Springel V., Di Matteo T., Hernquist L., 2005a, \mnras, {\bf 361}, 776
\item[] Springel V., Hernquist L., 2003, \mnras, {\bf 339}, 289
\item[] Springel V. {\it et al.},  2005b, \nat, {\bf 435}, 629
\item[] Springel V., Yoshida N., White S. D. M., 2001, \nat, {\bf 6}, 79
\item[] Stinson G. {\it et al.}, 2006, \mnras, {\bf 373}, 1074
\item[] Struve C., Conway J. E., 2010, aap, {\bf 513}, A10
\item[] Ueda J.{\it et al.}, 2014, apjs, {\bf 214}, 1
\item[] Van Wassenhove S., Volonteri M., Mayer L., Dotti M., Bellovary J., Callegari S., 2012, apjl, {\bf 748}, L7
\item[] Wadsley J. W., Stadel J., Quinn T., 2004, \nat, {\bf 9}, 137
\item[] White S. D. M., Frenk C. S., 1991, \apj, {\bf 379}, 52
\end{itemize}

\end{document}